\documentclass[useAMS,usenatbib]{mn2e}

\newcommand{\rgas}{\ensuremath{r_{\mathrm{gas}}}}
\newcommand{\rgasHI}{\ensuremath{r_{\mathrm{\hi}}}}
\newcommand{\fgas}{\ensuremath{f_{\mathrm{gas}}}}

\newcommand{\hi}{\ensuremath{\mathrm{H\,\textsc{i}}}}
\newcommand{\subhi}{\ensuremath{\mathrm{\scriptsize H\,\textsc{i}}}}
\newcommand{\mugas}{\ensuremath{\Sigma_{\mathrm{gas}}}}
\newcommand{\muHI}{\ensuremath{\Sigma_{\mbox{\scriptsize H\textsc{i}}}}}

\newcommand{\Mstar}{\ensuremath{M_*}}
\newcommand{\Msun}{\ensuremath{\mathrm{M}_\odot}}
\newcommand{\hmol}{\ensuremath{\mathrm{H}_2}}
\newcommand{\hii}{H\,\textsc{ii}}
\newcommand{\CO}{\ensuremath{\mathrm{CO}}}

\newcommand{\oii}[1]{[\textsc{O\,ii}]#1}
\newcommand{\oiii}[1]{[\textsc{O\,iii}]#1}
\newcommand{\nii}[1]{[\textsc{N\,ii}]#1}
\newcommand{\sii}[1]{[\textsc{S\,ii}]#1}
\newcommand{\ha}{\ensuremath{\mathrm{H}\alpha}}
\newcommand{\hb}{\ensuremath{\mathrm{H}\beta}}
\newcommand{\DGR}{\ensuremath{\delta_{\mathrm{DGR}}}}
\newcommand{\XCO}{\ensuremath{X_{\CO}}}
\newcommand{\alphaCO}{\ensuremath{\alpha_{\CO}}}
\newcommand{\sSFR}{\ensuremath{\mathrm{sSFR}}}
\usepackage{graphicx}

\begin{document}

\title[Estimating gas masses and dust-to-gas ratios from optical
spectroscopy]{Estimating gas masses and dust-to-gas ratios from
  optical spectroscopy} \author[Brinchmann et al]{Jarle
  Brinchmann$^{1,2}$\thanks{jarle@strw.leidenuniv.nl}, St{\'e}phane
  Charlot$^3$, Guinevere Kauffmann$^4$, \newauthor Tim Heckman$^5$,
  Simon D.\ M.\ White$^4$, Christy Tremonti$^6$
  \\
  $^{1}$Leiden Observatory, Leiden University, P.O. Box 9513, 2300 RA, Leiden, The Netherlands\\
  $^2$Centro de Astrofisica, Universidade do Porto, Rua das Estrelas,
  4150-762 Porto, Portugal \\
  $^3$UPMC-CNRS, UMR7095, Institut d'Astrophysique de Paris, F-75014, Paris, France  \\
  $^4$Max-Planck Institut f{\"u}r Astrophysik, 85741 Garching, Germany
  \\
  $^5$ Johns Hopkins University, Baltimore, Maryland 21218, USA \\
  $^6$ Department of Astronomy, University of Wisconsin-Madison, 1150
  University Ave, Madison, WI 53706, USA \\
  \\
}

\maketitle

\label{firstpage}

\begin{abstract}
  We present a method to estimate the total gas column density,
  dust-to-gas and dust-to-metal ratios of distant galaxies from
  rest-frame optical spectra. The technique exploits the sensitivity
  of certain optical lines to changes in depletion of metals onto dust
  grains and uses photo-ionization models to constrain these physical
  ratios along with the metallicity and dust column density.  We
  compare our gas column density estimates with \hi\ and CO gas mass
  estimates in nearby galaxies to show that we recover their total gas
  mass surface density to within a factor of 2 up to a total surface
  gas mass density of $\sim 75\; \Msun\;\mathrm{pc}^{-2}$. Our technique
  is independent of the conversion factor of CO to \hmol\ and we show
  that a metallicity dependent \XCO\ is required to achieve good
  agreement between our measurements and that provided by CO and \hi.
  However we also show that our method can not be reliably aperture
  corrected to total gas mass.  We calculate dust-to-gas ratios for
  all star-forming galaxies in the Sloan Digital Sky Survey Data
  Release 7 and show the resulting dependence on metallicity agrees
  well with the trend inferred from modelling of the dust emission of
  nearby galaxies using far-IR data.  We also present estimates of the
  variation of the dust-to-metal ratio with metallicity and show that
  this is poorly constrained at metallicities below 50 per cent solar.  We
  conclude with a study of the inventory of gas in the central
  regions, defined both in terms of a fixed physical radius and as a
  fixed fraction of the half-light radius, of $\sim 70,000$
  star-forming galaxies from the Sloan Digital Sky Survey. We show
  that their central gas content and gas depletion times are not
  accurately predicted by a single parameter, but in agreement with
  recent studies we find that a combination of the stellar mass and
  some measure of central concentration provides a good predictor of
  gas content in galaxies.  We also identify a population of galaxies
  with low surface densities of stars and very long gas depletion
  times.
\end{abstract}

\begin{keywords}
galaxies -- galaxies: evolution -- galaxies: individual -- galaxies: fundamental
parameters -- galaxies: ISM 
\end{keywords}

\section{Introduction}
\label{sec:intro}

The baryonic content of a galaxy exists in a range of states, from
cold molecular gas to diffuse hot gaseous haloes, yet for most
galaxies our knowledge of the baryon content is biased towards the
stars. Indeed, the last decade has seen a massive increase in our
understanding of the stellar content of galaxies over a wide range in
redshift \citep[e.g.][for a
review]{2000ApJ...536L..77B,Dickinson2003,2008MNRAS.388..945B,2009ApJ...701.1765M,2011ARA&A..49..525S}. Stellar
masses and star formation rates are regularly obtained from broad-band
photometry or optical spectroscopy for large samples of galaxies
\citep[e.g.][]{2001ApJ...550..212B,2003MNRAS.341...33K,2004MNRAS.351.1151B,2007ApJS..173..267S,2007AJ....133..734B,2012MNRAS.422.3285P,2012arXiv1207.6114M},
but to fully understand the baryon cycle in galaxies we need to also
understand the distribution of cold gas in galaxies.

The atomic gas content of galaxies can be traced using the 21cm line
of \hi, and integrated \hi\ masses are now available for large samples
of galaxies from \hi-selected surveys
such as HIPASS \citep{2001MNRAS.322..486B} and ALFALFA
\citep{2011AJ....142..170H} and targeted surveys such as the GASS
survey \citep{2012arXiv1206.3059C}, which has a uniform gas fraction
limit rather than a gas mass limit. Resolved \hi\ maps are available
for smaller, but still substantial samples of galaxies such as WHISP
\citep{2001ASPC..240..451V}, the Ursa Major Cluster survey
\citep{1996AJ....112.2471T,2001A&A...370..765V} and the THINGS survey
\citep{2008AJ....136.2563W}. 

The cold molecular gas is mostly composed of \hmol, but given its lack
of transitions that can be excited at low temperatures, molecular gas
measurements are mostly done using \CO\ as a proxy for
\hmol. Single-dish \CO\ surveys have generally been targeted
(e.g. Braine et al 1993; Boselli 1997) but for somewhat inhomogeneous
samples. A notable exception here is the COLDGASS
survey
\citep{2011MNRAS.415...32S} which obtained \CO\ observations for
galaxies from the GASS survey. As with \hi, resolved \CO\ studies have
a long history
\citep[e.g.][]{1991ARA&A..29..581Y,1995ApJS...98..219Y}, recent
examples are The Berkeley Illinois Maryland Association Survey of
Nearby Galaxies \citep[BIMA SONG; ][]{2003ApJS..145..259H}
interferometric maps of 44 spiral galaxies, the Nobeyama survey of 40
spiral galaxies \citet{2007PASJ...59..117K} and the HERA CO Line
Extragalactic Survey (HERACLES) by \citet{2009AJ....137.4670L} which
mapped 18 nearby galaxies with extensive multi-wavelength data.

These methods are powerful but they have important drawbacks: While
\hi\ traces the dominant ingredient in the atomic gas, it is very
challenging to detect significant numbers of individual galaxies in
\hi\ past $z\sim 0.3$ with current facilities
\citep[e.g.][]{2012ApJ...756L..28J,2010arXiv1009.0279V,2008ApJ...685L..13C}. In
contrast, \CO\ observations can be done out to significant redshifts,
but they do not trace the dominant component of the molecular gas and
a conversion factor between the column density of \CO\ and \hmol\
(\XCO) or CO luminosity and total molecular mass (\alphaCO) must be
used. Both \alphaCO\ and \XCO\ are believed to vary considerably with
metallicity and ionisation conditions
\citep[e.g.][]{1997A&A...328..471I,2002A&A...384...33B,2005A&A...438..855I,2009ApJ...702..352L},
although a quantitative understanding of this effect is only now
becoming available
\citep[e.g.][]{2011ApJ...737...12L,2012ApJ...746...69G}.

In part due to these constraints, and also because \hi\ and \CO\
surveys are very time-consuming, alternative methods have been
proposed to infer gas content in galaxies.  These can be roughly
divided into methods that make use of scaling relations to infer gas
content, and those that measure gas content using a more distant
tracer than \CO.  The former class has a long history and sprang out
of the realisation that there is a good correlation between galaxy
colour and the ratio of \hi\ mass to B-band luminosity
\citep{1969AJ.....74..859R,1984ApJ...277..532B}, or alternatively
between galaxy morphology and $M_{\subhi}/L_B$
\citep{1994ARA&A..32..115R}.  This technique was used by
\citet{2003ApJ...585L.117B} to infer a baryonic mass function for the
low redshift, $z<0.2$, Universe. Similarly
\citet[][K04]{2004ApJ...611L..89K} and
\citet[][Z09]{2009MNRAS.397.1243Z} have provided relations between
colour and \hi\ gas-fractions which can be applied to
large samples of galaxies. This has the advantage of being easy to
apply to large samples of objects
\citep[e.g.][]{2010MNRAS.409..491K,2012MNRAS.tmp..425Z}. A similar technique in
this class is to invert the empirically found relationship between
star formation rate and gas surface density, the Kennicutt--Schmidt
relation \citet[][and references therein]{1998ApJ...498..541K}. This
requires an estimate of the star formation rate which is then used to
infer a gas mass
\citep[e.g.][]{2006ApJ...646..107E,2004ApJ...613..898T,2010A&A...510A..68P}. These
relations do however rely on local calibration samples and are
therefore of questionable use for applications at higher redshift.
Even more importantly they are calibrated to give average trends and
are not suitable for detecting outliers or peculiar systems. We will
discuss these issues further in section~\ref{sec:comp-other-techn}.

The second class of methods aims instead to infer gas content via a
tracer of gas content. The most widely used technique in this class is
to make use of dust emission to infer gas content, building on the
observation by \citet{1985A&A...144L...9B} that a comparison of IRAS
and \hi\ maps implied that dust is a reliable tracer of cold gas. It
has been used to estimate gas content in the Milky Way
\citep[e.g.][]{2001ApJ...547..792D}, and in nearby galaxies by a
number of authors
\citep[e.g.][]{1997A&A...317...65I,1997A&A...328..471I,2005A&A...438..855I,2009ApJ...702..352L,2013arXiv1303.1178B}
and it has been extended to sub-mm observations
\citep{2002MNRAS.335..753J} and more recently to Herschel data
\citep[e.g.][see also Roman-Duval et al
2010]{2010A&A...518L..62E,2012arXiv1202.0547E}\nocite{2010A&A...518L..74R}. These
models depend on dust properties, such as the emissivity and
temperature, and on the typical beam sizes for sub-mm and far-IR
observations which are considerably larger than the resolution obtained in
the optical. Modelling the dust emission can, however, yield a plethora
of additional information
\citep[e.g.][]{2007ApJ...663..866D,2008MNRAS.388.1595D}. Of particular
interest to this paper is the ability to constrain the dust-to-gas
ratio of the interstellar medium \citep[e.g.][]{2012arXiv1207.4186A}.

Here we will discuss a related technique, but instead of exploiting
the dust \emph{emission}, we will make use of the dust
\emph{absorption}. As
we will discuss below this can be applied to large samples of
galaxies, is relatively insensitive to the detailed properties of the
dust grains, and like the dust emission measure, is sensitive to the
\emph{total} gas content and dust-to-gas ratio. At low redshift it
therefore offers a promising complement to direct detection of gas for
characterising its properties, while at higher redshift it offers a
simple method to measure the gas content and dust-to-gas ratios of
large samples of galaxies.  As we will outline below the method, being
based on optical data, generally has higher spatial resolution than
interferometric \hi\ and \CO\ maps, but it is naturally limited to the
stellar disk of galaxies since it is an absorption technique. Thus
most of our results in the final part of the paper will focus on the
gas content in the central few kpc of galaxies, nicely complementing
the integrated measurements gas measurements provided by single-dish
observations.

The plan of the paper is as follows: In sections~\ref{sec:method}
and~\ref{sec:fitt-spectr-feat} below we will discuss our method for
modelling gas content in some detail, including the importance of
priors on our model parameters. This will result in a model which
links dust content, metallicity and dust-to-metal ratio to produce an
estimate of gas column density in a given system. We review the data
we will use to validate our technique in section~\ref{sec:data}, and
emphasise the importance of the chosen attenuation curve.
Section~\ref{sec:test-gas-estim} is devoted to tests of our method and
we start with a study of how the dust-to-metal and dust-to-gas ratios
in galaxies depend on redshift, showing good agreement with modelling
of dust emission in nearby galaxies. We then show that our gas column
densities agree with those inferred for the same regions from \hi+\CO\
maps in nearby galaxies and exploit the aperture trends inherent in
Sloan Digital Sky Survey (SDSS) spectroscopic data to check that we can
recover well-known scaling relations. In section~\ref{sec:trends} we
discuss the gas content and gas depletion times in the central regions
of $\sim 70,000$ galaxies in the local Universe, selected from all
$\sim 200,000$ star-forming galaxies in the  DR7. We conclude in
section~\ref{sec:conclusions}. Throughout we adopt a cosmology with
$H_0 = 70 \mathrm{km}\;\mathrm{s}^{-1}\;\mathrm{Mpc}^{-1}$, $\Omega_m=0.3$ and
$\Omega_\Lambda = 0.7$. Where relevant we adopt a Kroupa initial mass
function \citep{2001MNRAS.322..231K}.  We will denote surface
densities by $\Sigma$.

\section{Estimating gas masses from dust attenuation}
\label{sec:method}

From observations it is clear that dust and gas trace each other
fairly closely, at least when averaged over sufficiently large
regions. Here we wish to build on this to construct a probe of gas
content using the dust in absorption as a probe of the gas column
density and exploiting the temperature dependence of emission lines to
place constraints on the dust-to-gas ratio; together they will then
provide an estimate of the effective gas column density in the region
probed by the spectrum.  We note that our focus here will be on the
attenuation of emission lines only, but if a separate constraint on
the absorption of the stellar continuum can be had
\citep[e.g.][]{2003MNRAS.341...33K}, further constraints on the model
can in principle be obtained.

We adopt the simple dust attenuation model by
\citet[][CF00]{2000ApJ...539..718C} and it is therefore convenient to
distinguish between attenuation in the stellar ``birth clouds''
(i.e.\ molecular clouds), $\tau_{\mathrm{BC}}$ and in the ``ambient''
(i.e. diffuse) ISM, $\tau^{\mathrm{\mathrm{ISM}}}$.
\begin{equation}
\tau_V = \tau_V^{BC} + \tau_V^{\mathrm{ISM}}
\label{eq:mugas_eq6}
\end{equation}
CF00 found a good fit to their sample of galaxies with the assumption
that the attenuation in the diffuse ISM is typically only about $1/3$
of the total attenuation affecting young stars \citep[see
also][]{2008MNRAS.388.1595D}, but the actual value turns out not to
matter for our argument here (but see the discussion in
section~\ref{sec:import-atten-curve}). 

We now follow \citet[][CL01]{2001MNRAS.323..887C} and introduce a new
variable, $\xi$, which is the dust-to-metal ratio of the ionised gas,
\begin{equation}
  \label{eq:11}
  \xi = \frac{M_d}{M_Z},
\end{equation}
where $M_d$ is the dust mass and $M_Z$ is the mass in
metals. Alternatively this can be written as
\begin{equation}
  \label{eq:12}
  \xi = \frac{\sum_i  m_i \delta(\mathrm{X}_i)
    \left(\frac{\mathrm{X}_i}{\mathrm{H}}\right)_c}{\sum_i m_i
    \left(\frac{\mathrm{X}_i}{H}\right)_c},
\end{equation}
where $\delta(\mathrm{X}_i)$ is the linear depletion of element
$\mathrm{X}_i$, and $m_i$ is the mass of element $\mathrm{X}_i$ and
where $\left(\mathrm{X}_i/\mathrm{H}\right)_c$ is the cosmic, ie.\
undepleted, abundance of element X$_i$.  We will return to the
question of expected values for $\xi$ below. Note that the combination
$\xi Z$, with $Z$ being the total metallicity, provides a measure of
the dust-to-gas ratio, and we will return to this in
section~\ref{sec:dust-gas-ratios}.

The surface mass density of gas contributed by the birth clouds
(assumed spherical) can thereby be derived using the definition of
$\xi=\Sigma_d/\Sigma_Z$ where $\Sigma_d$ is the surface density of
dust and $\Sigma_Z$ that of metals, and we have made the assumption
that the effective scale-length of metals and dust is the same. In
addition the metallicity can be written $Z=\Sigma_Z/\mugas$, which
taken together gives
\begin{equation}
\mugas^{BC} = \tau_V^{BC} m_d / (\sigma_d \xi Z),
\label{eq:mugas_eq9}
\end{equation}
where $m_d$ is the mass of the dust grain and $\sigma_d$ its optical
cross-section --- these should be viewed as effective, angle-averaged,
quantities in the sense adopted by CF00.

According to CF00, the model that produced the better fit to their
attenuation curve was on based on a Poisson distribution of discrete
clouds. Let $\tau_V^c$ be the optical depth per cloud and $\bar{n}$ be
the mean number of clouds encountered along different lines of
sight. The observations of starburst galaxies favour $0.1 < \tau_V^c <
0.5$ with a best-fit value of 0.3.  With these additional assumptions
about the optical properties and the spatial distribution of dust, we
can relate $\tau_V^{\mathrm{ISM}}$ to $\mugas^{\mathrm{ISM}}$. For a Poisson
distribution, the effective absorption optical depth in the diffuse
ISM is (CF00; Eq 25) simply
\begin{equation}
\tau_V^{\mathrm{ISM}} = \bar{n} [1 - \exp(-\tau_V^c)],
\label{eq:mugas_eq10}
\end{equation}
with typically $\tau_V^c=0.3$. Now, to go from a Poisson distribution
for the probability density of absorption, to a Poisson distribution
of gas clouds, we require a new assumption: that scattering is mostly
forward. This seems to be well supported by both models and
observations of interstellar dust in the Milky Way.
\citet{2003ApJ...598.1017D} compared predictions of $\langle \cos
\theta \rangle$ from models with a range of observations and found
that $\langle \cos \theta \rangle > 0.6$ in all environments at the
optical and UV wavelengths of interest to us (see also
Gordon (2004)\nocite{2004ASPC..309...77G} and references therein). In
this case, the mean 
surface mass density of gas along different lines of sight in the diffuse ISM is simply:
\begin{equation}
\mugas^{\mathrm{ISM}} = \bar{n} \mugas^c,
\label{eq:mugas_eq11}
\end{equation}
where $\mugas^c$ is the surface mass density of individual
clouds (note that, strictly, we have adopted here a Poisson
distribution of identical, face-on screens rather than one of
spherical clouds). Substituting $\bar{n}$ from
equation~(\ref{eq:mugas_eq10}), we then have
\begin{equation}
\mugas^{\mathrm{ISM}} = \frac{\tau_V^{\mathrm{ISM}} \mugas^c} {1 - \exp(-\tau_V^c)}.
\label{eq:mugas_eq12}
\end{equation}

The mass density of an individual cloud can be inferred as in
equation~(\ref{eq:mugas_eq9}) above as
\begin{equation}
\mugas^c = \tau_V^c  m_d / (\sigma_d  \xi  Z)
\label{eq:mugas_eq13}
\end{equation}

We can then substitute this in equation~(\ref{eq:mugas_eq12}) and
expand in powers of $\tau_V^c$. For optically thin clouds (recall
$\tau_V^c=0.3$), we can write
\begin{equation}
\mugas^{\mathrm{ISM}} \approx  \tau_V^{\mathrm{ISM}} m_d / (\sigma_d  \xi  Z). 
\label{eq:mugas_eq14}
\end{equation}
It is important to note that this is only an approximate formula,
which breaks down for more optically thick clouds, but for
$\tau_V^c=0.33$, the next term is only 17 per cent of the equation
above. With all the assumptions listed above, we can then write
\begin{eqnarray}
\mugas & =  & \mugas^{BC} + \mugas^{\mathrm{ISM}} \\
              & = &  \frac{(\tau_V^{BC} + \tau_V^{\mathrm{ISM}})  m_d} {\sigma_d
                \xi Z} \\
              & = & \frac{\tau_V m_d}{\sigma_d \xi Z}.
\label{eq:mugas_eq15}
\end{eqnarray}
For $m_d$ and $\sigma_d$ corresponding to Galactic-type dust, we adopt
a mass-weighted mean grain radius of $a=0.1\mu m$ (using a MRN
distribution between 0.005 and 0.25 micron), $\sigma_d \approx \pi
a^2$ (since we assumed forward scattering) and $\rho_d \approx 3
$g cm$^{-3}$ (the mean mass density of silicates and graphite grains).
This yields
\begin{equation}
\mugas \approx  0.2 \frac{\tau_V}{\xi Z} \mbox{M}_\odot\mbox{pc}^{-2}.
\label{eq:mugas_eq17}
\end{equation}
The numerical pre-factor in equation~(\ref{eq:mugas_eq17}) is fairly
robustly determined as it only depends on the mass-weighted radius to
the first power as well as the reasonably well-determined density of
carbonaceous and silicate grains. Using the
\citet[][WD01]{2001ApJ...548..296W} model we find that for $R_V=3.1$
the pre-factor ranges from 0.17 to 0.23 across the various C
abundances considered by those authors. For LMC and SMC dust from the
WD01 modelling, the pre-factor ranges between 0.21 and 0.26. Thus we
conclude that the overall scaling is robustly determined, at least
across a range of metallicities from the Milky-Way to the SMC. We will
adopt equation~(\ref{eq:mugas_eq17}) in the following. In passing we
comment that if we have a separate constraint on $\mugas^{\mathrm{ISM}}$
from e.g.\ the attenuation of the stellar continuum, it might be
possible to extend this method to provide an estimate of the relative
gas content in the ISM and in birth clouds and it would also allow
constraints to be placed on the gas column densities in galaxies
without significant on-going star formation.

Despite the overall robustness of the scaling factor to the dust grain
properties, the derivation above does depend on a number of uncertain
assumptions so it is essential to test the results carefully. In
particular it is not clear at what density the simplification made in
equation~(\ref{eq:mugas_eq14}) will break down, leading to an
underestimate of the gas density. We are also sensitive to the exact
extinction curve being adopted as well as the metallicity estimator
used and we will discuss this further in
section~\ref{sec:test-gas-estim}.  Finally, at very low column
densities our modelling will break down at some point. Since the
technique is based on absorption it might be expected to be more
sensitive than emission-based probes such as \CO\ or \hi, but we know
less about the properties of the attenuation curve at low gas
densities thus our systematic uncertainties could become important
here and the method as discussed here relies on a background source
with emission lines which might not be present at very low densities.

It is worth emphasising that this approach is sensitive to the total,
ie.\ atomic plus molecular, gas surface mass density in the observed
region. This is crucial for testing because it implies that using only
\hi\ is insufficient where significant contributions to the gas mass
are expected to be in molecular form.

\section{Derivation of \mugas\ through fitting to emission lines}
\label{sec:fitt-spectr-feat}

We now discuss briefly how we use equation~(\ref{eq:mugas_eq17}) to
constrain \mugas. We need a framework to constrain the metallicity,
the dust attenuation and the dust-to-metal ratio of the ionised
gas. Furthermore we have to take into account the fact that these
quantities likely will be correlated. To address these issues we
follow closely the approach outlined in \citet[][hereafter
B04]{2004MNRAS.351.1151B} and use a grid of the models derived by
\citet[][]{2001MNRAS.323..887C}. We will also discuss the choice of
priors and a number of uncertain ingredients in our modeling and
comparisons. For ease of reference this is summarised in
Table~\ref{tab:priors_etc} below.

The CL01 models provide predictions for emission line strengths,
combining evolving galaxy spectra from the \citet{1993ApJ...405..538B}
models with the Cloudy photoionisation code
\citep{1998PASP..110..761F}. The models are parametrised by five main
parameters and are calculated on a regular grid in all these
parameters (see table 2 in B04). For our work here we need four of
these parameters: the ionisation parameter, $U$, the dust-to-metal ratio
in the ionised gas, $\xi$, the dust attenuation in the $V$ band,
$\tau_V$, and the metallicity, $Z$, note that this is the total
metallicity including metals depleted onto dust grains and therefore
differs from the gas-phase metallicity. We then fit our grid of CL01
models to optical emission lines; when available we use the main
optical strong lines \oii, \hb, \oiii{4959,5007}, \ha, \nii{6584} and
\sii{6717,6731}, we will discuss their importance later. Note that
with the abundance patterns used in the CL01 models, the oxygen
abundance of the sun is $12 + \log \left(\mathrm{O/H}\right)_\odot = 8.82$.

The $\xi$ parameter, the dust-to-metal ratio of the ionised gas, is not
commonly used in photo-ionisation calculations and warrants a more
extensive discussion. It is a linear depletion parameter and its
average value, which is what we are concerned with, in the
interstellar medium (ISM) of galaxies is in general not known very
well from direct observations. However from the observed near-linear
correlation between metallicity and the dust-to-gas ratio
\citep[e.g.][also see below]{1990A&A...236..237I,1998ApJ...496..145L,2007ApJ...663..866D,2010MNRAS.403.1894D}
it is possible to conclude that it shows relatively little variation
in nearby galaxies. Chemical evolution models do however show that
$\xi$ is expected to show moderate evolution with time
\citep[e.g.][]{2001MNRAS.328..223E,2003PASJ...55..901I,2008A&A...479..669C},
justifying our decision to keep it as a free parameter.  Note that the
dust-to-gas ratio, $\DGR = \xi Z$, is a derived parameter in this
chosen parametrisation, we will return to this quantity in
section~\ref{sec:dust-gas-ratios} below.

In the Milky Way the depletion pattern in the ISM along certain
specific lines of sight is, however, fairly well-known. CL01 made
slight modifications to the default depletion pattern of Cloudy which
is based on \citet{1986ARA&A..24..499C} and
\citet{1987ASSL..134..533J}. The most recent in-depth discussion of
depletion patterns is the careful study of \citet{2009ApJ...700.1299J}
and while there are differences to previous works the overall
agreement is reasonable --- although unfortunately the most important
elements for our needs, C and O, are affected by rather significant
systematic uncertainties \citep[see the discussion in
][]{2009ApJ...700.1299J}.  The depletion patterns and abundances
adopted by CL01 results in an average $\xi$ of 0.3. This can be
contrasted with \citet{2009ApJ...700.1299J} whose modelling implies a
variation in $\xi$ from $\xi \sim 0.2$ to $\xi \sim 0.6$ along various
lines of sight in the Milky Way with both a different depletion
pattern and cosmic abundances than CL01. For reference, our $\xi=0.3$
corresponds to $F_*\approx 0.2$ in Jenkins' notation.

\begin{figure}
  \centering
  \includegraphics[width=84mm]{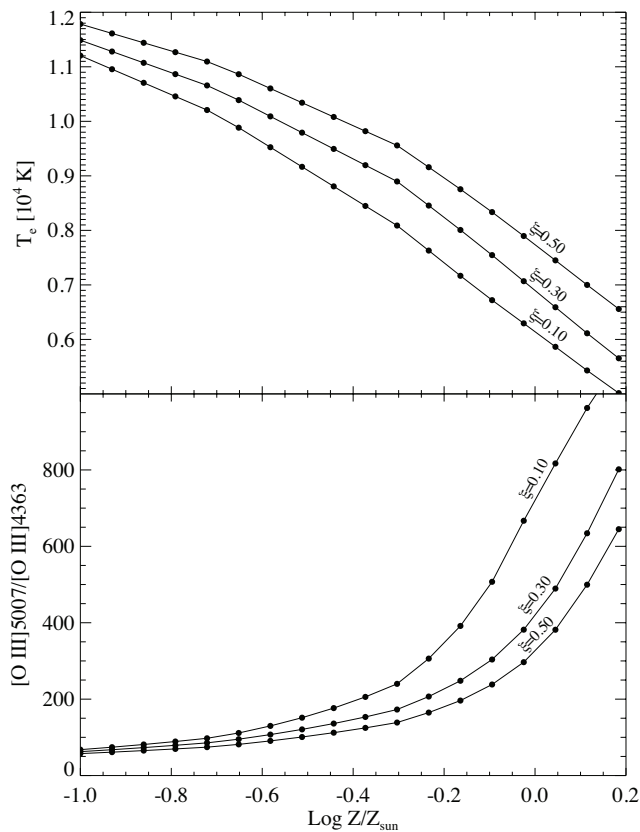}
  \caption{\emph{Top panel}: The mean electron temperature as a
    function of metallicity for three different values of $\xi$. Note
    that as more metals are removed from the gas-phase, the mean
    temperature goes up because the cooling is reduced. \emph{Bottom
      panel}: The ratio of the \oiii{5007} line to the auroral line,
    \oiii{4363}, as a function of metallicity. Note that the change in
    temperature has a much larger effect at high metallicity than at
    low --- this limits our ability to constrain $\xi$ at low
    metallicities. The curves in both plots were done at fixed
    ionisation parameter ($\log U=-3.4$) and dust attenuation
    ($\tau_V=0.17$). }
  \label{fig:te_o3ratio_vs_xsi_Z}
\end{figure}

Keeping other parameters fixed, the main effect of changing $\xi$ is
that removing more metals from the gas leads to an increase in the
electron temperature in the \hii-region. We show this explicitly for
the CL01 models in Fig.~\ref{fig:te_o3ratio_vs_xsi_Z} where the top
panel shows the electron temperature as a function of total
metallicity for three different values of $\xi$. As expected, a larger
depletion (larger $\xi$) leads to a higher temperature at fixed
metallicity. 

The impact of this temperature change is not uniform in observable
line ratios. The lower panel of the figure illustrates this with the
well-known temperature sensitive line ratio \oiii{5007}/\oiii{4363}
where we can see that the effect of the change in temperature is
substantial at high metallicity, but much less noticeable at low
metallicities.

\begin{figure*}
  \centering
  \includegraphics[width=184mm]{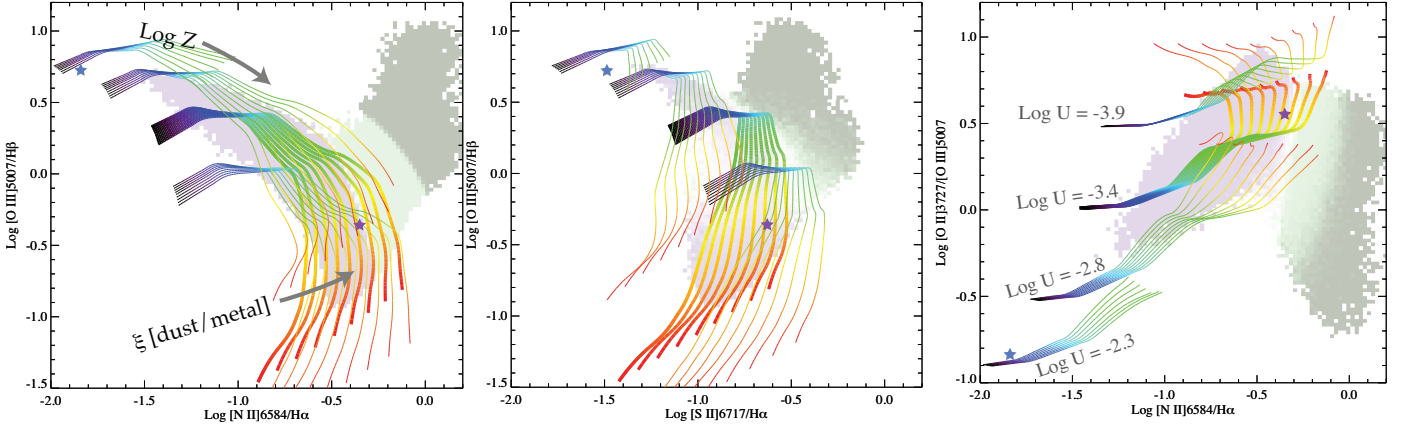}
  \caption{The behaviour of the CL01 models in three diagnostic
    diagrams. The background histogram shows the distribution of
    emission line galaxies in the SDSS DR7 and is coloured according
    to the mean emission line classification following B04 with
    star-forming (purple), composite (light green) and active galactic
    nuclei (AGN, dark green) classes shown.  There are four groups of
    model tracks for different values of $\log U$ as indicated in the
    rightmost panel. The colour changes proportionally to $\log Z$
    from $Z=0.1 Z_\odot$ in the top left to $Z=4Z_\odot$ at the
    bottom. For each $\log U$ value, except the highest, there are
    lines for nine $\xi$ values, linearly spaced from $\xi=0.1$ to
    $\xi=0.5$. The purple star shows the location of the object used
    for the illustrations of the fitting process in
    Figures~\ref{fig:example_fig_regular}
    and~\ref{fig:example_fig_reverse} and the blue star that of
    Figure~\ref{fig:xsi-prior-fit-example}.  The two leftmost panels
    show two BPT diagrams with slightly different sensitivity to
    $\xi$, while the rightmost panel plots \oii{3727}/\oiii{5007} on
    the y-axis against \nii{6584}/\ha\ to better separate the lines
    for different ionisation parameter.}
  \label{fig:models_xsi_variation}
\end{figure*}

An illustration of the effect of varying $\xi$ and $Z$ in the
\citet[][BPT]{1981PASP...93....5B} \nii{6584}/\ha\ versus
\oiii{5007}/\hb\ diagnostic diagram, frequently used
to classify galaxies based on their emission line properties
\citep[e.g.][]{2003MNRAS.346.1055K,2006MNRAS.372..961K} can be seen in
the leftmost panel of Fig.~\ref{fig:models_xsi_variation}. The
distribution of emission line galaxies in the SDSS DR7 is shown as the
underlying 2D histogram which is coloured according to the emission
line classification, star-forming (purple), composite (light green)
and active galactic nuclei (AGN, dark green). The middle panel
in the figure shows the \sii{6717}/\ha\ versus \oiii{5007}/\hb\
diagnostic diagram with the background 2D histogram again showing the
distribution of SDSS DR7 galaxies with the colour indicating the mean
classification based on the first diagram. The final panel shows
the ionisation parameter sensitive ratio \oii{3727}/\oiii{5007}
plotted against \nii{6584}/\ha. 

The lines show the model tracks for fixed values of $\xi$, increasing
from 0.1 to 0.5 from left to right. There are three sets of nine lines
corresponding to $\log U=-2.8, -3.4, -3.9$ respectively (see the
rightmost panel), going from the top downwards in all panels. The
colouring of the lines correspond to a change in 
$\log Z$ from $0.1 Z_\odot$ for black to $4 Z_\odot$ for red. The
effect of changing $\tau_V$ is not shown in this figure but is minor
in the first two panels and should be well known. 

As shown in Fig.~\ref{fig:te_o3ratio_vs_xsi_Z}, increasing $\xi$
increases the temperature, but it also reduces the amount of oxygen in
the gas-phase. At high metallicity the resulting increase in the
electron temperature, $T_e$, is more important than the reduced
abundance of oxygen, thus increasing the flux in \oiii{5007} relative
to H$\beta$, while nitrogen is assumed to be non-refractory and the
increased $T_e$ leads to an increase in the \nii{6584}/H$\alpha$ and
\sii{6717}/\ha\ ratios (see also the discussion in CL01).  This effect
of $\xi$ on the emission lines is the reason why we are able to place
some constraints on its value from observations. At low metallicity
and high $T_e$ the effect of $\xi$ is weak leading to poor
constraints on $\xi$, and we will return to this point below and in
Appendix~\ref{sec:impact-priors-mugas}. In the first two panels the
effect of the ionisation parameter is not easily isolated at high
metallicity, but the final panel shows that the inclusion of
\oii{3727} helps separate the model tracks. This highlights the need
to have a number of emission lines across the optical range to
constrain the various model parameters. We also see that the model
grid cover the emission-line properties of star-forming
galaxies in the SDSS.

\begin{figure*}
  \centering
  \includegraphics[width=184mm]{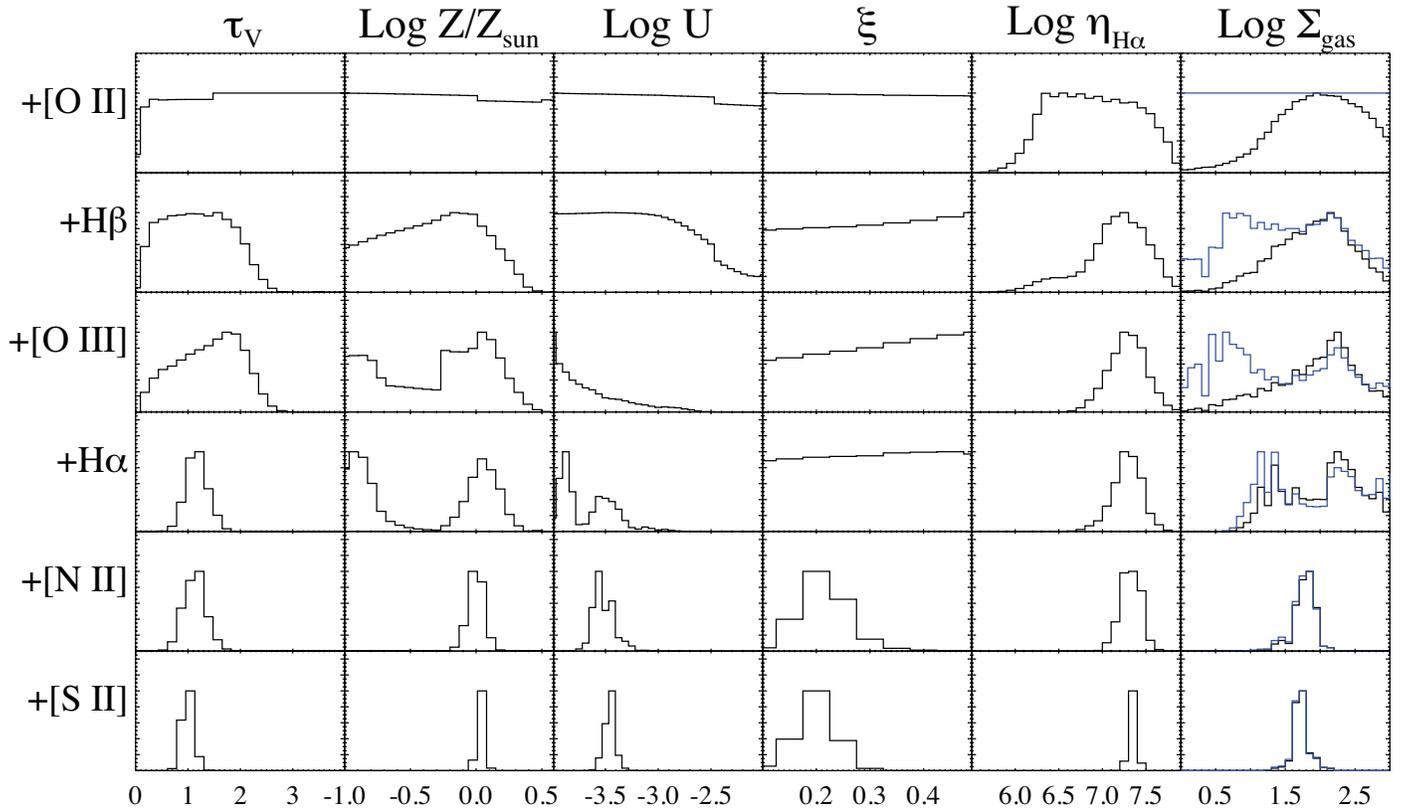}
  \caption{ An example of the result of the fit to the spectrum given
    by plateID-MJD-fiberID=736-52221-287. The location of this object
    in the BPT diagram is indicated by the purple square in
    Figure~\protect\ref{fig:models_xsi_variation}. Each column shows
    the PDF for one derived parameter as emission lines are
    progressively included to the fit (equation~\ref{eq:6}). The
    leftmost column shows the constraints on the total $V$-band
    optical depth, $\tau_V$. The following three columns show the
    metallicity, the ionisation parameter, $\log U$ and the
    dust-to-metal ratio of the ionised gas, $\xi$. The second-to-last
    column shows the log of the conversion factor between \ha\
    luminosity and star formation rate, $\eta_{\ha}$, in units of
    $\Msun\;\mathrm{yr}^{-1}\;\mathrm{L}_\odot^{-1}$. The final column
    shows $\log 
    \mugas$ in units of $\Msun\;\mathrm{pc}^{-2}$.  Given the
    construction of this plot, the top row essentially shows the prior
    adopted on the quantity in question in the model. In the last
    column we also show the PDF derived with a flat prior on $\mugas$
    in blue. See the text for a discussion.}
  \label{fig:example_fig_regular}
\end{figure*}

Given the CL01 model predictions we then adopt the Bayesian approach
outlined in B04 \citep[see also][]{2003MNRAS.341...33K} to derive the
probability distribution functions (PDFs) for the parameters of
interest. Given a particular set of lines, $\{L_i\}$, we use Bayes'
theorem to calculate the log likelihood of each model, ${\cal M}(U,
\xi, \tau_V, Z)$, as:
\begin{equation}
  \label{eq:6}
  \ln P({\cal M}|\{L_i\}) = -\frac{1}{2} \sum_{i \in \{L_i\}}
  \frac{ \left(f_i - A f_{\cal M}\right)^2}{\sigma_i^2} + \ln \mathrm{Pr},
\end{equation}
where $f_i$ is the flux in line $i$, $A$ is a scaling-factor and
$f_{\cal M}$ corresponds to the relevant model. $\mathrm{Pr}$ denotes
the prior on the model parameters and is shown in the top row of
Fig.~\ref{fig:example_fig_regular}, see also
Appendix~\ref{sec:impact-priors-mugas}, and we have ignored the
overall normalisation factor which is acceptable as long as the
$\sigma_i$'s do not depend on the model.

To calculate $\sigma_i$ for the SDSS galaxies we start with the formal
line flux errors calculated from the SDSS error spectrum. These errors
do not include continuum subtraction uncertainties and are therefore
likely to underestimate the true uncertainties. To adjust for this we
make use of duplicate observations of 33,794 emission line
galaxies. For these we calculate the difference in measured line flux,
normalised by the formal uncertainty of this difference. When the
errors are accurately estimated the distribution of these differences
should be a unit Gaussian and we use this to find the empirical
scaling factors for the uncertainties that would bring the duplicate
observations in agreement within the errors. These are given in
Table~\ref{tab:scale}. For the other data sources we use published
uncertainties. In both cases we furthermore follow B04 and apply an
approximate theoretical uncertainty by adding 4 per cent of the line
flux in quadrature to the uncertainties to get the final $\sigma_i$
that enters into equation~(\ref{eq:6}). This theoretical error is an
approximate adjustment to take into account the fact that each model
is characterised by a single set of parameters while in reality each
region likely has a distribution of physical properties (see B04 for
more discussion).

\begin{table}
  \centering
  \begin{tabular}{lc}
    \multicolumn{1}{l}{Line} & \multicolumn{1}{c}{Scaling factor} \\
    \oii{3727} & 2.199\\
    \hb & 1.882\\
    \oiii{4959} & 1.573\\
    \oiii{5007} & 1.566\\
    \ha & 2.473\\
    \nii{6584} & 2.039\\
    \sii{6716,6731} & 1.621\\
  \end{tabular}
  \caption{Factors to scale formal line flux errors by derived from a comparison of duplicate observations of SDSS galaxies.}
  \label{tab:scale}
\end{table}

To obtain constraints on a particular quantity we follow normal
practice and marginalise $P$ onto the parameter in question. For the
current discussion we are mostly interested in \mugas, and the
probability in bin $i$ is given by
\newcommand{\mugasi}{\ensuremath{\Sigma_{\mathrm{gas}, i}}}
\begin{equation}
  \label{eq:7}
  P(\mugasi) = \sum_{\cal R} P({\cal M}|\{L_i\}),
\end{equation}
where $\cal R$ is the region defined by
\begin{eqnarray}
  \label{eq:8}
  {\cal R} = & \left\{ {\cal M}(U, \xi, \tau_V, Z) |  \log
    (\mugasi-\delta_\Sigma) <  \phantom{\frac{\tau_V}{\xi Z}} \right. \\
   & \left. \log \frac{\tau_V}{\xi Z}<  \log (\mugasi + \delta_\Sigma) \right \}. 
\end{eqnarray}
We use a regular grid in $\log \mugas$ with bin width $\Delta_\Sigma =
0.05\, \mathrm{dex}$ between $\log \mugas = -1$ and $\log
\mugas=4$ for the calculation of \mugas.

For each PDF we also calculate the entropy
\begin{equation}
  \label{eq:26}
  S = \sum_i P_i \log_2 P_i,
\end{equation}
where the sum is over the bins of the PDF. This is a generalised
uncertainty estimate for the PDF and is useful when the PDF is
multi-peaked, or for determining whether a distribution is
significantly different from the prior distribution. For the latter
question we also calculate the Kullback--Leibler divergence between the
prior, $Q_i$, and the final PDF, $P_i$,
\begin{equation}
  \label{eq:27}
  d_{\mathrm{KL}}  = \sum_i P_i \log_2 \left(\frac{P_i}{Q_i}\right).
\end{equation}
Note that $d_{\mathrm{KL}}$ is often defined with natural logarithms
rather than base-2 logarithms, but we choose the latter for
consistency with the entropy definition. 

\begin{figure*}
  \centering
  \includegraphics[width=184mm]{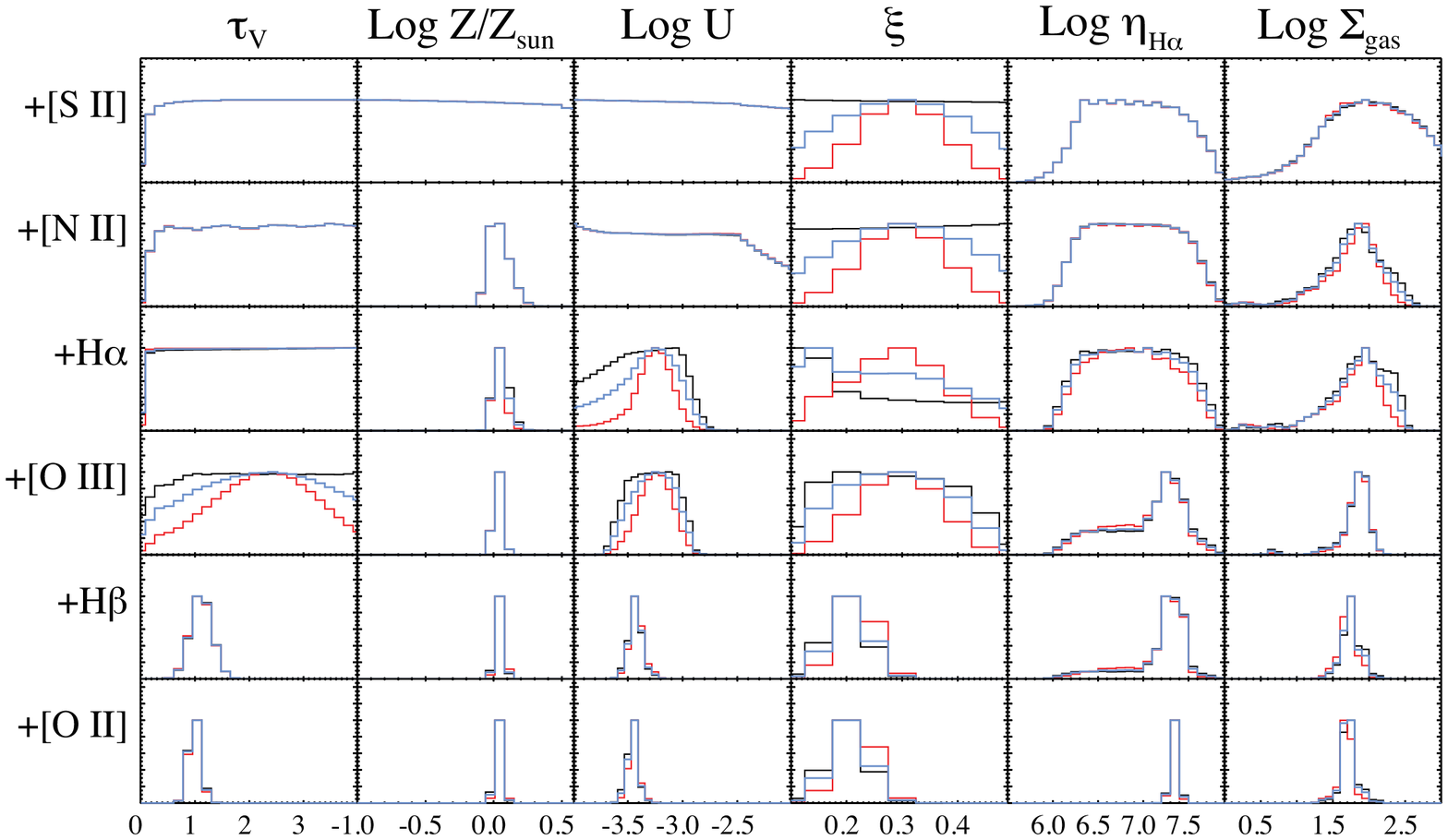}
  \caption{Similar to Fig.~\ref{fig:example_fig_regular}, but
    the emission lines are added in reverse order, providing a
    clearer illustration of the importance of \oii{3727} and \hb\ for
    the fitting procedure. The effect of applying a broad (blue) and
    narrow (red) prior on $\xi$ is also indicated here. Since this is
    a relatively metal rich galaxy, the effect is modest on most
    quantities as long as enough emission lines are present.}
  \label{fig:example_fig_reverse}
\end{figure*}

The process is illustrated in Figures~\ref{fig:example_fig_regular} \&
\ref{fig:example_fig_reverse} where we show how the likelihood
distribution of five different quantities changes as an increasing
number of lines is included in the summation in
equation~(\ref{eq:6}). Between one row and the next we add the line
indicated in the left-hand edge to the sum.  The first row in the
figure corresponds to the fit when only one line is included; this
obviously provides no constraints so returns approximately the priors
applied to the different quantities. From left to right the columns
correspond to the total dust attenuation in the \textit{V}-band,
$\tau_V$, the metal abundance in units of solar metallicity, $\log Z$,
the ionisation parameter, $\log U$, the dust-to-metal ratio of the
ionised gas, $\xi$, the log of the conversion factor between
\ha-luminosity and star formation rate, $\eta_{\ha}$ in units of
$\Msun\;\mathrm{yr}^{-1}\;\mathrm{L}_\odot^{-1}$ and finally the log
of the gas column density, $\mugas$, in units of
$\Msun\;\mathrm{pc}^{-2}$.

Focusing first on Fig.~\ref{fig:example_fig_regular}, in the first
column we see the well-known fact that to get a good constraint on the
dust attenuation at least two recombination lines must be available
(here \ha\ and \hb). Some (minor) improvement on the attenuation
estimate is possible if a better constraint on the electron
temperature, and thence the intrinsic \ha/\hb\ Case B ratio, can be
had through the addition of \nii{6584} and
\sii{6717}. Fig.~\ref{fig:example_fig_reverse} shows the same but
with the lines added in the reverse order which illustrates the
usefulness of \hb\ and \oii\ --- for this particular object \oii\ does
not add much but \hb\ is essential.

The second column shows the constraints on the metallicity --- in this
case this is very similar to what is obtained for the oxygen
abundance, and shows the well-known fact
\citep[e.g.][]{1980MNRAS.193..219P,1991ApJ...380..140M} that an
abundance indicator based on \oii{3727}, \oiii{5007} and \hb\ is
double-valued in some range of parameter space. The addition of
\nii{6584} serves to break this degeneracy and we obtain a tight
constraint on $Z$.

The ionisation parameter is shown in the third column and is only
well-constrained in this case when a good dust correction can be
obtained and lines with a range in ionisation potential are
included. The dust-to-metal ratio of the ionised gas, $\xi$, is only
well constrained when lines from non-depleted elements, such as N and
S, are included in the fit.  Note that while CL01 assumed that S is
not depleted in the ISM, it is not entirely clear that this is correct
\citep[e.g.][and references
therein]{2009ApJ...693.1236C,2009ApJ...700.1299J}, it is however
outside the scope of this paper to study this further.
As should be clear from Fig.~\ref{fig:models_xsi_variation}, the
quality of the constraint on $\xi$ is a strong function of abundance
and at a low abundance we are not able to place any constraints on
$\xi$ at all. The conversion factor between \ha-luminosity and SFR was
discussed in CL01 and B04 and we refer the readers to those articles
(note that the similarity between the PDF for $\log \eta_{\ha}$
and $\log \mugas$ is not a general feature).

The PDF for \mugas\ is shown in the last column. The first point to
notice is that the prior on \mugas\ is not flat. This is a natural
consequence of it being a ratio of three quantities with flat
priors. Thus we show the impact of the prior by also showing the
PDF if one assumes a flat prior on \mugas\ (blue line). This indicates
that as long as the PDF is well constrained, the impact of the prior
is minor.

This figure is for a ``typical'' galaxy in the SDSS and hence has
close to solar metallicity. The figure is representative for such
galaxies and the main points to note are: One of \nii{6584} or
\sii{6717} should be available in addition to \hb, \oiii{5007} and
\ha\ to break the abundance
degeneracy. Fig.~\ref{fig:example_fig_reverse} shows the impact of
lines in the reverse order and while in the shown case \oii{3727} does
not improve the constraints significantly in general it is beneficial
to have this line as well.

At lower metallicity we can still constrain the column density of dust
using the model above, but as we are unable to place strong
constraints on $\xi$, our \mugas\ estimates have larger error-bars and
we are somewhat sensitive to the prior placed on $\xi$, thus we turn to
look at this next.

\subsection{The impact of priors on $\boldmath{\xi}$ on  the \mugas\ determinations}
\label{sec:impact-priors-mugas}

An advantage of the Bayesian approach to parameter estimation is that
it makes it necessary to state prior beliefs explicitly. In general
the priors adopted will not be important if the evidence outweighs the
prior belief.  However for quantities that are not well constrained by
the evidence, the derived constraints are usually affected by the
priors adopted. 

By default the model parameters have flat priors over the range
covered by the models. Note that this \emph{does} of course imply a
preferred median value in the case of no information which is set by
the mid-point of the range covered by the parameters. However as long
as most of the optical lines are present with good S/N this is
unimportant and most quantities we are interested in, such as the
oxygen abundance, O/H, the star formation rate, SFR, and the dust
attenuation, $\tau_V$ are unaffected by the choice of prior --- at
least as long as we do not adopt strongly peaked priors on the model
parameters.

This is not always the case for $\xi$. This parameter has a more
subtle effect on emission line fluxes and can only be reliably
determined at high metallicity where its impact on cooling is
significant (c.f.\ Fig.~\ref{fig:te_o3ratio_vs_xsi_Z}). At low
metallicity it is not possible to put strong constraints on $\xi$ and
its derived value is thus sensitive to the prior adopted.

Since $\xi Z$ provides a constraint on the dust-to-gas ratio that we
need to convert dust column density to gas column density, it is
natural to ask how sensitive \mugas\ is to the adopted priors. In
Fig.~\ref{fig:example_fig_regular} we show one example where
adopting a flat prior on \mugas\ for a metal rich system made little
difference on the final result, but for a general study it is better
to focus our attention on $\xi$.

To explore this dependence we therefore explore three different priors
for $\xi$. The \emph{default} prior is flat between $\xi=0.1$ and
$\xi=0.5$, and then we have two priors taken to be Gaussians in
$\xi$. We centre both on $\xi=0.3$ because, as we mentioned above,
this is the average value for the SDSS and also in good agreement with
theoretical models when adjusted to the same abundances as used in the
CL01 models
\citep[e.g.][]{2001MNRAS.328..223E,2008A&A...479..669C}. For the
\emph{wide} prior we take a width of $\sigma_\xi=0.15$ while the
\emph{narrow} prior has $\sigma_\xi=0.08$.

\begin{figure*}
\centering
\includegraphics[width=184mm]{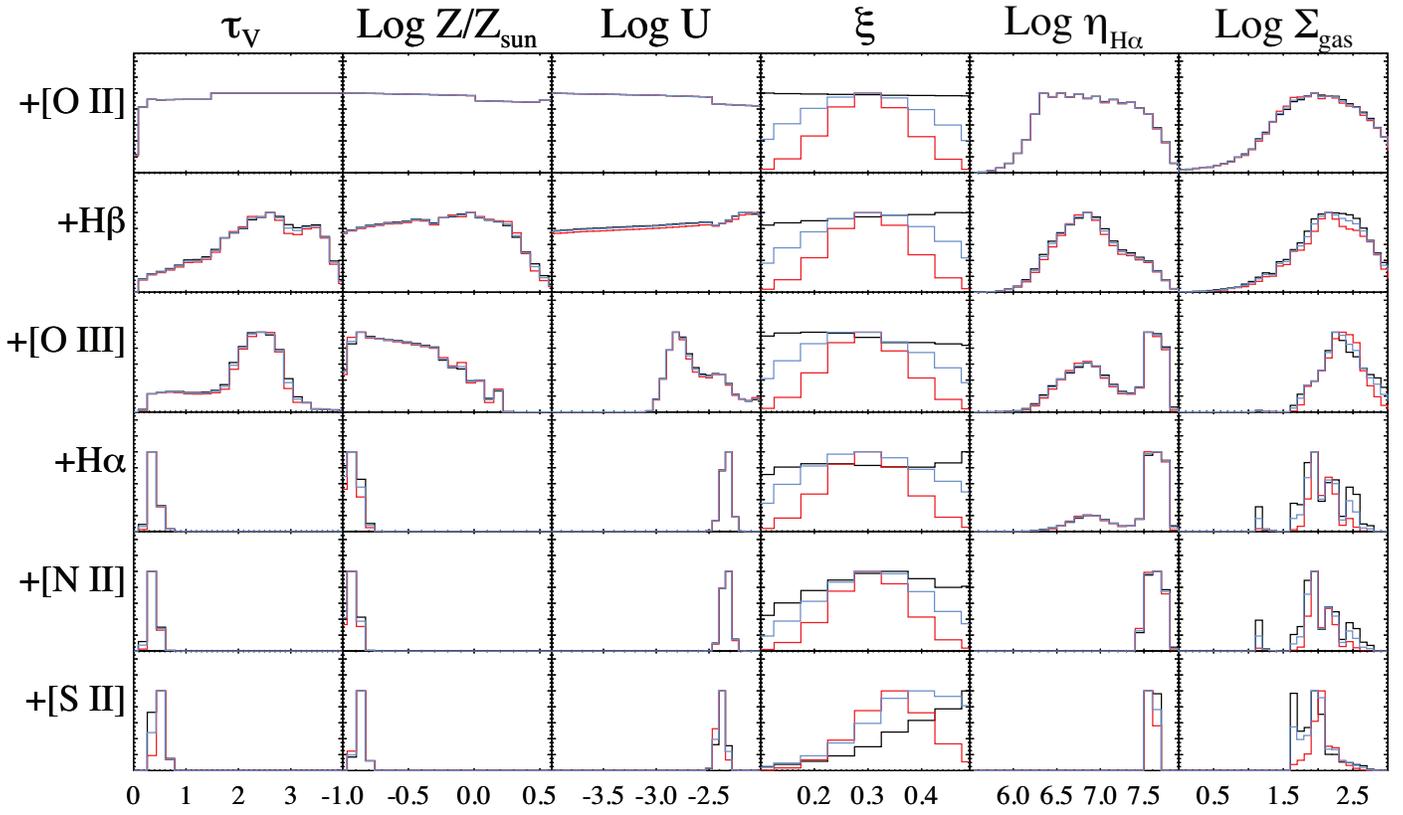}
\caption{Similar to Fig.~\ref{fig:example_fig_regular}, but this time
  showing a low metallicity galaxy (691-52199-0636 at $z=0.038$).  The
  results with a flat prior on $\xi$ is shown in black, overplotted on
  this in red we show the results for the \emph{narrow} prior on $\xi$
  and in blue that of a \emph{broad} prior on $\xi$. The effect of the
  prior is striking for $\xi$ but rather modest for everything else.}
\label{fig:xsi-prior-fit-example}
\end{figure*}

\begin{figure}
\centering
\includegraphics[width=84mm]{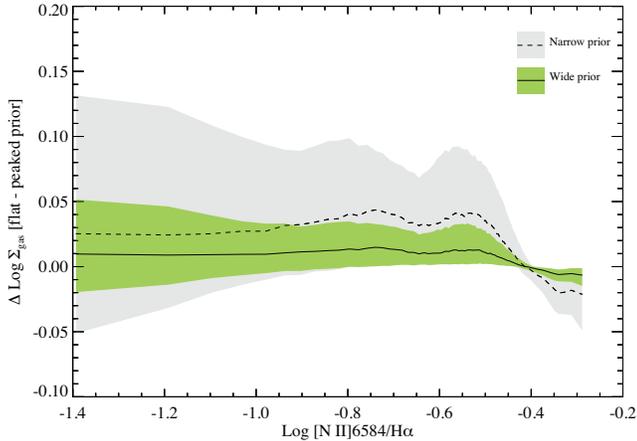}
\caption{The figure shows the effect on our estimates of $\mugas$ for
  all star-forming galaxies in the SDSS DR7 from changing the prior on
  $\xi$ from flat to a Gaussian with width $\sigma=0.08$ (grey
  shading/dashed line) or to a Gaussian with width $\sigma=0.15$
  (green shading/solid line). The dashed and solid lines show the
  median trend while the shaded area encloses 68 per cent of the scatter.}
\label{fig:xsi-prior-change-mugas}
\end{figure}

Fig.~\ref{fig:xsi-prior-fit-example} compares the results of fitting
with a flat prior on $\xi$ (black line) to that obtained with the
narrow prior on $\xi$ (red line) and that of a broad prior (blue
line). The object is a fairly low-metallicity galaxy ($12 + \log
\mathrm{O/H} \approx 7.8$) so that the constraints on $\xi$ are weak,
the format is otherwise the same as in
Fig.~\ref{fig:example_fig_regular}.  It is quite clear from the figure
that the effect of changing the prior on $\xi$ is negligible for most
parameters, but using a peaked prior on $\xi$ does lead to somewhat
tighter constraints on \mugas. This improvement in the constraints is
of course prior dependent, thus the most crucial question is whether
changing the prior leads to systematic differences in the gas mass
estimates.

Fig.~\ref{fig:xsi-prior-change-mugas} shows that the effect of the
prior on $\xi$ on the gas mass estimates is modest. The figure shows
the difference between $\mugas$ estimated using a flat prior and that
obtained using a peaked prior. The dashed line shows the median trend
for the narrow prior and we plot it against the \nii{6584}/\ha\ ratio
as an approximate proxy for metallicity. We can see that the
difference is typically less that 0.05 dex even for the most peaked
prior on $\xi$. When viewed as relative changes, we find that even at
the lowest metallicities less than 5 per cent of the galaxies show
changes in excess of 30 per cent. 

It is important to note that the choice of prior also has an effect at
high metallicity. This is because although at high metallicity we can
constrain the value of $\xi$, applying a peaked prior on $\xi$ will
bias its estimate towards the peak of the prior, with an effect that
is stronger for a more peaked prior, as seen in the figure, because
stronger evidence is then required to overcome the prior
belief. Figure~\ref{fig:example_fig_reverse} shows, however, that as
long as the emission lines are detected at sufficiently high
signal-to-noise, the effect of a prior on $\xi$ is very modest at high
metallicity. Thus employing a $\xi$-prior primarily makes sense in
low-metallicity systems, at high metallicity there is enough
information to constrain $\xi$ independently and applying a peaked
prior could bias the estimate of $\xi$ and \mugas.

While we do not show it explicitly here, the effect of a prior on
$\xi$ on other parameters is very small, at least at low
metallicity. The oxygen abundance, $12 + \log \mathrm{O/H}$, for
instance, changes by less than 1 per cent when applying a narrow prior on
$\xi$. Thus while applying a prior on $\xi$ does lead to slight, but
systematic, offsets in our \mugas\ estimates, at least these are
independent of other parameters to first order. We have however, not
discussed in detail what an appropriate prior should be --- the
simplest choice is indeed a prior that is metallicity independent but
as we will see later, in section~\ref{sec:dust-metal-ratio}, it might
be better to implement a metallicity dependent prior.

\begin{table*}
  \begin{minipage}{160mm}
    \caption{Sections where the influence of non-standard priors is explored throughout the paper} 
  \label{tab:priors_etc}
  \begin{tabular}{ll} 
    \hline
    Flat prior on \mugas\ versus the default peaked prior on \mugas &
    Figures~\ref{fig:example_fig_regular} 
    and~\ref{fig:example_fig_reverse}  \\
    The effect of peaked priors on $\xi$ relative to the default flat
    prior & Sections~\ref{sec:impact-priors-mugas} and
    \ref{sec:comp-resolv-hi} \\
    & Results in section~\ref{sec:trends} are checked \\
    & for sensitivity to the choice of $\xi$ prior. \\
    The impact of changing the dust attenuation laws & Section~\ref{sec:import-atten-curve} \\
    & All figures use $\tau(\lambda) \propto \lambda^{-1.3}$ \\
    The importance of a variable \XCO\ on the tests of the method &
    The comparisons in section~\ref{sec:comp-resolv-hi} use \\
    &  \XCO\ from equation~(\ref{eq:XCO_B02}) \\ 
    \hline
  \end{tabular}
\end{minipage}
\end{table*}

\subsection{The importance of the attenuation curve}
\label{sec:import-atten-curve}

Equation~(\ref{eq:mugas_eq17}) depends on the absolute value of the
V-band optical depth, which implies a dependence on the attenuation
curve adopted. While the shape of the optical attenuation curve is
more constrained than, say, the UV, it must still be treated as an
uncertain ingredient in the models. 

For a simple attenuation curve $\tau(\lambda) \propto \lambda^{-n}$,
we have
\begin{equation}
  \label{eq:21}
  \tau_V = \frac{\lambda_V^{-n}}{\lambda_{\hb}^{-n} - \lambda_{\ha}^{-n}} \ln r 
 \end{equation}
 where $r$ is the ratio of the Balmer decrement to the Case B
 ratio. Thus a change from $n=-0.7$ to $n=-1.3$ leads to an change in
 $\log \mugas$ of $-0.28$ for the same Balmer ratio and corresponding
 changes to the SFR estimate. When applying the fitting methodology
 described above, the final differences are slightly different because
 of slight changes in the best-fit model. Overall however, parameters
 that do not depend directly on $\tau_V$, such as O/H, $\xi$, and
 $\log U$, are very weakly affected by a change in $n$.

More complex attenuation laws do not in general lead to constant
offsets. We therefore have therefore derived estimates of $\log
\mugas$ for four different attenuation laws as follows:
\begin{enumerate}
\item $\tau(\lambda) \propto \lambda^{-1.3}$. CF00 observed that a
  slope of $-1.3$ corresponds to the middle range of the optical
  properties of dust grains between the Milky Way, the Large and the
  Small Magellanic Cloud. This was argued to be appropriate for
  attenuation of birth clouds in \citet{2008MNRAS.388.1595D}
  and~\citet{2011MNRAS.417.1760W}. We adopt this as our reference law.
\item $\tau(\lambda) \propto \lambda^{-0.7}$. This is the attenuation
  law recommended by CF00 for absorption ultraviolet to infrared
  continua by the ambient ISM in star-forming galaxies, and used in
  e.g.\ \citet{2004MNRAS.351.1151B}. However the attenuation of
  emission lines is likely less grey than this.
\item $\tau(\lambda)/\tau(\lambda_V) = (1-\mu)
  (\lambda/\lambda_V)^{-1.3} + \mu (\lambda/\lambda_V)^{-0.7}$. This
  is the combination of (1) and (2), as suggested observationally
  by~\citet{2007MNRAS.381..543W} and motivated theoretically
  by~\citet{2008MNRAS.388.1595D}. $\mu$ here is the relative
  contribution to the total $\tau_V$ contributed by the ambient ISM. A
  value of $\mu=0.3$ was suggested by CF00, but it is also possible to
  fit this keeping $\mu$ a free parameter.
\item $\tau(\lambda)/\tau(\lambda_V) = (1-\mu)
  (\lambda/\lambda_V)^{-1.3} + \mu \lambda^{-n(\lambda)}$ with
  $n(\lambda) = 2.8/(1+3\sqrt{\mu \tau_V}) + (0.3-0.05 \mu \tau_V)
  (\lambda - 0.55)$, with $\lambda$ measured in microns. This is the
  effective attenuation curve recommended by Chevallard et al (2013,
  submitted) based on an analysis of a range of radiative transfer
  models.
\end{enumerate}

These all give broadly speaking the same $\log \mugas$ estimates, but
with systematic shifts between the determinations. To make this
quantitative, we fit all SDSS DR7 star-forming galaxies with each of
the four different attenuation curves. The median effect is such that
with option 2 above, $\log \mugas$ is increased by 0.30 dex in the
mean relative to our reference curve, while option 3 leads to $\log
\mugas$ estimates higher by 0.19 dex in the mean than the reference
values. The fourth option produces gas column densities that are 0.06
dex higher than our reference value.

In the following we will adopt option 1 above as it is a simple law
and it is also very similar to the widely used
\citet{1979MNRAS.187P..73S} extinction curve at $\lambda >
4200$\AA. Furthermore option 1 gives results that are very similar to
that of option 4 which is the most rigorously justified curve, but
which depend on one further parameter, $\mu$, which is not well
constrained by emission lines alone.

\subsection{The importance of the abundance estimator}
\label{sec:import-abund-estim}

It is well-known that different abundance estimators can give
significantly different oxygen estimates for the same \hii\ regions
\citep[see][ for a comparison]{2008ApJ...681.1183K}. Since our
estimator depends on the absolute value of the metallicity, we are
affected by this uncertainty and it is therefore appropriate to
revisit this issue for our technique.

Abundance comparisons are generally made between \emph{direct}
estimators where auroral lines are used to determine electron
temperatures appropriate for different ionisation stages of elements
\citep[e.g.][]{2006A&A...448..955I} and other techniques. The basic
assumption is that the abundances provided by the direct method are
the most reliable and thus provide a reference abundance estimate. 

\begin{figure}
  \centering
  \includegraphics[width=84mm]{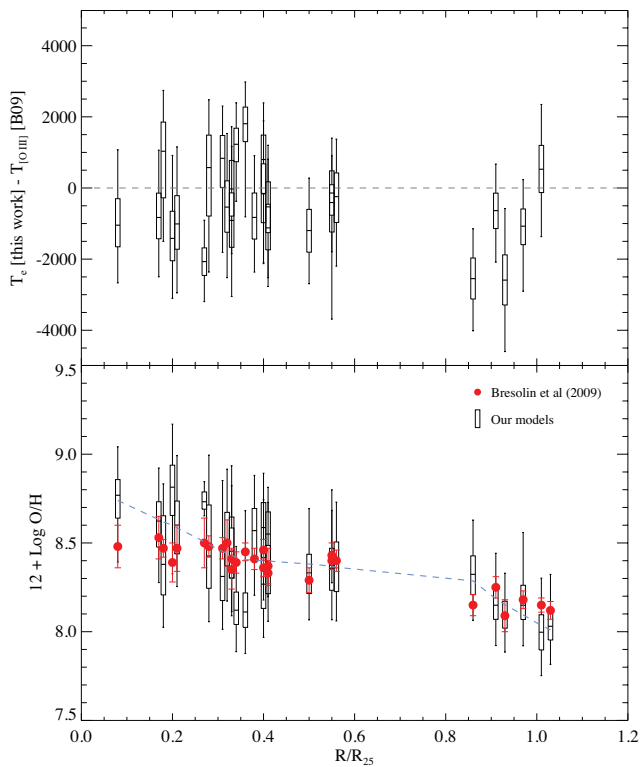}
  \caption{Top panel: The difference in effective $T_e$ from our fits
    and the electron temperature in the \oiii{} zone from the direct
    method calculation by~\citet{2009ApJ...700..309B}.  The agreement
    is satisfactory given that our fits do not include any
    particularly temperature sensitive lines. Bottom panel: The
    abundance as a function of the radius normalised by $R_{25}$ for
    NGC 300. The solid red circles show the oxygen abundance derived
    in~\citet{2009ApJ...700..309B} using the temperature sensitive
    method. The box plots show the oxygen abundances estimated using
    our model fits, using only the \oii{3727}, \hb, \oiii{5007}, \ha,
    and \nii{6584} lines in the fit. The horizontal line shows the
    median and the box encloses 50 per cent of the likelihood while the error
    bars enclose 95 per cent of the likelihood. The blue dashed line is a
    local regression, or loess, smooth of the medians of our estimates.}
  \label{fig:ngc300-comparison}
\end{figure}

For our comparison here we wish to extend this comparison to as high
metallicity as possible. To do so, we took the comprehensive study of
NGC 300 by~\citet{2009ApJ...700..309B} who published line fluxes for
\hii\ regions across the galaxy as well as abundances derived using
the direct method. We removed the de-reddening applied to the published
fluxes and ran our models. The resulting comparison of oxygen
abundance estimates is shown in Fig.~\ref{fig:ngc300-comparison}.

The top panel compares the mean electron temperature in our models to
that estimated from the direct method by Bresolin et al against the
radius in units of $R_{25}$. The difference between the $T_e$
predictions from our models and the published $T_{[\mbox{\scriptsize O\textsc{iii}}]}$ estimates,
including the published uncertainty on the $T_{[\mbox{\scriptsize O\textsc{iii}}]}$  values, are
shown as box-plots where the horizontal line shows the median, the
boxes enclose 50 per cent of the likelihood, and the vertical bars
enclose 95 per cent, or $\pm 2\sigma$ for Gaussian errors. The
agreement is quite satisfactory, in particular given that our fits,
in contrast to the direct method, do not include any lines that are
strongly sensitive to $T_e$, such as \oiii{4363}.

In the bottom panel we show the resulting oxygen abundance gradient.
The solid circles show the data from Bresolin et al, while the box
plots show the results of our fits, with the dashed line indicating a
local regression, or loess, smooth of the medians from our fits. The
immediate point to notice is that the abundances are in very good
agreement at nearly all radii --- this is in contrast to the
comparisons in~\citet{2009ApJ...700..309B} which used fitting formulae
for the strong line abundance estimates.

In general we see similar agreement at lower metallicities although
there has been some indication of a poorer fit in systems with
significantly enhanced nitrogen relative to oxygen
\citep{2007A&A...462..535Y}. This is likely caused by the fixed
abundance ratios adopted at a given metallicity in the current CL01
models. These systems typically have a very high
SFR/M$_*$ and make up only a very small fraction of the sample used in
the present work.

Most of our galaxies are more metal rich than the range probed in
Fig.~\ref{fig:ngc300-comparison}, so it is natural to ask whether
the agreement continues or not. This is uncertain but here we will
take the approach that we will a) test our gas mass estimates and look
for offsets there and b) make the reasonable assumption that although
there might be uncertainties in the \emph{absolute} abundances, the
\emph{relative} abundances should be robustly determined. In passing
we note that this uncertainty also leads to uncertainties in
the determination of \hmol\ masses due to the metal dependence of the
\XCO\ conversion factor. Indeed with current parametrisations of this
dependency, e.g.\ equations~(\ref{eq:XCO_B02}) and~(\ref{eq:5}) below,
the uncertainty on molecular masses due to metallicity uncertainties
is as large, or larger than those intrinsic to our method.

%
%
%
%

\section{Data used}
\label{sec:data}

In the next section we will discuss the tests we have made to verify
the present method for gas column density and dust-to-gas ratio
estimation. To carry out these tests we need both optical
spectroscopy to run our method on, as well as resolved and integrated
gas mass measurements from \CO\ and \hi\ observations to compare our
results with.

We use optical spectroscopy from the Sloan Digital
Sky Survey \citep[SDSS][]{2000AJ....120.1579Y} Data Release
7\footnote{\texttt{http://www.sdss.org/dr7}}
\citep[DR7][]{2009ApJS..182..543A}. The SDSS obtained 5-band imaging
using a drift-scanning camera \citep{1998AJ....116.3040G} on the 2.5m
telescope at Apache Point \citep{2006AJ....131.2332G}. Spectroscopic
targets were selected on the basis of the imaging and a tiling
algorithm described in \citet{2003AJ....125.2276B}. The spectra were
obtained using a fibre spectrograph with 640 fibres per plate, with
3'' fibre apertures. The photometric system is discussed by
\citet{1996AJ....111.1748F} with the photometric calibration being
covered in more detail in \citet{2002AJ....123.2121S}. Where relevant
we use the photometry from the Catalogue Archive
Server\footnote{\texttt{http://cas.sdss.org/dr7/en}}. We classify
galaxies into six classes following B04 based on their emission line
properties. In particular we classify galaxies with S/N$>3$ in
\hb, \oiii{5007}, \ha\ and \nii{6584} into star-forming, composite and
AGN based in the leftmost diagnostic diagram in
Figure~\ref{fig:models_xsi_variation}. In the following we will focus
on the star-forming (SF) class.

\begin{figure*}
  \centering
  \includegraphics[width=184mm]{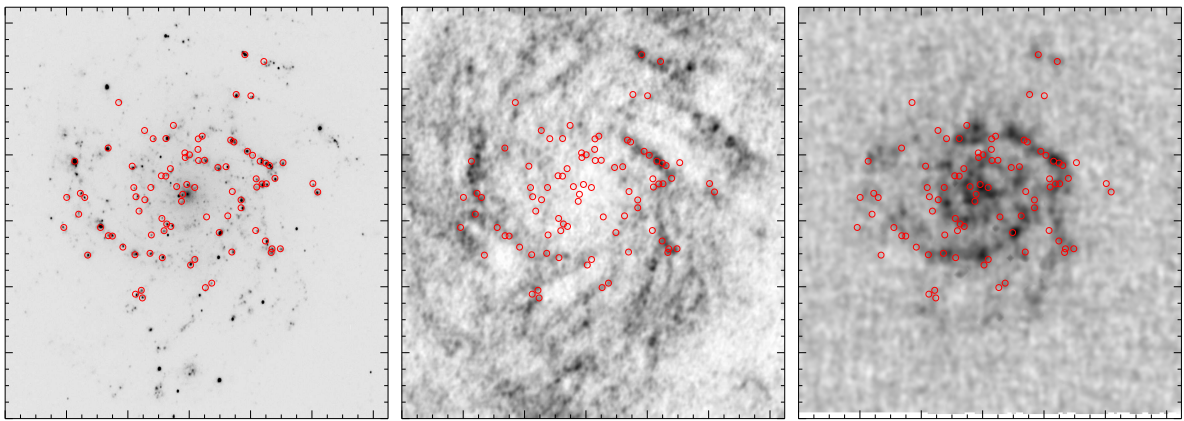}
  \caption{Three images of NGC 628, on the left the SINGS \ha\ image,
    in the middle the \hi\ image from the THINGS survey, and on the
    right the \CO\ image from the HERACLES survey. The locations of
    the 87 regions in NGC 628 where we calculate gas column densities
    from the IFU, \hi\ and \CO\ maps are overplotted as red
    circles. Note that the regions sample a range of \hi/\CO\ ratios.}
  \label{fig:ngc628-illustration-of-regions}
\end{figure*}

We will also make use of the massive Integral Field Spectroscopy
(IFS) mosaic of NGC 628 published by \citet{2011MNRAS.410..313S} as
part of the PINGS
project\footnote{\texttt{http://www.ast.cam.ac.uk/ioa/research/pings/html/index.html}}. This
was obtained with the PPAK spectrograph on the 3.5m telescope at
Calar Alto. We use the reduced data provided on the PINGS project
page and selected a total of 87 regions across the galaxy (see
Fig.~\ref{fig:ngc628-illustration-of-regions}). For each region we
sum up line fluxes within circular apertures of diameters of 6'' to
24'', we used the range of aperture sizes to test the sensitivity of
our results to aperture size, but as the results are only weakly
sensitive to aperture size we will focus on results for 12''
apertures in this paper unless otherwise stated. To ensure we have a
consistent alignment of the images, we tie the IFU mosaic world
coordinate system to that of the \ha\ image of NGC 628 provided
by the SINGS survey \citep{2003PASP..115..928K}.

We will also make use of resolved \hi\ maps from the THINGS survey
\citep{2008AJ....136.2563W} as well as \CO\ maps from the BIMA
SONG\footnote{Downloaded from\\
  \texttt{http://nedwww.ipac.caltech.edu/level5/March02/SONG/SONG.html}}
\citep{2003ApJS..145..259H}, Nobeyama\footnote{Downloaded from\\
  \texttt{http://www.nro.nao.ac.jp/\~nro45mrt/COatlas/}}
\citep{2007PASJ...59..117K} and HERACLES
\citep{2009AJ....137.4670L}\footnote{Downloaded from\\
  \texttt{http://www.cv.nrao.edu/\~aleroy/HERACLES/Overview.html}}
surveys --- we use HERACLES data whenever possible but use the other
data sets for reference.

 \begin{table*}
   \caption{An overview of regions with \hi\ and \CO\ interferometric
     data, while at the same time have spectroscopy from SDSS or from
     the NGC 628 IFU mosaic consistent with being ionised by stars. The
     complete table is provided electronically. The first column gives
     the name of the galaxy with 
     a running index for the region appended. The second column gives
     the SDSS spectroscopic identification, followed by the position
     of the fibre the spectrum was obtained through. The final column
     gives the zero-offset index of this object in the MPA--JHU value-added
     catalogues\citep{2004MNRAS.351.1151B} found at \texttt{http//www.mpa-garching.mpg.de/SDSS}.} 
\label{tab:multi-spec}
\begin{tabular}{@{}lcccc} \hline
Galaxy & PlateID-MJD-FiberID & \multicolumn{2}{c}{Ra (J2000) Dec} & IDR7\\ \hline
NGC2403-00 &  2944-54523-237 &  +07:36:51.23 &  +65:41:01.65 & 920120\\
NGC2903-00 &  2289-53708-615 &  +09:32:10.09 &  +21:30:08.26 & 794801\\
NGC2903-01 &  2292-53713-310 &  +09:32:09.68 &  +21:31:06.60 & 795658\\
NGC2903-02 &  2289-53708-616 &  +09:32:00.35 &  +21:27:40.40 & 794802\\
NGC2976-00 &  1879-54478-291 &  +09:47:15.31 &  +67:55:00.05 & 664806\\
NGC2976-02 &  1878-54474-601 &  +09:47:07.99 &  +67:55:53.14 & 664573\\
NGC2976-03 &  1879-54478-281 &  +09:47:07.99 &  +67:55:53.14 & 664798\\
\end{tabular}
\end{table*}

 \begin{table*}
 \caption{Measurements for the regions indicated in
   Table~\ref{tab:multi-spec}, the complete table is provided
   electronically. }
\label{tab:multi-spec-measurements}
\begin{tabular}{@{}lrrrrrrrrrr} \hline
  Galaxy &
  \multicolumn{1}{c}{$\log \Sigma_{\mathrm{\hi}}$}   &
  \multicolumn{1}{c}{$\log \Sigma_{\mathrm{mol}}$}   &
  \multicolumn{1}{c}{$\log \Sigma_{\mathrm{spec}}$}  & 
  \multicolumn{1}{c}{$\log \Sigma_{\mathrm{spec}}$}  & 
  \multicolumn{1}{c}{$\log \Sigma_{\mathrm{spec}}$}  & 
  \multicolumn{1}{c}{EW(\ha)} & 
  \multicolumn{1}{c}{EW(\hb)} & 
  \multicolumn{1}{c}{\ha/\hb} & 
  \multicolumn{1}{c}{$g-r$} & 
  \multicolumn{1}{c}{$u-r$} \\
 & & 
 \multicolumn{1}{c}{16\%}  &  \multicolumn{1}{c}{median}  & 
 \multicolumn{1}{c}{84\%}  & 
 \multicolumn{1}{c}{\AA} &
 \multicolumn{1}{c}{\AA} & & & \\ \hline
NGC2403-00 &   1.07 &   1.07 &   0.59 &   1.25 &   1.92 &   17.9 &    3.1 &  3.19 &  0.16 &  0.99\\
NGC2903-00 &   0.58 &   2.00 &   1.09 &   1.21 &   1.31 &   73.0 &   13.2 &  5.36 &  0.65 &  1.58\\
NGC2903-01 &   1.21 &   1.53 &   1.07 &   1.11 &   1.18 &  164.7 &   21.0 &  5.81 &  0.49 &  0.94\\
NGC2903-02 &   0.93 &   0.95 &   0.96 &   1.29 &   1.54 &   44.6 &   10.8 &  3.51 &  0.21 &  0.64\\
NGC2976-00 &   0.87 &   1.45 &   0.93 &   1.04 &   1.15 &   58.1 &   15.8 &  3.81 &  0.91 &  2.21\\
NGC2976-02 &   1.33 &   1.87 &   1.61 &   1.78 &   1.91 &  207.9 &   32.3 &  4.24 &  0.47 &  0.94\\
NGC2976-03 &   1.33 &   1.87 &   1.67 &   1.80 &   1.91 &  178.6 &   27.3 &  4.22 &  0.47 &  0.94\\
\end{tabular}
\end{table*}

We match these datasets to the SDSS to find regions where we have
resolved \hi\ and (optionally) \CO\ maps as well as SDSS spectra which
we can use to obtain total gas column densities using our method. We
searched the SDSS for all RC3 galaxies with multiple SDSS spectra
across the face of the galaxy and then matched this to the THINGS and
HERACLES surveys. This resulted in the identification of 100 SDSS
spectra in 12 galaxies, see Table~\ref{tab:multi-spec} for an overview
of these. All these galaxies have \hi\ maps from the THINGS survey and
\CO\ maps from the HERACLES survey, cross-checked against the BIMA
SONG and Nobeyama surveys when possible. For these maps we also
measure gas surface mass densities in apertures with diameters of 6",
12'' and 24''', but the spectroscopic aperture is of course fixed to
3'' by the SDSS fibre diameter.

We will also make use of integrated \hi\ data and for
this we make use of the ALFALFA survey $\alpha$-40 release
\citep{2011AJ....142..170H}, the second data release (DR2) for the
GASS survey \citep{2012arXiv1206.3059C} and the compilation of \hi\
data by \citet[][hereafter S05]{2005ApJS..160..149S}. For GASS DR2 we
use the catalogues provided by \citet{2012arXiv1206.3059C}, and for
ALFALFA we use the match to SDSS DR7 provided by
\citet{2011AJ....142..170H}. We match S05 to the SDSS DR7 using a 5
arcseconds match radius, which is small but our aim is not to be complete
here. We furthermore require that the velocity offset between the SDSS
and \hi\ redshifts is less than 250 km s$^{-1}$. When there are \hi\
observations from more than one survey, our order of priority is GASS,
ALFALFA and S05. In total this gives 11,253 unique matches, 416 from
GASS, 9,762 from ALFALFA and 1,259 from S05.

We calculate \hi\ masses in solar masses from the fluxes using
\begin{equation}
M_{\subhi}  [\Msun] = 2.36 \times 10^5 D^2  S
\label{eq:m_hi_def}
\end{equation}
where $S$ is the flux in Jy, or Jy/beam for resolved data, and $D$ is
the distance in Mpc. For the S05
dataset, which is abundant in edge-on galaxies, we adopt the flux
corrected for self-absorption, while for the other datasets we use the
observed flux. Note that we do not correct for inclination effects
because these would affect the \mugas\ derived from our method in
the same way as it would the \hi\ and \CO\ maps. Thus throughout we
will quote observed surface mass densities, not deprojected. 

For the molecular gas data, we follow \citet{2008AJ....136.2782L} and
calculate molecular surface mass density maps from CO $J=1\to 0$ maps
using 
\begin{equation}
  \label{eq:4}
  \Sigma_{\hmol} \left[\Msun \mathrm{pc}^{-2}\right] = 4.4 \frac{X_{\CO}}{2\times 10^{20}} I_{\CO},
\end{equation}
with $I_{\CO}$ being the \CO\ intensity in $\mathrm{K}\; \mathrm{km}\;
\mathrm{s}^{-1}$. We convert the HERACLES $J=2\to 1$ maps to $J=1\to
0$ maps using the conversion $I_\CO (2\to 1) = 0.8 I_\CO (1\to 0)$
recommended by \citet{2008AJ....136.2782L}.  $X_{\CO}$ is the
\CO-to-\hmol\ column density conversion factor in units of
$\mathrm{cm}^{-2}\,\mathrm{K}^{-1}\,\mathrm{km}^{-1}\,\mathrm{s}$
\citep[see for instance][]{2008AJ....136.2782L}. For Milky Way
conditions it appears that a constant value of $X_{\CO}=2\times
10^{20}\,
\mathrm{cm}^{-2}\,\mathrm{K}^{-1}\,\mathrm{km}^{-1}\,\mathrm{s}$ is
appropriate \citep[e.g.][]{2001ApJ...547..792D}, but as the galaxies
considered here span a range in physical properties it is likely
incorrect to adopt a fixed $X_{\CO}$
\citep[e.g.][]{1996PASJ...48..275A,1997A&A...328..471I}. Hence we also
make use of a variable $X_{\CO}$.  For this we use the metallicity
dependent \XCO\ from \citet{2002A&A...384...33B} as our reference
parametrisation,
\begin{equation}
  \label{eq:XCO_B02}
  \log X_{\CO} = -1.01 \log \mathrm{O/H} + 29.28,
\end{equation}
but we have also explored the stronger trend advocated by
\citet{1997A&A...328..471I} viz.\
\begin{equation}
  \label{eq:5}
  \log X_{\CO} = -2.6 \log \mathrm{O/H} + 11.6.
\end{equation}

Calibrations of $X_{\CO}$ against metallicity in the literature show a
large variation between different authors \citep[see][for
summaries]{2008ApJ...686..948B,2011ApJ...737...12L,2012ARA&A..50..531K}.
They range from finding no trend with metallicity, ie.\ a constant
$X_{\CO}$ via a logarithmic slope of $-1$ to $-2.6$. As
\citet{2008ApJ...686..948B} points out, there are systematic
variations between the techniques used to infer $X_{\CO}$ but the
implication is that one must view $X_{\CO}$ as being significantly
uncertain at low metallicity and a similar behaviour is also found at
$z>1$ \citep{2012ApJ...746...69G}. We remark that although the
majority of the effort lately has focused on the metallicity
dependence of \XCO, it is possible that other factors, such as the
intensity of the radiation field also could have an effect
\citep[e.g.][]{1997A&A...328..471I} but we ignore this complication
here.

\section{Tests of the methodology}
\label{sec:test-gas-estim}
 
We will test the method outlined in section~\ref{sec:method} in two
steps that logically follow from viewing
equation~(\ref{eq:mugas_eq17}) as having two components: The dust
column density and the dust-to-gas ratio, obtained through a
combination of the metallicity and the dust-to-metal ratio, of the
gas. We commence our testing by checking our dust-to-metal and
dust-to-gas ratio estimates. We do not test our metallicity estimates
explicitly here, see section~\ref{sec:import-abund-estim} for a brief
discussion of this. To test the gas column densities we take three
different approaches that have different inherent systematic
uncertainties and by combining these we should be able to provide a
robust test of our method. The first technique compares \mugas\
determined from SDSS spectra and the NGC 628 IFU mosaic with that of
\hi+\CO\ gas maps. The beams of the resolved maps are only larger than
the SDSS fibres by a factor of 2--3 so we anticipate only moderate
aperture effects in this comparison. We will find below that our
technique gives gas mass estimates that are in very good agreement
with the \hi+\CO\ maps.

However the derivation in section~\ref{sec:method} is only strictly
valid for angle-averaged quantities, not resolved properties as used
in this comparison. Furthermore, the sample for which we have good
quality \hi\ and \CO\ maps is relatively modest in size, and the
galaxies are fairly close to face-on. This leads us to the second
test which looks at how our gas estimates change as the aperture
sampled changes --- we will show that this way we can recover trends
of \hi\ content in galaxies found in recent \hi\ surveys when the SDSS
fibre samples a significant fraction of the stellar disk.

The natural final step is to compare our gas content measurements from
SDSS spectroscopy to measurments of total \hi\ masses which are
available for large samples of galaxies today. This requires careful
adjustments for the difference in apertures and it turns out that
although this comparison works well, systematic uncertainties in the
aperture corrections weaken any conclusion reached with this
test. This comparison is therefore discussed in
Appendix~\ref{sec:aperture_corr}.

Note that in the following we will also define
\begin{equation}
 \rgas = \frac{\Sigma_{\mathrm{gas}}}{\Sigma_*} \approx \frac{M_{\mathrm{gas}}}{M_*},
 \label{eq:rgas_def}
\end{equation}
as the gas ratio. Note that the second equality which we generally
will assume holds, assumes that the scale-lengths of the gas and the
stars are at least comparable. This is closely related to the gas fraction,
\begin{equation}
 \fgas = \frac{\Sigma_{\mathrm{gas}}}{\Sigma_{\mathrm{gas}} + \Sigma_*} = \frac{1}{1 + 1/\rgas}.
 \label{eq:fgas_def}
\end{equation}

\subsection{The dust-to-metal ratio as a function of metallicity}
\label{sec:dust-metal-ratio}

\begin{figure}
  \centering
  \includegraphics[width=84mm]{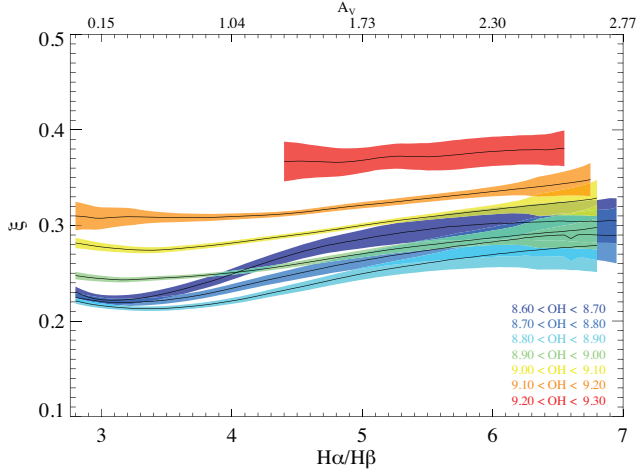}
  \caption{The dust-to-metal ratio of the gas as a function of the
    Balmer decrement in seven bins in oxygen abundance; the inferred
    $A_V$ using a \citet{1979MNRAS.187P..73S} extinction curve is
    shown on the top axis for reference. The shading indicate the
    uncertainty on the mean --- the spread around the mean is
    considerably larger.  It is clear that the trend with dust
    attenuation is similar at different metallicities although there
    is a trend for an increasing depletion at higher metallicity. } 
  \label{fig:xsi_vs_hahb}
    \end{figure}

\begin{figure}
  \centering
   \includegraphics[width=84mm]{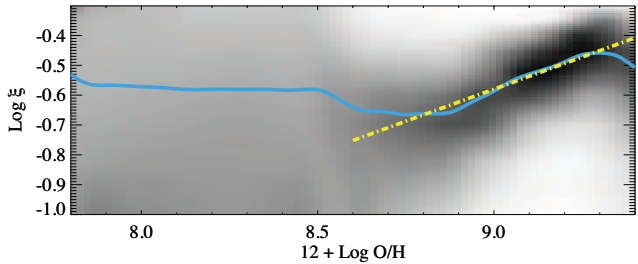}
  \caption{The dust-to-metal ratio of the gas as a function of
    the oxygen abundance. The grey scale shows the conditional
    likelihood of $\log \xi$ in bins of O/H. The solid blue line shows
    the median trend and the yellow dashed-dotted line shows the
    linear fit at high metallicity given in equation~\ref{eq:17}. Note
    the lack of constraints on $\xi$ below an oxygen abundance of
    8.6.}
  \label{fig:oh_vs_xsi}
\end{figure}

We discussed the properties of $\xi$ in
section~\ref{sec:fitt-spectr-feat} but we did not discuss what values
our fitting procedure provides. In the Milky Way we know there is a
correlation between depletion strength and (local) density
\citep[e.g.][]{2009ApJ...700.1299J} and Fig.~\ref{fig:xsi_vs_hahb}
shows that we find a similar result for our fits to star-forming
galaxies in SDSS DR7. This figure plots our $\xi$ values against
\ha/\hb\ as a proxy for line-of-sight attenuation.  We indicate the
V-band attenuation, $A_V$, obtained using a
\citet{1979MNRAS.187P..73S} extinction curve on the top axis for
reference.  The figure was constructed by co-adding the individual
PDFs of $\xi$ and \ha/\hb. The PDF of the latter was calculated on a
grid with bins of 0.043 while the PDF for $\xi$ was interpolated onto
a grid with bin size 0.004. We calculate the mean and uncertainty
on the mean on the resulting combined two-dimensional PDF and only
show the trends where at least 10 galaxies contribute to the bin in
\ha/\hb. In this and the following figure we use all SF class galaxies
in the SDSS since we are comparing two quantities derived within the
same region, the figure does not change significantly except in the
metallicity ranged spanned when limiting attention to a specific
redshift range.

The solid line shows the mean and the shading the uncertainty on the
mean for each trend. The results are shown for seven bins in oxygen
abundance, from $0.5 Z_\odot$ to $\sim 3 Z_\odot$. We do not include
lower abundance bins because, as we discuss next, we are unable to
constrain $\xi$ accurately at low metallicity.

The first point to notice is that we do see stronger depletion in more
dusty systems, similar to the trend seen along different sight-lines
in the Milky Way although the significant difference in methodology
bars us from making a much stronger statement, and the trend is fairly
weak. The second point to note is that although the depletion appears
to vary systematically with oxygen abundance, the dependence on
attenuation appears to be the same for most metallicities. 

This indicates a significant dependence of depletion on metal
abundance and we make this explicit in Fig.~\ref{fig:oh_vs_xsi} which
shows the trend of $\xi$ with oxygen abundance for 
star-forming galaxies from the SDSS DR7. This uses the full likelihood
distributions but is shown conditional on the oxygen abundance, with
the solid line indicating the median to guide the eye. This shows two
noticeable features. Firstly we can see a clear correlation between
$\xi$ and $12 + \log \mathrm{O/H}$ at high metallicity --- in the
sense that we predict a stronger depletion of elements in galaxies
with higher metallicity. For reference, the behaviour at high
metallicity is well described by
\begin{equation}
  \label{eq:17}
  \log \xi = -4.45 + 0.43 \mathrm{OH},
\end{equation}
where $\mathrm{OH} = 12 + \log \mathrm{O/H}$. This is a good fit to the data
for $12 + \log \mathrm{O/H}>8.7$ and is overplotted as a dashed-dotted
yellow line in the figure. But note that this includes the trend with
attenuation as well thus the intrinsic trend at fixed attenuation is
slightly shallower but we ignore this here.

The second point to note in this figure, is that we cannot place any
constraints on $\xi$ at metallicities lower than $12 + \log
\mathrm{O/H} \approx 8.6$, or $\sim 50$ per cent solar, recall that the solar
oxygen abundance for the CL01 models is 8.81. This appears to be a
consequence of the higher electron temperature in low metallicity gas
and subsequent lower sensitivity to $\xi$ as indicated by
Fig.~\ref{fig:te_o3ratio_vs_xsi_Z}. Thus we are unable to say whether the
dust-to-metal ratio continues to decrease, as an extrapolation using
equation~(\ref{eq:17}) would predict, or whether the relationship at
low metallicity is flatter, as for instance the chemical evolution
models by \citet{2008A&A...479..669C} suggest and which the fit
results hint at.

In the following section we will therefore refer to two different
estimates of $\xi$. Our first estimate will be that derived from the
fitting method and hence which shows a flattening at low metallicity,
the second will use the fit in equation~(\ref{eq:17}) to adjust $\xi$
after the fact --- we can do this because the model fits at low
metallicity are approximately independent of $\xi$. In the rest of the
paper we will for simplicity focus on the values provided by the
fitting method with and without a prior on $\xi$.

\subsection{Dust-to-gas ratios as a function of metallicity}
\label{sec:dust-gas-ratios}

\begin{figure}
  \centering
  \includegraphics[width=84mm]{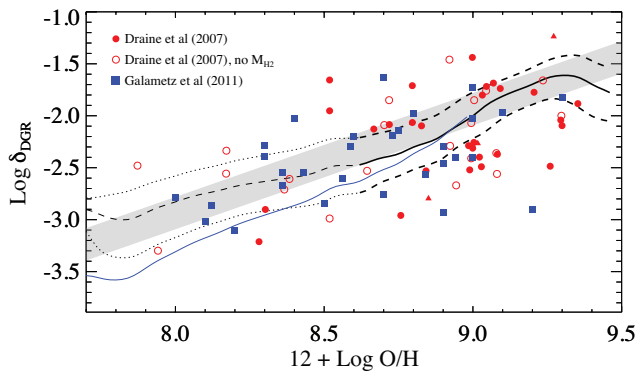}
  \caption{The solid/dashed line shows the median dust-to-gas ratio
    for the SDSS DR7 as a function of gas-phase oxygen abundance, with
    the dashed/dotted lines indicating the 68 per cent spread around
    this. The change in dashing at 12 + $\log \mathrm{O/H}=8.6$
    indicates the region where constraints on $\xi$ are poor. To
    illustrate the effect this has, the blue solid line shows the
    effect of assuming that $\xi$ follows equation~(\ref{eq:17}).  The
    open and filled circles show the \DGR\ estimates from Draine et al
    (2007) for SINGS galaxies --- filled circles for galaxies with
    \CO\ detections, empty circles for galaxies without this.  The
    solid blue squares are from the study of nearby galaxies
    by~\citet{2011A&A...532A..56G}. The grey shaded region shows the
      expected locus if the dust-to-gas ratio is scaled linearly with
      metallicity from the Milky Way value inferred by~\citet{2009ApJ...700.1299J}.}
  \label{fig:dgr_vs_draine}
\end{figure}

The dust-to-gas ratio in galaxies is a crucial parameter for a wide
range of models, and has been extensively modelled with focus on dust
emissivity \citep[see e.g.\ the compilation in][]{2011A&A...532A..56G}. As
remarked above a key ingredient in our approach to estimate gas
content is that we can place constraints on the dust-to-metal, and
by extension the dust-to-gas, ratio
\begin{equation}
  \label{eq:16}
  \DGR = \xi Z,
\end{equation}
using our emission line modelling. In this equation \DGR\ is the ratio
of dust mass density to that of the total gas mass. For comparison to
other work, which focuses on the mass ratio of dust relative to
hydrogen, we adjust our estimates here by a factor of 1.38 to take out
the contribution of helium to the gas mass density, a factor which
uses the solar abundance ratios from~\citet{2009ARA&A..47..481A}. 

To test our estimates, we focus our attention on the \DGR\ versus
oxygen abundance relation. We use the full likelihoods of these
parameters and show the resulting distribution in
Fig.~\ref{fig:dgr_vs_draine}. The solid/dashed black line shows the median
\DGR\ at the given oxygen abundance, while the dashed/dotted lines
show the 68 per cent spread around the median. In
section~\ref{sec:dust-metal-ratio} above, we showed that at $12 + \log
\mathrm{O/H}<8.6$, our method is not able to place strong constraints
on $\xi$ and as a consequence our estimate of \DGR\ is much less
certain. This region is therefore indicated with a different linestyle
in the figure.  To indicate the effect of modifying $\xi$, we overplot
the result of modifying our \DGR\ estimates using
equation~(\ref{eq:17}) as a thin blue line. At low metallicity the
effect can be almost 0.5 dex. 

We compare our estimates to those of~\citet{2007ApJ...663..866D}, who
modelled the dust emission of SINGS galaxies using a range of MIR data,
and those of~\citet{2011A&A...532A..56G} who used a similar approach
but a different modelling technique. These are overplotted in
Fig.~\ref{fig:dgr_vs_draine} as red circles (open for galaxies
lacking a CO detection) and filled squares respectively. For the
Draine et al data we have calculated oxygen abundances using our
models using the data described in~\citet{2010ApJS..190..233M}, kindly
provided by J.\ Moustakas. For the Galametz et al data we use the
metallicities provided in their table 1.  For consistency with
Galametz et al, we have adopted the \XCO\ dependency in
equation~(\ref{eq:XCO_B02}) for all data shown here. For the Draine et
al sample we use the gas masses measured within the region where dust
emission is detected --- see the discussion
in~\citet{2007ApJ...663..866D} for why this is the reasonable choice.

It is also interesting to compare our results with the simple assumption
that \DGR\ scales linearly with metallicity
\citep[c.f.][]{2007ApJ...663..866D}. To calculate this we have made
use of the depletion patterns derived by~\citet{2009ApJ...700.1299J},
adjusted to the CL01 abundances. This leads to a prediction for \DGR\
of
\begin{equation}
  \label{eq:19}
  \DGR = \mbox{($5.3\times 10^{-3}$--$1.1\times 10^{-2}$)} 
  \left(\frac{\mathrm{O/H}}{(\mathrm{O/H})_\odot}\right).
\end{equation}
Where the range is given for $F_*=0$ to $F_*=1$. The resulting region
is shaded in grey in Fig.~\ref{fig:dgr_vs_draine}. It is clear that
the linear scaling appears to be a decent overall description but our
data differ from this somewhat in detail, having a steeper slope down
to $1/2 Z_\odot$, and if we extrapolate with equation~(\ref{eq:17}) we
differ rather more at low metallicity.

The main conclusion from this test is that there is a good general
agreement between the measurements in the literature and our values
despite the very different techniques used. A notable difference
between our method and the other efforts in the literature is that our
measure of \DGR\ is a \emph{local} measure in contrast to methods that
use data obtained over different areas and with different spatial
resolution.

It is also worth asking whether a scaled dust-to-gas ratio would be
sufficient, since that appears to capture much of trend in
Fig.~\ref{fig:dgr_vs_draine}. This does of course depend on the
question asked, but it does rely on $\xi$ being approximately constant
and Fig.~\ref{fig:xsi_vs_hahb} and~\ref{fig:oh_vs_xsi} show that this
is a questionable assumption in detail. Ignoring the variation in
$\xi$ would lead to systematic errors in the dust-to-gas ratio that
could amount to as much as a factor of two across the bulk of the
galaxies in the SDSS and this would be a significant source of error
in many studies.



\subsection{Comparisons to resolved gas masses}
\label{sec:comp-resolv-hi}

\begin{figure*}
  \centering
  \includegraphics[width=184mm]{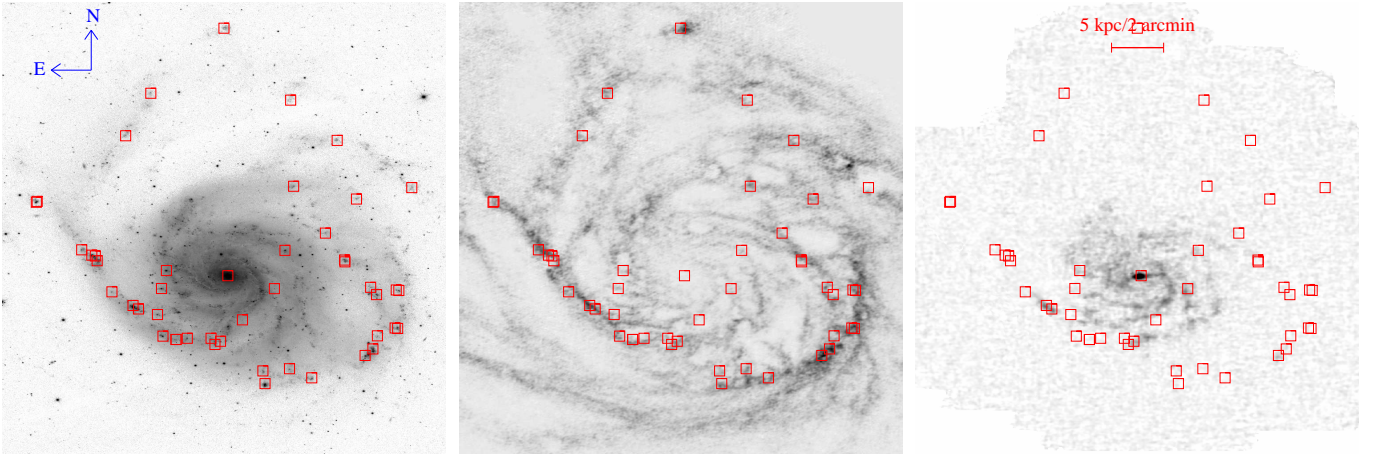}
  \caption{Left: SDSS r-band image of M101 with the location of SDSS
    spectra overlaid as red boxes. Middle: \hi\ map of M101 from the
    THINGS survey.  Right: The \CO\ gas mass map of M101 from the
    HERACLES survey.}
  \label{fig:M101-images}
\end{figure*}

We can carry out a direct test of our gas measurements when we have
spectroscopic observations of galaxies where we also have
interferometric maps in \hi\ and \CO. We will here use two sources of
such spectra: fibre spectra from the SDSS and spectra from the IFU
mosaic of NGC 628 by~\citet{2011MNRAS.410..313S}. We show these using
different colours below to verify that the method works equally well
for both classes of spectra.

As an illustration of our resolved maps, Fig.~\ref{fig:M101-images}
shows the $r$-band, \hi, and \CO\ images of M101 (NGC 5457) with the
location of spectra from SDSS indicated as red squares and NGC 628 was
shown earlier in Fig.~\ref{fig:ngc628-illustration-of-regions}. The
$r$-band image is a mosaic of SDSS $r$-band images, the \hi\ map is
from the THINGS survey and the \CO\ map from the HERACLES survey. We
have explored apertures from 6'' to 24'' when measuring \muHI\
and $\Sigma_{\CO}$, corresponding to 0.2--0.8 kpc at the adopted
distance for M101 of 7.4 Mpc \citep{2004AJ....127.2031K}, and
comparable sizes for the rest of the sample. Here we will only show
results for apertures of 12'' since this is close to the angular
resolution of the HERACLES data \citep{2009AJ....137.4670L}. Note that
this aperture is four times that of the SDSS data, for which all
\mugas\ measurements are obtained in an aperture of 3''. For the NGC
628 data we can synthesise spectra in arbitrary aperture sizes but
since we work with the emission line maps rather than the data cube,
this is not an exact sum of the emission flux within the aperture.

For each spectrum we run our model fits as detailed above and inspect
each resulting fit in detail. The SDSS spectra do not sample
\oii{3727} and generally have fairly low S/N which occasionally leads
to poorly constrained fits, while for the NGC 628 mosaic, the \sii{}
line fluxes are occasionally significantly off and we therefore do not
include \sii{} in the fits shown here. We exclude fits that give
multi-modal PDFs and very low S/N spectra --- quantitatively we make a
cut at a Kullback--Leibner divergence of 15 for the \mugas\ PDF. In
practice this corresponds to spectra with such weak emission lines
that no constraint can be had for \mugas. This results in a total of
104 (59 from SDSS, 45 from NGC 628) high-quality measurements. The
results do not depend much on these cuts because these poor fits
merely add scatter to the relation and since the uncertainty is large
their effect on any fit to the data is minor and we have verified that
including them does not significantly affect the conclusions below.

\begin{figure*}
  \centering  
  \includegraphics[width=184mm]{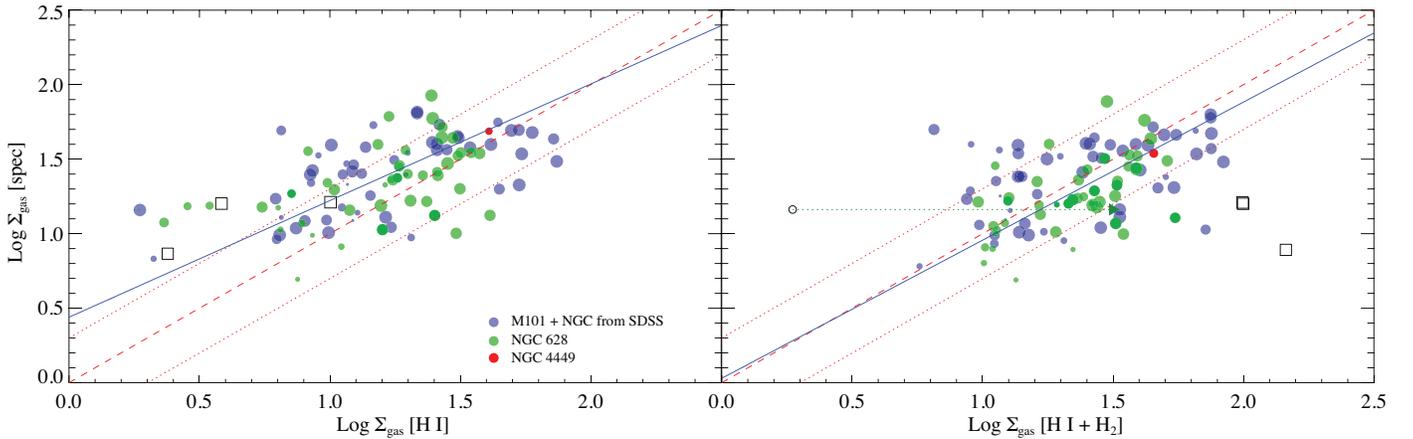}
  \caption{\emph{Left:} Surface-mass density of \hi\ measured in a
    12'' aperture versus our spectroscopic gas estimates for 59
    locations in 11 nearby galaxies with SDSS spectroscopy (blue) and
    45 locations in NGC 628 (green). The red solid symbol is for NGC
    4449. The points have been scaled according to the uncertainty
    estimate so that the largest circles correspond to objects with
    1$\sigma$ uncertainty on $\log \mugas$ less than 0.1 dex. The
    smallest symbols correspond to uncertainties larger than 0.6
    dex. The dashed red line shows the 1--1 relation, with the dotted
    lines either side showing a factor of 2 spread. The solid blue
    line shows the best-fit bisector fit to the data. \emph{Right:}
    Like the right-panel but now plotting the total gas content on the
    x-axis. We show this for a metallicity dependent \XCO\ and using
    the broad $\xi$ prior. The green line with an arrow shows the
    contribution of $\hmol$ to the central aperture in M101 with the
    location for \hi\ only shown as an open circle. From left to right
    the open squares indicate the position of NGC2903--00,
    NGC3627--01, and, NGC3351--00 see the text for a discussion.}
  \label{fig:mass_comp_M101}
\end{figure*}

We compare the \mugas\ from our models with the values extracted from
the \hi+\CO\ maps in Fig.~\ref{fig:mass_comp_M101}. The left panel
plots \muHI\ for these regions against the spectroscopic \mugas\
estimates. The point sizes in this figure are inversely proportional
to the uncertainty estimate on $\log \mugas$ between 0.1 and 0.6
dex. The blue line shows an unweighted bisector fit to the data which
has a slope of 0.79 and an intercept of 0.43. The dashed red line
shows the 1--1 relation with a factor of 2 offset indicated by the
dotted lines. While the data clearly correlate with each other, the
slope is noticeably different from 1.

The right panel shows the same but now using the total gas mass on the
abscissa. We calculated the molecular mass using the metallicity
dependent \XCO\ from equation~(\ref{eq:XCO_B02}) with the metallicity
from our model fit. We have also applied the broad prior on $\xi$
although the effect of both of these corrections are moderate as the
majority of the spectra are close to solar metallicity. For NGC 4449
(red filled circle) we use the observations
of~\citet{2000AJ....119..668H}, whose region B overlaps with several
of our spectra and when adjusted to our default \XCO\ has
$M_{\hmol}/M_\subhi=0.225$. The agreement is better in this panel with
the bisector fit having a slope of 0.93 with an intercept of 0.03.

The overall agreement must be said to be good as long as the total gas
mass is used, showing explicitly that the method traces the total gas
surface mass density. The slope and intercept does depend somewhat on
the aperture used for the gas measurements, whether a metallicity
dependent \XCO\ is used and whether or not a $\xi$ prior is
applied. For apertures of either 12'' or 24'', the slope of the best
unweighted bisector fit varies from 1.06 to 0.93 and the intercept
from -0.15 to 0.06, thus fully consistent with a 1--1 relation.

The scatter around the best linear fit varies from 0.30 to 0.36 dex,
which is mostly explained by the uncertainties on the variables. It is
difficult to accurately quantify the intrinsic scatter in this
relation as it depends sensitively on the uncertainty estimates on the
CO measurements which are hard to quantify due to the aperture
mismatch between the SDSS and the \CO\ data, as well as uncertainties
in the \CO-\hmol\ conversion factor, but fitting a linear relation
with intrinsic scatter following \citet{2007ApJ...665.1489K} the data
are consistent with a linear relation with intrinsic scatter of
0.1--0.15 dex.

Thus we can conclude that at least over the range of gas surface
densities from 10--100 $\Msun\;\mathrm{pc}^{-2}$, our spectroscopic
method gives results in agreement with those inferred from \hi+\hmol\
maps to within a factor of two. Although shown only for an aperture of
12'', this conclusion is robust to the aperture size chosen. Given the
lack of suitable comparison data, we are unable to extend this to
lower densities but when applied to SDSS data the method appears well
behaved down to at least a surface gas mass density of $\sim 2
\Msun\;\mathrm{pc}^{-2}$ where our sample starts to peter out.

The figure also shows that our estimator provides a measure of the
\emph{total} gas mass. As an example to highlight this, the nuclear
spectrum in M101 (1323-52797-012) corresponds to a location with
essentially no \hi\ gas and the resulting mass density is plotted as an
open circle in the right panel of Fig.~\ref{fig:mass_comp_M101} and
connected to the total (\hi+\hmol) mass density by a green line with
an arrow.

\begin{figure*}
  \centering
  \includegraphics[width=184mm]{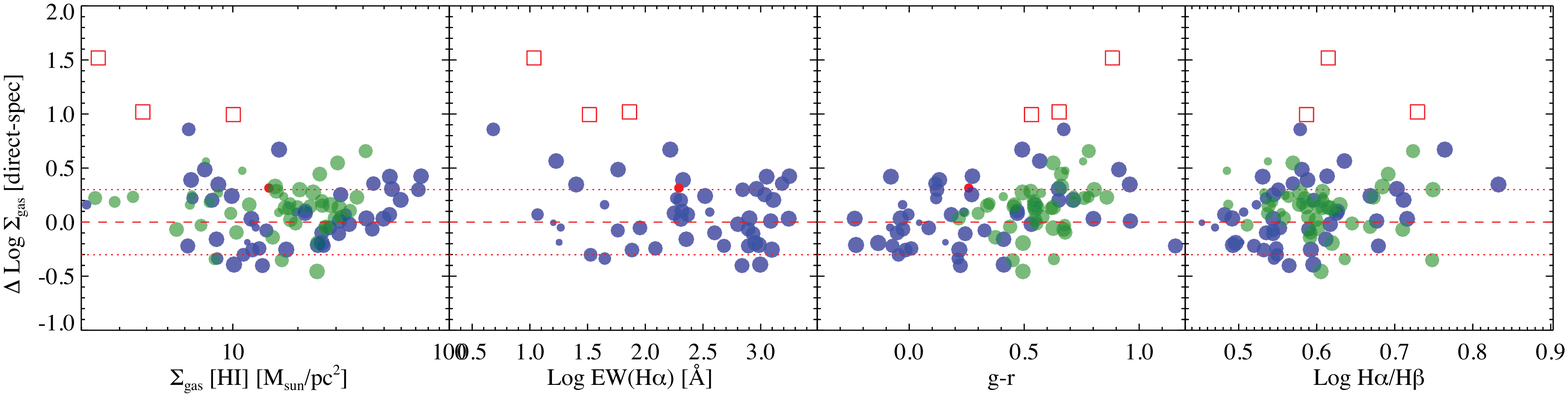}
  \caption{The difference between the spectroscopic and direct gas
    mass estimates in 12'' apertures plotted against the \hi\ surface
    mass density (left panel), equivalent width of \ha\ (second
    panel), the local $g-r$ colour (third panel) and \ha/\hb\ flux
    ratio (rightmost panel). The SDSS data are shown in blue and the
    NGC 628 data in green, the other symbols are as in
    Fig.~\ref{fig:mass_comp_M101}. The red horizontal line shows the
    zero level and the dashed lines around this shows a factor of 2 of
    scatter. We do not have equivalent width measurements for \ha\ for
    NGC 628.}
  \label{fig:residual_plot}
\end{figure*}

There are however some cases where the method fails. The three main
outliers are indicated with open squares in the figure. From left to
right these are: the nuclear bar in NGC 2903 (2289-53708-615), which
is well-known to have a particularly high molecular gas density
\citep[e.g.][]{2009AJ....137.4670L}, an off-centre region in NGC 3627,
and a nuclear spectrum in NGC 3351. The off-centre region in NGC 3627
has strongly discrepant \hmol\ estimates between BIMA and HERACLES and
if BIMA were used instead of HERACLES it would fall close to the
1--1 line, but the nuclear spectra have consistently high mass
estimates from BIMA and HERACLES and the disagreement with the
spectroscopic estimates appears to be real. Quantitatively we start to
see a systematic disagreement between our spectroscopic gas estimator
and the \CO+\hi\ maps at surface mass densities of gas above 
$\sim 75 \Msun\;\mathrm{pc}^{-2}$.

It is important to test whether the agreement between the two gas mass
measurements depend on other properties of the regions sampled in a
systematic manner. We therefore calculate the difference between $\log
\mugas$ inferred from \hi+CO data, the ``direct'' measure, and the
spectroscopic estimate. In Fig.~\ref{fig:residual_plot} we plot this
difference as a function of four observables. On the left we show it
as a function of \muHI\ and we see no clear evidence of a systematic
mismatch between the two as a function of \hi\ surface mass density.

The second panel shows the difference as a function of equivalent
width of \ha. No significant trend is seen, although we note that the
systems with high total gas mass density tend to have low EWs. This is
due to a combination of these being nuclear spectra, where the
background stellar light is significant, and to the star formation
here typically being obscured. There are no NGC 628 symbols here as
the data release does not provide equivalent widths.

In the third panel we plot the residuals against local colour (fibre
magnitudes for SDSS and aperture photometry on the SDSS images of NGC
628 otherwise). Here we see a slight systematic offset at the red end,
which is due to the nuclear spectra, where the very high \hmol\
densities are found. This could conceivably be used to identify
possible cases where our method fails, but we do not attempt to do
that here.

The final panel shows the residuals as a function of the Balmer
decrement. Here we see no trend showing that our method works well for
a wide range of optical depths. We have further verified that no
systematic trends are visible as a function of oxygen abundance or the
signal-to-noise of the spectroscopic gas estimate.  

We conclude from this that the method appears to work well for surface
densities of gas $<50\; \Msun\;\mathrm{pc}^{-2}$ but appears to break down
at surface densities $>100\; \Msun\;\mathrm{pc}^{-2}$.  This limitation is
expected, because at high optical depth the emission lines we see will
typically come from a skin outside the high optical depth region. 

While this is a real limitation, we should emphasise that it is a
problem only for nuclear spectra.  The scale sampled by a 3'' fibre
varies from 42 pc (NGC 4214) to 200 pc (NGC 3198), while the median
fibre in the SDSS $(z<0.2)$ subtends $\sim 5$kpc.  On larger scales we
will not be dominated by these extreme conditions in the majority of
galaxies.

\subsection{Moving from smaller to larger scales}
\label{sec:moving-from-smaller}

The preceding section showed that our method provides gas mass
measurements in good agreement with resolved gas measures for all but
the highest gas densities. Formally, however, our derivation in
section~\ref{sec:method} is only valid for angle-averaged quantities,
thus we would like to carry out our checks for larger samples and
since most of the SDSS fibre spectra sample much larger scales than
the $\sim 100\;\mathrm{pc}$ our tests above explored, we would also
like to extend our tests to larger spatial scales.

\begin{figure}
  \centering
  \includegraphics[width=84mm]{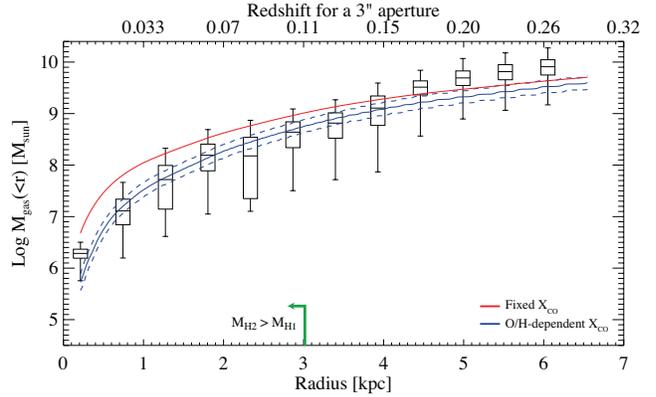}
  \caption{The total gas mass within apertures with different radii
    for NGC 628. The results are plotted as box-plots with the box
    enclosing 50 per cent around the median of the PDF, and the error bars
    enclosing 95 per cent of the likelihood. The solid red line shows the
    total mass estimate from the \hi\ and CO maps assuming our
    reference \XCO. The blue line with dashed error lines shows the
    result of assuming a metallicity dependent \XCO, using the full
    PDF of $12 + \log \mathrm{O/H}$ from our fits. The flux
    uncertainties for each radius were scaled to have the same S/N as
    the second radial bin before running the fits. The top axis shows
    the redshift at which a 3'' SDSS fibre would enclose a physical
    aperture with radius given on the lower axis. Note that as this is
    a cumulative plot, the error bars are not independent.}
  \label{fig:concentric_ngc628}
\end{figure}

In lieu of large samples of more distant galaxies with \hi\ maps, we
have two options. We start by synthesising apertures of increasing
size using the IFU mosaic of NGC 628. This allows us to test how well
we can recover the \hi\ content within the optical radius of the
galaxy.  The result of this exercise is shown in
Fig.~\ref{fig:concentric_ngc628}. The box plots summarise the PDFs
from our fits, while the blue line shows the total gas mass estimated
from the \hi\ and \hmol\ maps using the metallicity dependent \XCO\ in
equation~(\ref{eq:XCO_B02}).  The red line shows the result for a
constant \XCO.

The overall agreement is very good as long as we take into account the
metallicity dependence of $\XCO$, with the main uncertainty coming on
the largest scales. The agreement is typically better than a factor of
2 across a wide range of sampled sizes. It is also worth noting that
the agreement appears to be equally good regardless of whether the gas
is dominated by \hmol\ (within a radius of 3 kpc), or atomic
gas. There does seem to be a tendency for systematic offsets at very
large radii. The reason for this appears to be that only regions of
relatively high star formation activity have sufficient S/N to have
their line fluxes measured in the IFU data. These regions are
typically more gas rich than the average at those radii and hence the
spectroscopic estimates are biased high, but this should not be taken
to be a failure of the method, rather it is a sign that we are not
comparing like with like here.


We can extend this kind of test to the full SDSS sample by exploiting
the fact that the SDSS fibre aperture spans different physical sizes
at different redshifts, thus as redshift increases we should trace a
larger and larger fraction of the gas content of the galaxy. This will
\emph{not} provide total gas masses for the galaxies, but at $z\sim
0.2$ the SDSS fibre will subtend $\sim 10\;$kpc and encompass $\approx
50$ per cent of the light of the galaxy in our sample, so one might
expect that the gas content is more representative for the total.

To test this we wish to compare our $\rgas$--$M_*$ relation against
the $M_{\subhi}/M_*$--$M_*$ relation.  This has been well
characterised locally
\citep{1984ApJ...277..532B,2004ApJ...611L..89K,2010MNRAS.403..683C,2012arXiv1206.3059C},
and within large enough apertures, the dominant gas phase might
expected to be atomic \citep[but see][]{2011MNRAS.415...61S}. As an
example, ignoring \hmol\ for NGC 628 underestimates the total gas mass
by $<0.2$dex as long as we sum in apertures with radius
$>4.5$kpc. Thus the $M_{\subhi}/M_*$--$M_*$ relation should be a good
proxy for the $\rgas$--$M_*$ relation as long as the \hi\
determinations are from a region larger than this.

\begin{figure*}
\centering
\includegraphics[width=180mm]{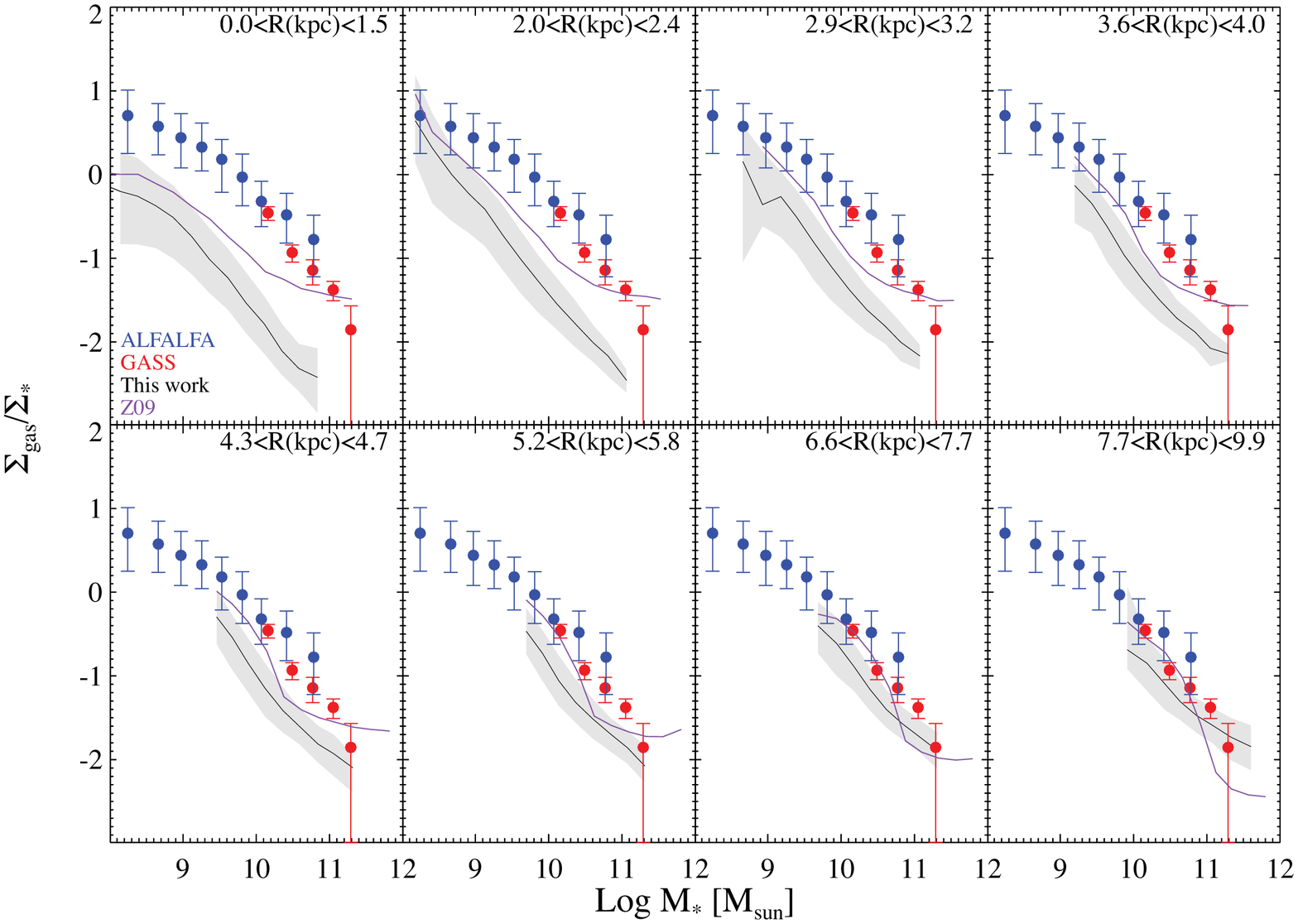}
\caption{Each panel compares the scaling law between stellar mass and
  \rgas\ in for star forming galaxies in the SDSS (solid line with
  $1\sigma$ scatter shown as shaded grey region) to the scaling law
  between $M_{\subhi}$ and stellar mass found using the ALFALFA survey
  (blue points), the GASS survey \citep{2010MNRAS.403..683C} (red
  points) and the $\rgas$ values predicted using the Z09 photometric
  gas content estimator (purple solid line). The different panels
  correspond to different redshifts, here indicated by the size
  subtended by the 3" fibre used by the SDSS spectrograph. At the
  lowest redshifts the gas fractions are clearly below the mean
  gas fractions from the \hi\ surveys because we sample the central parts
  of galaxies which are known to be less gas dominated, but as we
  sample more than $\sim 5\;\mathrm{kpc}$ the trends found approach
  each other.}
\label{fig:rgas_vs_mstar_vs_z}
\end{figure*}

The result of this comparison is shown in
Fig.~\ref{fig:rgas_vs_mstar_vs_z}. The solid black line here shows
the $\rgas-M_*$ trend found for star-forming galaxies in the SDSS
using the fibre spectra. Each panel corresponds to a range in physical
size subtended by the SDSS 3'' fibre, as indicated in each panel. The
blue and red points are from the ALFALFA and GASS \hi\ surveys
respectively, note that for GASS we only use the \hi\ detections as
the gas mass estimates using our method only includes objects with
measured gas content. Since only \hi\ detections are included these
are likely to be biased high, so we also include the median \rgas\
predicted using the photometric gas content technique introduced by
\citet[][Z09]{2009MNRAS.397.1243Z}, discussed further below, plotted
as the solid purple line. This is calculated for all galaxies falling
in the indicated radius/redshift bin.

When sampling small scales, the gas content is low relative to the
stellar content, as one would expect in most optically bright galaxies
(see for instance the atlas
in~\citet{2008AJ....136.2782L}). Nevertheless when the 3'' fibre
samples regions with diameter in excess of 5 kpc, we approach the same
scaling law found by the integrated \hi\ measurements but we always
fall below the locus of ALFALFA and GASS detections. This is not
unreasonable both because \hi-detected galaxies form a gas-rich subset
of all galaxies, and because gas disks generally are more extended
than the stellar disks of galaxies. We do, however, fall above the
locus predicted by the photometric estimator at high mass, indicating
that the sample for which we can estimate gas masses is a biased
sample and we will return to this topic in
section~\ref{sec:completeness} below.

While this is an indirect test of our method, it is an important
consistency check and taken together with the results for NGC 628 we
argue that it is consistent with the assumption that our method can
give reliable estimates of \mugas\ within the optical disk of a
galaxy. Taken together with the results of the previous
section we conclude that the technique works well from $\sim 100\;$pc
scales up to $\sim 10\;$kpc scales.  Our method will, however, not work
outside the optical disk of a galaxy because it relies on a background
source that the attenuation can be measured against.

We have also verified, but do not show, that we can recover the
scaling law between stellar surface mass density and the star
formation efficiency (SFR/$M_{\mathrm{gas}}$) found
by~\citet{2008AJ....136.2782L} using our method.

%
%
%
%


In theory we could make the preceding comparison much more direct by
exploiting the large samples of galaxies with integrated \hi\
masses. In practice, however, a direct comparison is problematic
because of the very large difference in apertures between the 3''
aperture of the SDSS spectrum and the 3.5' Arecibo beam.  We have
carried out this test and the details are discussed in
Appendix~\ref{sec:aperture_corr} but although the results are
encouraging, and consistent with the conclusions of the preceding
tests, the strength of this test is limited by the significant
systematic uncertainties on the aperture corrections.  One might
wonder if we could improve on this by focusing on a comparison for the
more distant objects where the SDSS fibre subtends a considerably
larger spatial size. However at those redshifts we hardly have any
\hi\ data to compare to so at present this is not a practical approach
to take.


\subsection{Comparison to other techniques}
\label{sec:comp-other-techn}

In closing this section, it is useful to compare our method to a
selection of indirect gas predictors in the literature. These are
generally calibrated on a sample of nearby galaxies and applied to
much larger samples.  Early work on this made use of the fact that
there is a good correlation between galaxy colour and the ratio of
\hi\ mass to B-band luminosity
\citep{1969AJ.....74..859R,1984ApJ...277..532B}.  More recently,
\citet[][K04]{2004ApJ...611L..89K} found a tight relationship between
galaxy colour, k-corrected to $z=0$, and $\rgasHI = M_{\subhi}/M_*$:
\begin{equation}
  \label{eq:13}
  \log \rgasHI = 1.46 - 1.06 (u-r),
\end{equation}
and a somewhat tighter relation with $u-K$ which we will not use
here. She showed that the scatter in this relation was only 0.42 dex
and could therefore be used to give reasonable estimates of a galaxy's
gas content on the basis of its colour.

An improved version of this was presented by Z09 who showed that an
even tighter relationship could be had by including the $i$-band
surface brightness, $\mu_i$:
\begin{equation}
  \label{eq:14}
  \log \rgasHI = -1.73238 (g-r) + 0.215182 \mu_i - 4.08451,
\end{equation}
with a scatter of only 0.31 dex. A similar scatter was also found for
massive galaxies based on the GASS survey by
\citet{2010MNRAS.403..683C}, incorporating instead the near-UV to $r$
colour and stellar mass surface density, $\mu_*$.  For simplicity we
will use the Z09 relationship in equation~(\ref{eq:14}), as this will
suffice for our needs here.

We calculated these gas mass estimates for the SDSS and NGC 628, when
necessary calculating k-corrected colours using version 4.1.4 of the
\texttt{kcorrect} code \citep{2007AJ....133..734B}. In contrast to Z09
we also use the stellar mass estimates from the \texttt{kcorrect}
code. For NGC 628 we obtained photometry in the 6'', 12'' and 24''
apertures on the basis of SDSS imaging of the object.

\begin{figure}
  \centering
  \includegraphics[width=84mm]{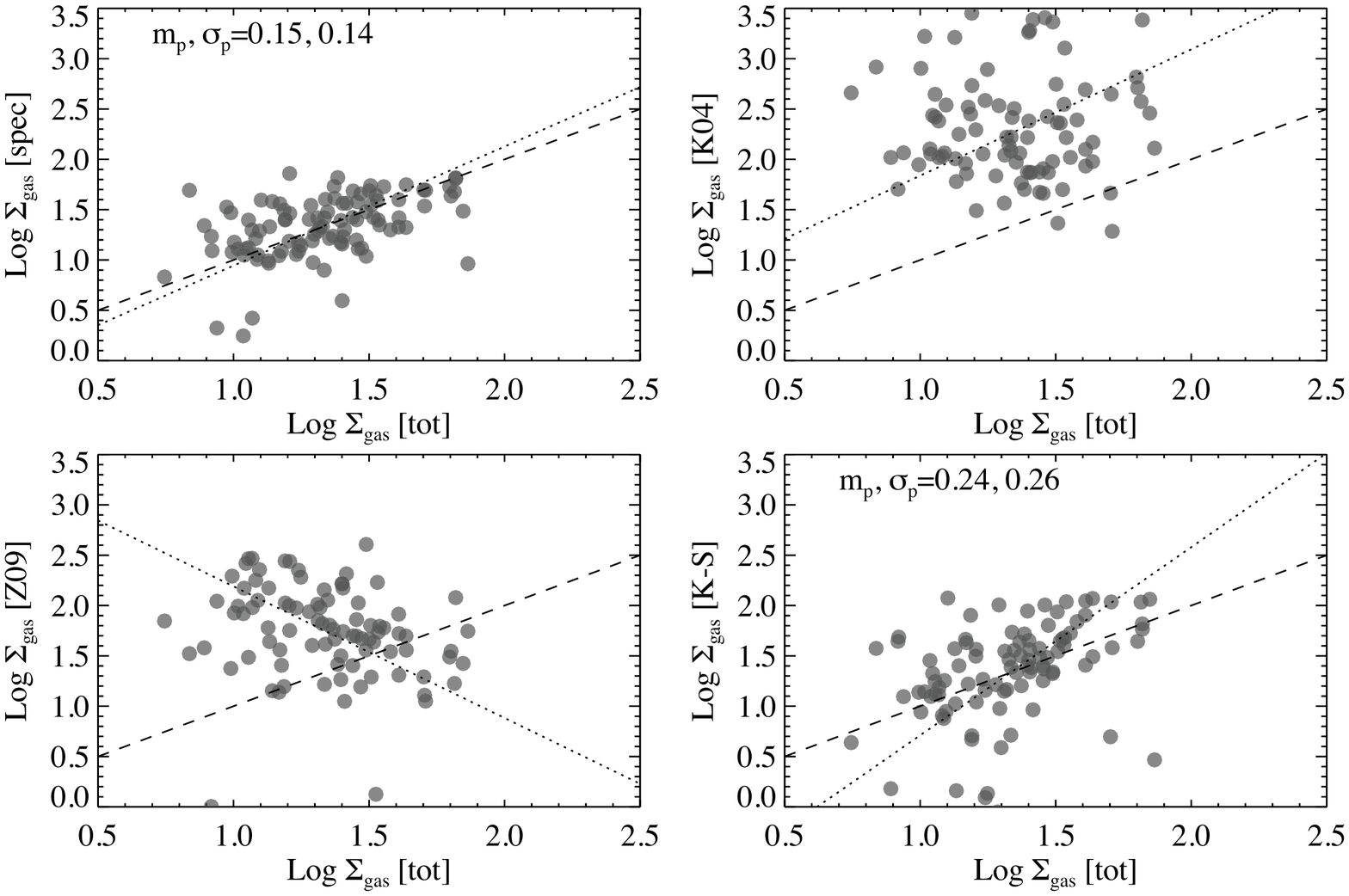}
  \caption{A comparison of four different gas mass estimators plotted
    against the total (\hi\ + molecular) gas mass using a metallicity
    dependent \XCO\ on the x-axis. The top left panel shows our
    spectroscopic gas mass estimates, the comparison in the top right
    panel uses the $\rgasHI$ -- $(u-r)$ calibration from K04, the
    bottom left the calibration from Z09 and the bottom right shows
    the results of inverting the Kennicutt--Schmidt (K--S) relation from
    \citet{1998ApJ...498..541K}. The dashed line is the 1-1 line,
    while the dotted line shows the best linear bisector fit to the
    data shown. Our estimator clearly does best, while the K--S
    relation does well but has more scatter and a slope differing
    somewhat from 1.}
  \label{fig:m101-other-estimators}
\end{figure}

An alternative technique, widely used at high redshift
\citep[e.g.][]{2006ApJ...646..107E}, is to invert the
Kennicutt--Schmidt (K--S) scaling law
\citep{1959ApJ...129..243S,1998ApJ...498..541K} between the surface
mass density of gas and star formation rate per unit area:
\begin{equation}
  \label{eq:15}
  \Sigma_{\mathrm{SFR}} = 2.5 \times 10^{-4} \,\Sigma_{\mathrm{gas}}^{1.4},
\end{equation}
where $\Sigma_{\mathrm{SFR}}$ is in units of
$\Msun\;\mathrm{yr}^{-1}\;\mathrm{kpc}^{-2}$ and $\Sigma_{\mathrm{gas}}$ is in
units of $\Msun\;\mathrm{pc}^{-2}$, and where uncertainties on the
parameters are suppressed. We have written the equation here for a
Salpeter IMF. To follow common usage in the literature we derive the
SFR from the \ha-luminosity and correcting for dust attenuation using
\ha/\hb, assuming a fixed Case B value of 2.86 and $\tau(\lambda)
\propto \lambda^{-1.3}$, but we have checked that we get similar
results if we use the more sophisticated modelling of B04.

Fig.~\ref{fig:m101-other-estimators} shows the comparison of these
three gas estimators together with our spectroscopic method against
the total gas content for the regions in nearby galaxies and NGC
628. We have excluded the three regions marked with a red square in
Fig.~\ref{fig:mass_comp_M101} for clarity as these are outliers for
all the estimators.  The dashed line in each panel is a 1--1 relation,
while the dotted line is the best unweighted bisector fit to the
data. For the K--S scaling relation and our method we also indicate the
median perpendicular distance to the 1-1 line, $m_p$, as well as the
perpendicular scatter around this line, $\sigma_p$.

We can immediately see that the K04 and Z09 estimators correlate
poorly with the local gas content. This is not unexpected because they
were calibrated on the \emph{integrated} gas content of galaxies and
they should not be expected to work for resolved properties. While a
slightly unfair comparison, it does highlight very clearly the problem
of applying a relation calibrated on local data outside of its range
of validity --- a similar caution would be valid for an application at
high redshift as well. For statistical studies of total gas content in
low redshift galaxies, they are however well suited.

The Kennicutt--Schmidt relation, in contrast, is sensitive to local
properties and it clearly does a much better job than the colour-based
techniques in Fig.~\ref{fig:m101-other-estimators}. Nevertheless,
the scatter around the 1-1 line is 85 per cent larger then with our method
and the correlation has a slope different from one. This is not
surprising as it is known that the K--S relation shows significant
variations
\citep[e.g.][]{2007ApJ...671..333K,2008AJ....136.2846B}. But the main
problems with this method are not seen in this figure: a) It is
calibrated on nearby galaxies so it is not a given that it would work
for all galaxies and at all redshifts and b) it is a transformation of the
star formation rate so it would be meaningless to plot it against a
star formation rate indicator.

\section{The central gas content of local galaxies}
\label{sec:trends}

As an example application of the methodology outlined above we now
turn to look at how the central gas content in star-forming galaxies
in the local Universe varies with the physical properties of the
galaxies. We will concentrate on galaxies from the SDSS DR7 for which
we have fibre spectra.

\begin{figure}
  \centering
  \includegraphics[width=84mm]{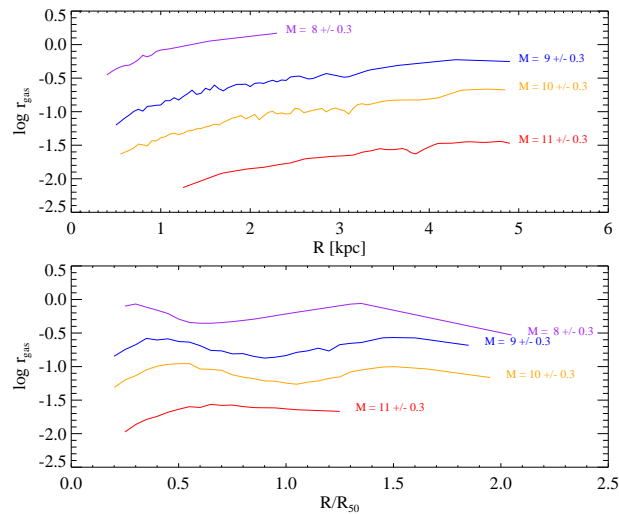}
  \caption{Top panel: The median $\log \rgas$ as function of radius
    for four ranges in stellar mass. Note that in each range we
    see a similar trend towards higher \rgas\ when we sample larger
    radii but that the typical \rgas\ increases with decreasing
    stellar mass. Bottom panel: The median $\log \rgas$ as a function
    of the size of the fibre in units of the half-light radius of the
    galaxy. Note the lack of correlation between the normalised radius
    and the gas ratio. }
  \label{fig:rgas_vs_r}
\end{figure}

Since the fibre spectra subtend a fixed size in observer space we are
faced with the fact that we sample different parts of the galaxies at
different redshifts. As discussed in appendix~\ref{sec:aperture_corr}
aperture corrections suffer from substantial systematic uncertainties
thus we prefer here to focus on galaxies with for which the aperture
samples similar scales. Given the quantities we have available this
could imply focussing either on a fixed physical scale or on a fixed
fraction of the galaxy size. The former ensures that the galaxies
considered all lie in a narrow redshift range, while the latter option
mixes galaxies across a range of redshifts but might provide a more
physically meaningful result.

To illustrate the variation in properties with these two different
definitions of central regions as well as to explore the impact of
different aperture sizes, we calculate \rgas\ in bins of stellar mass
as a function of aperture size in physical units, as well as in bins
of the size of the aperture relative to the half-light radius of the
galaxy. The results are shown in Fig.~\ref{fig:rgas_vs_r} and show
that \rgas\ is effectively independent of the aperture size in units
of the half-light radius of the galaxy out to $\sim 2\times R_{50}$.
If we contrast this with Fig.~\ref{fig:ap_corr_vs_T_type}, we see some
differences which in part is due to that figure using \hi\ only,
adding \hmol\ will flatten the central profiles, but it is also likely
that part of this flat structure is a sign of the biased nature of our
sample relative to the total galaxy population and we will return to
this below.

Since both these definitions of ``central'' are defendable, we will
show results for both in the following. Specifically we will look at
gas content in the central $2\pm 0.5$ kpc, corresponding to
$0.05<z<0.09$, when we fix the physical scale. For simplicity we do
not attempt to correct for aperture effects within this redshift
interval, but based on Fig.~\ref{fig:rgas_vs_r} these would be $\pm
0.1$dex in median. We also will show results when the sample is
limited to $0.7<r_n<1.4$, with $r_n \equiv R/R_{50}$. The samples
resulting from these two cuts both include $\sim 7\times 10^4$
galaxies. These definitions should be considered a central, but not
nuclear, gas content of star-forming galaxies in the nearby
Universe. The qualitative results in the following section are only
weakly sensitive to this choice of aperture but the quantitative
results do of course change. The ultimate consequence of this is that
the results we derive below are not representative for the
\emph{total} gas content of galaxies, for which suitable studies are
available in the literature
\citep[e.g.][]{2012arXiv1206.3059C,2010arXiv1006.5447S,2011MNRAS.415...32S}.

\subsection{The central gas-to-stellar ratio in galaxies in the nearby
  Universe}
\label{sec:gas-content-as}

\begin{figure*}
\centering
\includegraphics[width=175mm]{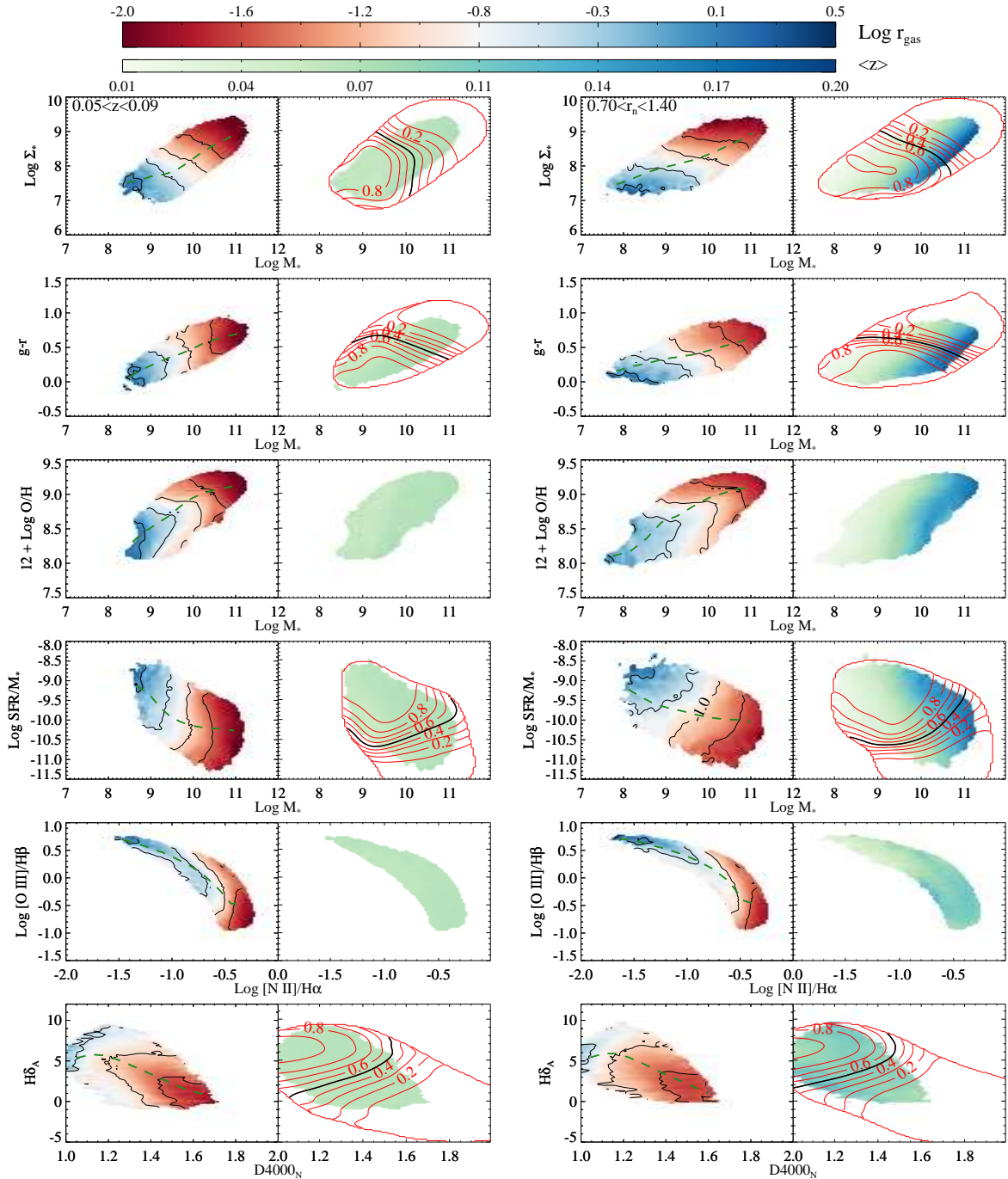}
\caption{The gas content of star-forming galaxies within $2\pm 0.5\;$
  kpc (left column) or within $0.7<r_n < 1.4$ (right column) in
  the SDSS as a function of various physical parameters. Each quantity
  is shown with the average $\log \rgas$ in the left panel with blue
  to red colour scale, while the panel of the right shows the mean
  redshift in the colour scale (green to blue) and overlaid a contour
  map of the fraction of all SDSS galaxies that we have gas
  measurements for (see text for details).  The top row shows the
  trend with stellar mass and stellar surface mass density, while the
  following rows show \rgas\ in the $\log M_*$ versus rest-frame $g-r$
  colour plane, the $\log M_*$--oxygen abundance plane, the $\log M_*$
  versus specific SFR plane, the BPT line ratio diagram and finally,
  the D4000$_N$ versus H$\delta_A$ plane. Contour lines are overlaid
  in steps of 0.5 dex from $\log \rgas = -2$. In the mean redshift
  panel, the completeness contours are not shown for the BPT plane or
  the $\log M_*$--oxygen abundance plane since essentially all
  galaxies that we can plot in these planes have gas mass
  estimates. Notice the very regular trend with various parameters but
  note in particular that the trends are \emph{not} a function of
  stellar mass only, or indeed any other single parameter
  considered. The dashed green line in each panel shows the median
  value of the quantity plotted on the y-axis as a function of the
  parameter on the x-axis.}
\label{fig:rgas_vs_various}
\end{figure*}

We show our inferred variation in \rgas\ in the central regions of
galaxies in Fig.~\ref{fig:rgas_vs_various}. The details of the construction
of these maps are given in appendix~\ref{sec:pract-creat-2d} but they
are basically 2D histograms of the average \rgas\ combined with a
kernel estimator where there are few galaxies.  The figure shows two
main columns, on the left we show the results when limiting ourselves
to a small range in physical aperture size, while the right hand
column shows the same for a limited range in aperture size relative to
half-light radius. In each row we have $2+2$ panels with the left
panel in each pair showing the average $\log \rgas$ in each bin, while
the right panel shows the mean redshift in each bin as a colour-scale
and the completeness in each bin as a contour plot. We will discuss
these further below but focus on the mean $\rgas$ maps first.

The top row shows the variation in \rgas\ in the stellar mass versus
stellar mass surface density plane. The top colour scale gives the
median $\log \rgas$ at a given position and for ease of reading we
have also overlain contour lines that are spaced 0.5 dex apart
starting at $\log \rgas=-2$. What is particularly noticeable here is
that the contour lines are not parallel to either coordinate axis
which implies that neither $M_*$ nor $\Sigma_*$ alone are sufficient
to predict the central gas content of these galaxies.  This was also
pointed out by Z09 and was one of their reasons for including a
surface brightness term in their calibration for \rgas\ versus colour,
but note that their conclusion was relevant for the \emph{total} gas
content, here we show that a similar result is appropriate also for
the central regions of galaxies. Also note that while the slope of the
contour lines differ between the cut in physical size and in $r_n$, as
one would expect from Fig.~\ref{fig:rgas_vs_r}, the basic conclusion
is still the same. 

The second row shows the mean \rgas\ in the plane of stellar mass
versus rest-frame total $g-r$ colour. We see that galaxies with a
particular \rgas\ can have a range of colours but here there is a
qualitative difference between a limit in physical size and in
$r_n$. When we limit ourselves to the central $2\pm 0.5\;$kpc, the gas
content at fixed stellar mass does not vary much with integrated
colour. In contrast, when we limit ourselves to a range in $r_n$, we
find a clear correlation so that redder galaxies have less central gas
at a fixed stellar mass. 

The mass-metallicity plane is shown in the following row. This was
examined also by Z09 and we find a similar diagram to those
authors. What is particularly notable here is that there is a
significant trend in \rgas\ across the mean $M_*-Z$ relation
\citep{2004ApJ...613..898T}. Also note the constant \rgas,
particularly at $12 + \log \mathrm{O/H} > 8.6$ and in the right
column. This is consistent with the expectations of a closed-box
model, but we note that this does not appear to hold everywhere in the
$M_*-Z$ plane and indeed a full exploration of this result is outside
the scope of the present paper.

The subsequent row shows the plane of stellar mass versus specific
star formation rate, $\mathrm{sSFR}=\mathrm{SFR}/M_*$. We see clearly
that the gas fractions of galaxies with the same SFR/$M_*$ decline
steadily with mass but that there is also a slight trend for galaxies
at fixed $M_*$ to have declining gas fractions as the specific SFR
declines, although the decline is not very strong. This is more
visible when we focus on a range in $r_n$. 

The penultimate row shows the gas fraction distributions as a function
of position in the BPT diagram. The main point to note here is that
the trend is smooth with position in the BPT diagram but with some
trend across the main star forming ridge
\citep[c.f.][]{2008MNRAS.385..769B} and decreases towards the
composite/AGN branch.

The final row shows how the gas fraction change with position in the
D4000$_N$ versus H$\delta_A$ diagram. Recall that galaxies with high
H$\delta_A$ have undergone a significant burst in the last $\sim 1$
Gyr, while the average age of the stellar population typically
increases with increasing D4000$_N$. The first point to note that
along the median line the gas fractions decline steadily towards
larger D4000$_N$. This median line is close to that expected for a
slow declining star formation rate
\citep[e.g.][]{2003MNRAS.341...33K}, while above it we expect to find
post-burst systems.

While these 2D distributions really capture the interdependencies of
various parameters, it is useful to simplify these into scaling
relations. Focusing on the relationship with stellar mass and stellar
surface mass density, we get
\begin{eqnarray}
  \label{eq:20}
  \log r_{\mathrm{gas}} & = & (8.76 \pm 0.03) - (0.474 \pm 0.004)
  \log M_* \\ 
  &  & - (0.619\pm 0.004) \log
  \Sigma_*
\end{eqnarray}
with a 1$\sigma$ scatter of 0.27 dex. If one only includes the stellar
mass in the fitting, the best-fit relation is 
\begin{equation}
  \label{eq:23}
  \log r_{\mathrm{gas}} = (8.17 \pm 0.03) - (0.925 \pm 0.003) \log M_*
\end{equation}
which has a 1$\sigma$ scatter of 0.33 dex. Qualitatively these
relationships are similar to that found for total \hi\
\citep[e.g.][]{2012arXiv1206.3059C} and CO
\citep[e.g.][]{2011MNRAS.415...32S}, but since the trends found here
are for the central regions, they differ quantitatively, primarily in
amplitude.


\subsection{The gas depletion time of galaxies in the nearby Universe}

\begin{figure*}
\centering
\includegraphics[height=184mm]{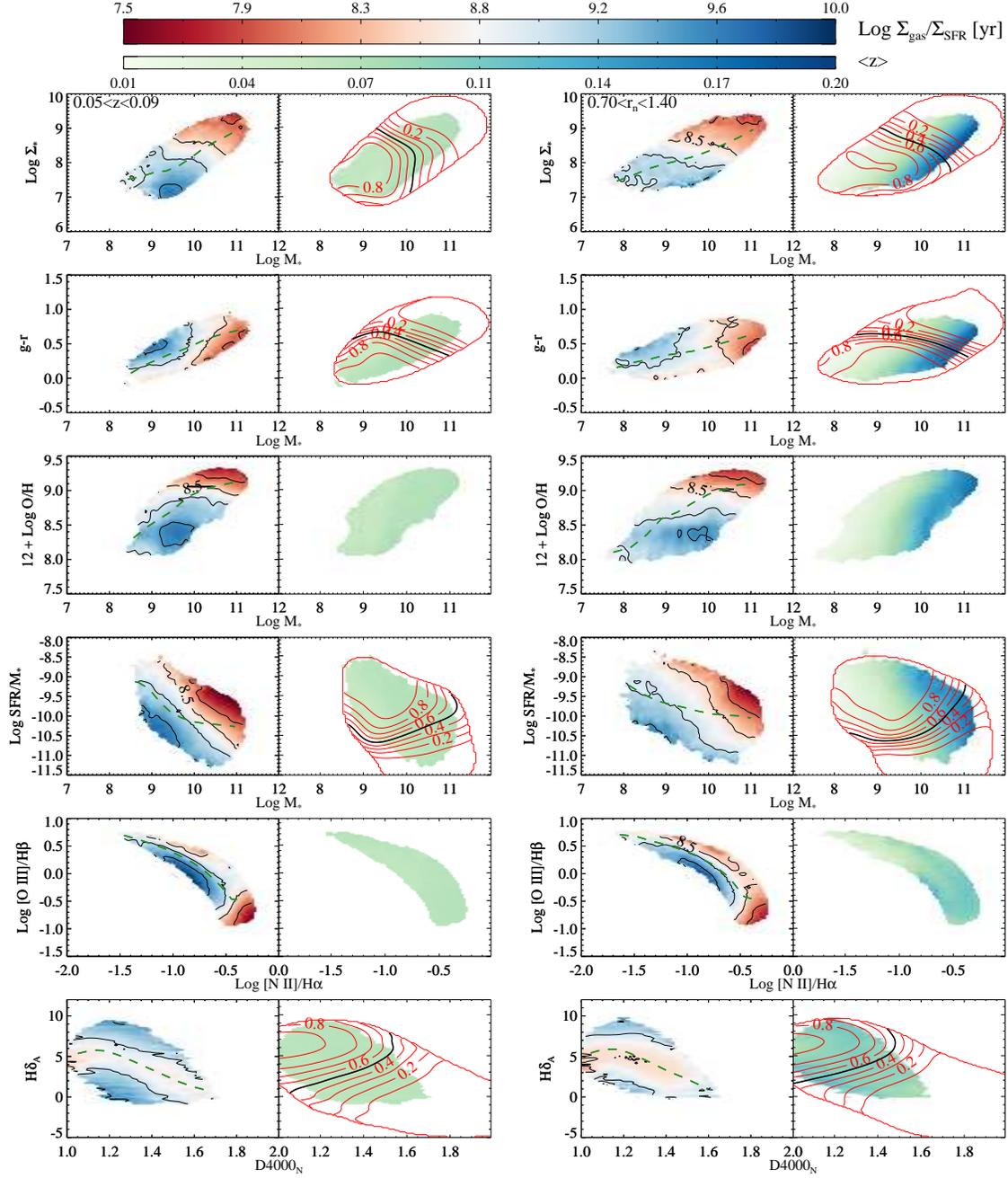}
\caption{Similar to Fig.~\ref{fig:rgas_vs_various} but this time
  showing the median gas depletion time in each bin.}
\label{fig:tr_vs_various}
\end{figure*}

Besides the fact that our method provides a measure of the gas content
of the galaxies, one key advantage of our methodology over estimating
gas content using the Schmidt--Kennicutt relation is that we can
compare our gas content to the star formation rate. Since we have star
formation estimates for exactly the same regions we have gas estimates
for, the calculation of the gas depletion time,
\begin{equation}
  t_R = \frac{\Sigma_{\mathrm{gas}}}{\Sigma_{\mathrm{SFR}}} \approx
  \frac{M_{\mathrm{gas}}}{\mathrm{SFR}}, 
\label{eq:roberts-time}
\end{equation}
is straightforward in principle. In practice we rewrite
equation~(\ref{eq:roberts-time}) to give
\begin{equation}
  \label{eq:25}
  t_R = \frac{\mathrm{Area}}{L_{\ha}} \; \eta_{\ha} \mugas\equiv
  \frac{\mathrm{Area}}{L_{\ha}} f_{t_R},
\end{equation}
where Area is the area subtended by the 3'' fibre in square parsecs
and $\eta_{\ha}$ is the conversion factor between \ha\ luminosity and
the SFR \citep{2001MNRAS.323..887C} defined through $\mathrm{SFR} =
L_{\ha}/\eta_{\ha}$. We then calculate the likelihood distribution for
$f_{t_R}$ and insert this in equation~(\ref{eq:25}) to calculate a PDF
for $t_R$. This accounts for intrinsic correlations between the
parameters and we use this in the following. None of the conclusions
below would change significantly if we were to use point estimates of
the SFR and $\mugas$ and assume they were independent but the
uncertainties would be more difficult to estimate and there is a
slight overall bias in $t_R$ --- this is expected, see appendix A
in~\citet{2004MNRAS.351.1151B} for a discussion.

Fig.~\ref{fig:tr_vs_various} is similar to
Fig.~\ref{fig:rgas_vs_various} and shows how the median $t_R$ varies
across the same parameter spaces.  Comparing these two figures it is
immediately obvious that long depletion times do not directly follow
high gas fractions.  Another immediate result is that galaxies that
have very long depletion times are located in very particular parts of
most diagrams. They have low $\Sigma_*$, are red for their mass, lie
well below the median mass-metallicity relation but also have very low
SFR/$M_*$ for their mass as well as a low \oiii{5007}/\hb. We note in
passing that a similar class of objects can be identified if one uses
the Z09 calibration for gas mass estimation.

The D4000$_N$--H$\delta_A$ diagram is particularly intriguing --- it
shows that along the median trend, shown by the green dashed line, the
galaxies have near constant depletion time of $\sim 10^8$ years in
their central regions. However Fig.~\ref{fig:rgas_vs_various} shows
that the gas fractions decrease steadily moving along the median curve
from low to high D4000$_N$. This strongly hints at a regulatory
mechanism for the central star formation rate, similar to that
proposed to explain the same result for the total depletion times of
galaxies by \citet{2010arXiv1006.5447S}, see also the following
section.

The median (average) depletion time we are finding for the central 2
kpc is 2.2 (3.6) Gyr and a similar result if were to limit ourselves
in $r_n$ instead. We note that this is very close to the mean
molecular depletion time, 2.35 Gyr, found on $\sim 1$kpc scales
by~\citet{2011ApJ...730L..13B} in nearby spiral galaxies with CO and
\hi\ maps. Since the inner 2 kpc is likely to have a very significant
molecular component, these numbers should be comparable.

In contrast to~\citeauthor{2011ApJ...730L..13B}, however, we find a
significant variation in depletion time with galaxy properties in the
central regions. Instead our results match closely the strong
variation in molecular depletion time with galaxy properties found
by~\citet{2011MNRAS.415...61S} in galaxies from the COLDGASS
survey and we will return to this in a future paper.

The distribution of $t_R$ values in the $\log M_*$--$\sSFR$ diagram
for the inner $2\pm 0.5$ kpc is
well fit  by a simple linear function
\begin{eqnarray}
  \label{eq:24}
  \log t_R & = & 8.99 \pm 0.37 - (0.969 \pm 0.003)\log M_* \\ 
  & & - (0.948  \pm 0.004) \log  \sSFR
\end{eqnarray}
with a scatter of only 0.31 dex. In this equation, and in
Figures~\ref{fig:rgas_vs_various} and \ref{fig:tr_vs_various} the
stellar mass is the \emph{total} stellar mass, while the specific star
formation rate is calculated for the region sampled by the fibre, as
is the depletion time. If the stellar mass is measured within the
fibre, the coefficients would change to $(10.06, -1.04, -0.85)$ and if
both the stellar mass and the specific SFR are the galaxy integrated
values, the coefficients would be $(12.38, -1.10, 0.73)$.  It is
important to emphasise that the gas content in these equations are for
the central 2 kpc, and not for the galaxy as a whole; these are rather
different quantities and we do not attempt to correct $t_R$ for
aperture effects.

\subsection{Sample completeness and selection biases}
\label{sec:completeness}

In the preceding section we have given an overview of the gas content
and gas depletion times in the central regions of star-forming
galaxies from the SDSS. However we are unable to measure the gas
content of all galaxies and it is reasonable to ask to what extent our
results are representative for the galaxy population and not
significantly biased.

The main bias is that for a large number of galaxies we do not have
sufficient emission lines detected in the spectra to apply our
technique. There is also an additional biases for some galaxies with
emission lines where the ionisation source is not likely to be
stellar. In this case the modelling outlined in
section~\ref{sec:fitt-spectr-feat} is not applicable. Thus there is a 
set of galaxies which likely have gas, but for which we cannot measure
$\mugas$ using our method. 

\begin{figure*}
  \centering
  \includegraphics[width=184mm]{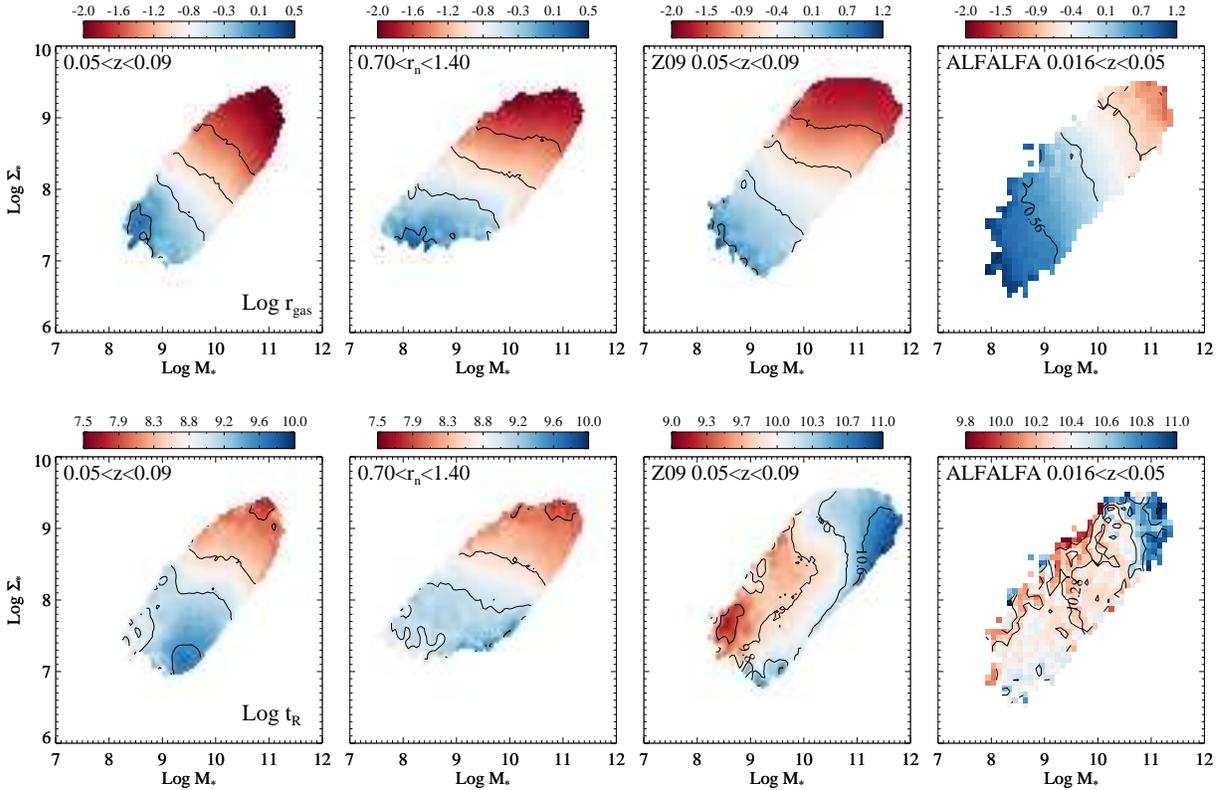}
  \caption{The variation of $\rgas$ (top row) and $t_R$ (bottom row)
    in the $\log M_*$ versus $\log \Sigma_*$ plane. The two left-most
    panels on each row are the same as shown in
    Figures~\ref{fig:rgas_vs_various}
    and~\ref{fig:tr_vs_various}. The third panel shows the mean $\log \rgas$ and
    $\log t_R$ in this plane using the Z09 photometric gas estimator,
    while the final panel shows the same using total gas masses from
    the ALFALFA $\alpha.40$ data release. Thus the two right-most
    panels show galaxy integrated quantities. Note that the behaviour
    for $t_R$ is clearly different for the total versus central
    measurements, while the \rgas\ trends are qualitatively similar,
    but the quantitative levels are somewhat different as the colour
    bars above shows} 
  \label{fig:logm_logmu_check_vs_a40_z09}
\end{figure*}

To illustrate the effect of the former bias, the second panel in each
pair of panels in Figures~\ref{fig:rgas_vs_various}
and~\ref{fig:tr_vs_various} shows a contour map of the fraction of
galaxies for which we could determine \mugas\ with our method. This
shows more or less the expected distribution: For blue and actively
star forming galaxies we have \mugas\ estimates for most galaxies,
while for red, massive and low sSFR galaxies we only have \mugas\
measurements for a very small subset. 

A reasonable way to think about this is that where we have \mugas\ for
more than 50 per cent of the galaxies our mean \rgas\ estimates likely
to be fairly representative for the galaxy population as a whole. This
level is indicated by a thicker black contour line in the plot. At
lower completeness levels we do not know exactly how representative
our measurements are, but as the main cause of incompleteness is a
lack of star formation those missed galaxies most likely have lower
gas content than those included in Figures~\ref{fig:rgas_vs_various}
and~\ref{fig:tr_vs_various}. In this case $\rgas$ will be an upper
limit to the true mean, but it is not immediately possible to say what
the implications are for the depletion time.

\begin{figure*}
  \centering
  \includegraphics[width=184mm]{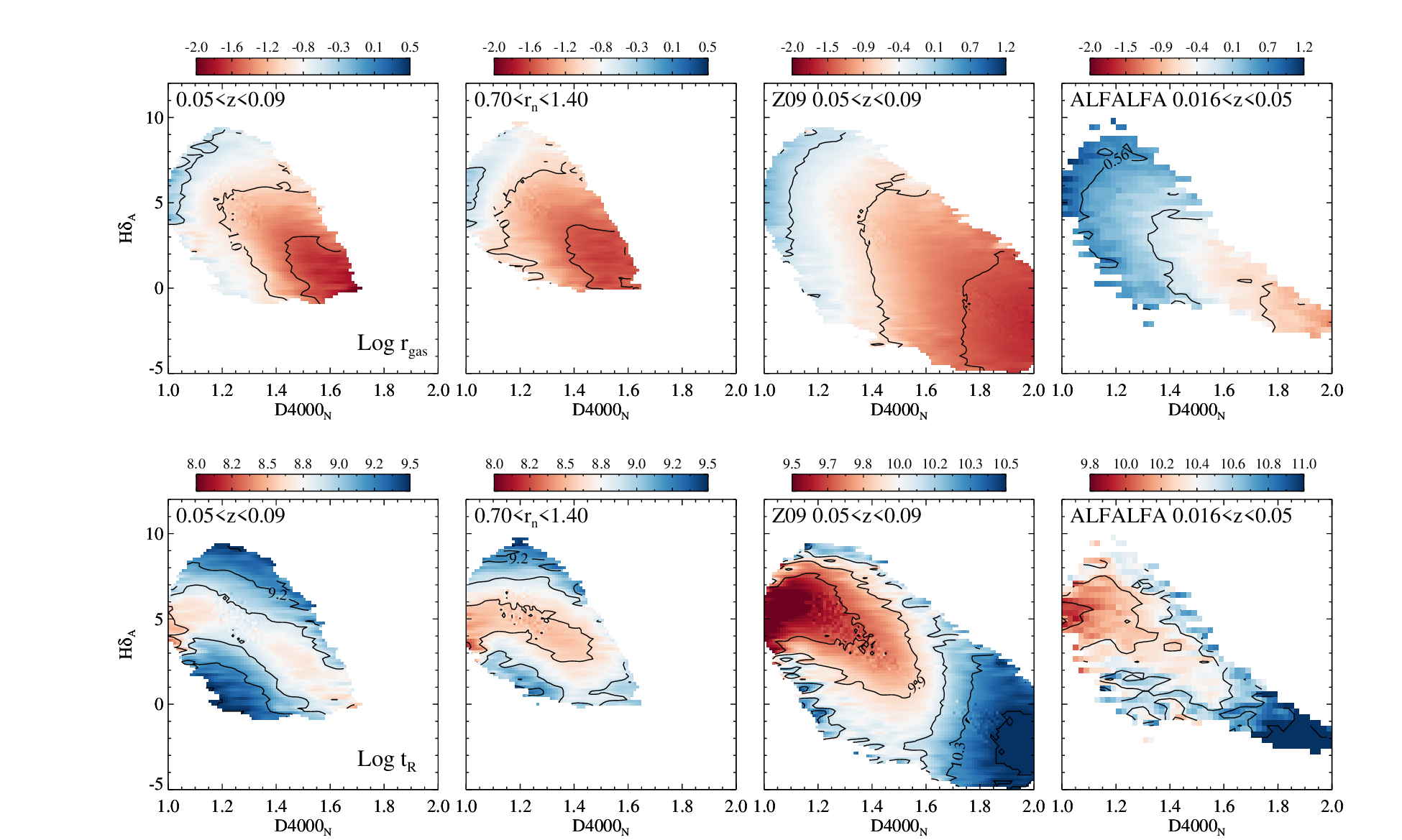}
  \caption{The same as Fig.~\ref{fig:logm_logmu_check_vs_a40_z09}
    but for the D4000$_N$ versus H$\delta_A$ plane. Note that there
    are clear similarities between the total and central distributions
    here, in particular a high depletion time both for high and low
    H$\delta_A$ at fixed D4000$_N$.} 
  \label{fig:d4_hda_check_vs_a40_z09}
\end{figure*}

A complementary method to test for the effect of incompleteness is to
use a different technique to estimate the gas content and gas
depletion time. We have two such methods available to us: gas content
measurements from the ALFALFA survey and gas estimates from fitting
formulae, in particular here the photometric method introduced by
Z09.  Unfortunately we can only use these methods to get total gas
content so we cannot do a direct comparison here, but they are still
very useful as a sanity check on the results. 

We focus on two distributions, and in
Fig.~\ref{fig:logm_logmu_check_vs_a40_z09} we show the variation of
\rgas\ (top row) and $t_R$ (bottom row) in the $\log M_*$ versus $\log
\Sigma_*$ plane for our method in the two leftmost panels in each
row. These are the same displayed in Figures~\ref{fig:rgas_vs_various}
and~\ref{fig:tr_vs_various} and are therefore estimates of central gas
properties. These can be contrasted to the total estimates using the
Z09 estimator in the third panel of each row and ALFALFA in the final
panel in each row.

The first point to note is that the $\rgas$ distributions are
qualitatively similar in each panel. In particular we would expect
that a 2 kpc radius (leftmost panel)  for the lowest mass galaxies
would approach the total gas content and indeed we see a fairly close
correspondence to the total estimates using the Z09 estimator.  We
also note that the ALFALFA sample is biased to more gas-rich galaxies
overall so differs significantly from the rest. 

The other striking result is that the distribution of $t_R$ is
substantially different for the total estimator and the central. Again
at low mass we can see a correspondence to the Z09 estimator and
possibly also the ALFALFA measurement, but note that the colour scales
are quite different.

A slightly different picture is offered by the D4000$_N$ versus
H$\delta_A$ plane shown in
Fig.~\ref{fig:d4_hda_check_vs_a40_z09}. The run of \rgas\ is again
qualitatively similar in each case. However in this case the $t_R$
distribution is also comparable and in particular we note that the
trend for higher depletion times at both low and high H$\delta_A$ at
fixed D4000$_N$ that we noted above, can also be seen, albeit at a
lower S/N. 

This is a somewhat surprising result. The normal interpretation of
strong Balmer absorption, here a high H$\delta_A$, is that it
indicates a recent cessation of or strong drop in star formation
activity \citep[e.g.][]{2009MNRAS.395..144W}. What we are finding is
that in these galaxies $\rgas$ is elevated (top row), particularly on
a 2 kpc scale and that this increase is not fully taken out by the
increase in star formation, thus $t_R$ increases. What
Fig.~\ref{fig:d4_hda_check_vs_a40_z09} implies is that many of these
systems still retain significant amounts of gas but that as the star
formation has dropped $t_R$ is high.

The population falling below the median line is less studied, but when
matched in D4000$_N$ and stellar mass to other star forming galaxies
they are seen to have a significantly, $\sim 0.7\;$dex, lower star
formation rate than galaxies on the median line. They are then
concieveably rejuvenated post-starburst galaxies.

These results are qualitatively consistent with a picture where most
galaxies have a steady decline in gas content as they evolve and where
inflow of gas leads to a star burst episode. In some cases when this
ends, the star formation is quenched but all the gas is not removed
from the system. The spread in $t_R$ on the mean
D4000$_N$-H$\delta_A$ track is quite substantial so the eventual fate
of these systems cannot be well constrained from just this plot but
this might offer an interesting constraint on the gas cycle in
galaxies.

\section{Conclusions}
\label{sec:conclusions}

In the preceding we have presented a method for obtaining
dust-to-metal and dust-to-gas ratios, as well as gas surface mass
densities in galaxies using optical spectra. We showed that the
dust-to-gas ratios provided by our method give consistent results to
those obtained from modelling of the dust emission although a detailed
one-to-one comparison was impossible with current data. We also showed
that our method provides gas content estimates that agree with
measurements of \hmol\ + \hi\ to within a factor of $\approx 2$. This
method requires emission lines that constrain the dust column density
and lines of elements that are both (mostly) unaffected by depletion
onto dust grains as well as elements that are depleted onto dust
grains and that are important coolants in the interstellar medium. In
practice this means that most strong optical emission lines from
\oii{3727} to \sii{6716,6731} should be measured.

The method provides total gas column densities unaffected by the
substantial uncertainty on \XCO, and because optical spectroscopy can
provide much higher spatial resolution than most interferometric
studies can, while at the same time providing star formation rates in
the same region, this offers potentially a great complement to \CO\
and \hi\ studies of nearby galaxies. It would for instance be possible
to compare our gas masses to \hi+\hmol\ masses to infer the
\CO-to-\hmol\ conversion factor for large samples, similar to what was
done for a sample of nearby galaxies by~\citet{2011ApJ...737...12L}. 

At the same time, rest-frame optical spectroscopy is fairly easy to
carry out for star-forming galaxies over a wide range in redshift. As
long as the spectra can be properly flux-calibrated, this opens up the
possibility to trace the \emph{central} gas content of galaxies over a wide
range in redshift.

However we saw that there are a few limitations of the method: At
metallicities below 50 per cent solar, our constraints on the dust-to-metal
ratio are very weak and we need to apply a prior on this
quantity. Overall this appears not to be a major issue as the effect
of applying the prior is small and in comparison to the uncertainty in
going from \CO\ to \hmol\ mass estimates, it is negligible.

The second issue is that at high gas column densities, $\mugas/\Msun\; 
\mathrm{pc}^2 > 75$--100, the method breaks down. This is simply
because the emission lines must probe through the gas-rich region in
order to provide information on it, but at very high column densities
the regions become opaque to optical radiation. This is a fundamental
limitation of this technique and does limit the usefulness of the
technique when studying the very central regions of nearby
galaxies. It is also likely that it will limit its usefulness to some
classes of more distant galaxies although any firm conclusions on the
latter must await a more careful assessment in future works. The
method is also likely to break down at very low surface mass densities
but we have not found this limit with our tests. From our tests we are
confident that the method works well in the regime $10 <
\mugas/\Msun\; \mathrm{pc}^{-2} < 50$--$70$, but the \mugas\ estimates
are well behaved at least down to $\mugas \sim 2
\Msun\;\mathrm{pc}^{-2}$ below which only 1 per cent of the SDSS DR7 galaxies
fall, but we are unable to test this directly.

The final issue is that our method can only provide information on gas
in regions where there is star formation going on. This is a key
reason why this method cannot be a replacement for CO and \hi\
measurements of galaxies.  However this is offset by the fact that,
particularly at higher redshifts, the method provides a much higher
spatial resolution than \hi\ and CO observations can provide, and it
can account for atomic gas whose detection at significant redshifts is
infeasible with current facilities, and thus complements the methods
very nicely.

We demonstrated the usefulness of this method by applying it to the
star-forming galaxies in the SDSS and presented a summary of the
variation in gas content and gas depletion times in the central $2\pm
0.5$ kpc of these galaxies. These trends in general follow the
trends for the overall gas content found in recent \hi\ and CO surveys
of nearby galaxies, so that the gas content of galaxies decreases with
stellar mass and stellar mass surface density. However the large
sample size has made it possible to highlight the complex variation of
gas content and gas depletion time with physical parameters.  In
particular we found that galaxies with high and low Balmer absorption
at a fixed D4000$_N$ appear to have high gas depletion times both in
the central regions as well as in integrated quantities, possibly
reflecting a feedback cycle associated with star bursting galaxies.

\section{Acknowledgements}
\label{sec:acknowledgements}

We thank Marijn Franx, Frank van den Bosch, Karl Glazebrook, Amelie
Saintonge, Vivienne Wild, Richard Ellis, Kevin Bundy and Marc
Verheijen, for discussions and input to this project.  We thank J.\
Moustakas for providing emission line fluxes for SINGS galaxies in a
machine readable form. We thank the anonymous referee for a
constructive report that has helped improve the clarity of the paper. 

This work has made use of SDSS data.  Funding for the SDSS and SDSS-II
has been provided by the Alfred P. Sloan Foundation, the Participating
Institutions, the National Science Foundation, the U.S. Department of
Energy, the National Aeronautics and Space Administration, the
Japanese Monbukagakusho, the Max Planck Society, and the Higher
Education Funding Council for England. The SDSS Web Site is
http://www.sdss.org/.

The SDSS is managed by the Astrophysical Research Consortium for the
Participating Institutions. The Participating Institutions are the
American Museum of Natural History, Astrophysical Institute Potsdam,
University of Basel, University of Cambridge, Case Western Reserve
University, University of Chicago, Drexel University, Fermilab, the
Institute for Advanced Study, the Japan Participation Group, Johns
Hopkins University, the Joint Institute for Nuclear Astrophysics, the
Kavli Institute for Particle Astrophysics and Cosmology, the Korean
Scientist Group, the Chinese Academy of Sciences (LAMOST), Los Alamos
National Laboratory, the Max-Planck-Institute for Astronomy (MPIA),
the Max-Planck-Institute for Astrophysics (MPA), New Mexico State
University, Ohio State University, University of Pittsburgh,
University of Portsmouth, Princeton University, the United States
Naval Observatory, and the University of Washington.

The colour maps in
Figures~\ref{fig:rgas_vs_various}--~\ref{fig:d4_hda_check_vs_a40_z09}make
use of ColorBrewer colour schemes from
\texttt{http://www.ColorBrewer.org }\citep{ColorBrewer}.

We gratefully acknowledge the software utilised at various stages of
this work: The Interactive Data Language (IDL), including the
Astronomy Library maintained by Wayne Landsman at GSFC, the GAIA image
analysis program and the Topcat table processing software
\citep{2005ASPC..347...29T}, both fruits of the now-defunct Starlink
project. At times we have also made use of the Perl Data Language
(\texttt{http://pdl.perl.org}), and much of the statistical analysis
has used or built on the R programming language for statistical
computing version 2.15.0 (\texttt{http://www.R-project.org}).

This research has made use of the NASA/IPAC Extragalactic Database
(NED) which is operated by the Jet Propulsion Laboratory, California
Institute of Technology, under contract with the National Aeronautics
and Space Administration.  This research has made use of the VizieR
catalogue access tool, CDS, Strasbourg, France. The original
description of the VizieR service was published in
\citet{2000A&AS..143...23O}.

\bibliographystyle{mn2e}
\bibliography{mxp_references}

\appendix

\section{Aperture corrections}
\label{sec:aperture_corr}

A key goal for our aperture correction scheme is to allow a direct
comparison of spectroscopic gas mass measurements using the 3'' SDSS
fibre and \hi\ mass estimates obtained with Arecibo, whose typical
beam size is 3.5' full-width half-power. Since the SDSS spectra rarely
sample even 30 per cent of the total light of the galaxies and it is
well-known that the \hi\ disks often extend significantly outside the
stellar disk, a direct comparison without taking into account aperture
effects would be very misleading.

In order to compare the total \hi\ gas masses to those derived from
optical spectroscopy we have to account for differences in
aperture. To carry out these aperture corrections we need \hi\ and
optical radial profiles for a wide range of galaxies. We take these
from two samples observed using the Westerbork array: the Ursa Major
sample (UMa) of Verheijen, Tully et al
\citep{1996AJ....112.2471T,2001A&A...370..765V}, and the WHISP survey
\citep{2005A&A...442..137N,2007MNRAS.376.1513N,2007MNRAS.376.1480N}.

For the UMa survey we take surface photometry data from
\citet{1996AJ....112.2471T} who presented results for \textit{B},
\textit{R}, \textit{I} and \textit{K}. We match this data to the \hi\
profiles from \citet{2001A&A...370..765V}, kindly supplied in
electronic form by M.\ A.\ W.\ Verheijen. For the WHISP survey we take
the \hi\ profile fits reported by \citet{2005A&A...442..137N} and
complement these with \textit{B}, \textit{R} \& \textit{I} profiles
derived from surface photometry by \citet{2007MNRAS.376.1480N}. For
both surveys we ignore uncertainties in the fits as other
uncertainties dominate.

Given the 2D distributions of \hi\ and light we can calculate the
gas-to-stellar mass ratio, \rgas:
\begin{equation}
  \label{eq:r_gas}
  \rgas(r<R) = \frac{\mugas(r<R)}{\mu_*(r<R)},
\end{equation}
the gas fraction, \fgas:
\begin{equation}
  \label{eq:fgas}
  \fgas(r<R) = \frac{\mugas}{\mu_*(r<R) + \mugas(r<R)},
\end{equation}
and the surface gas mass density: $\mugas(r<R)$. Here we indicate
explicitly that these values are integrated values inside a radius, $R$. 

This allows us to calculate aperture effects for these quantities as a
function of radius. Note that \fgas\ and \rgas\ are of course closely
related since
\begin{equation}
  \label{eq:1}
  \fgas = \frac{\rgas}{1+\rgas},
\end{equation}
but we will treat them as two separate quantities as \fgas\ depends
on the absolute value of the stellar mass-to-light ratio, $M_*/L$,
within the aperture, while \rgas\ only depends on the relative value
between the small and large aperture.

\begin{figure}
  \centering
  \includegraphics[angle=90,width=84mm]{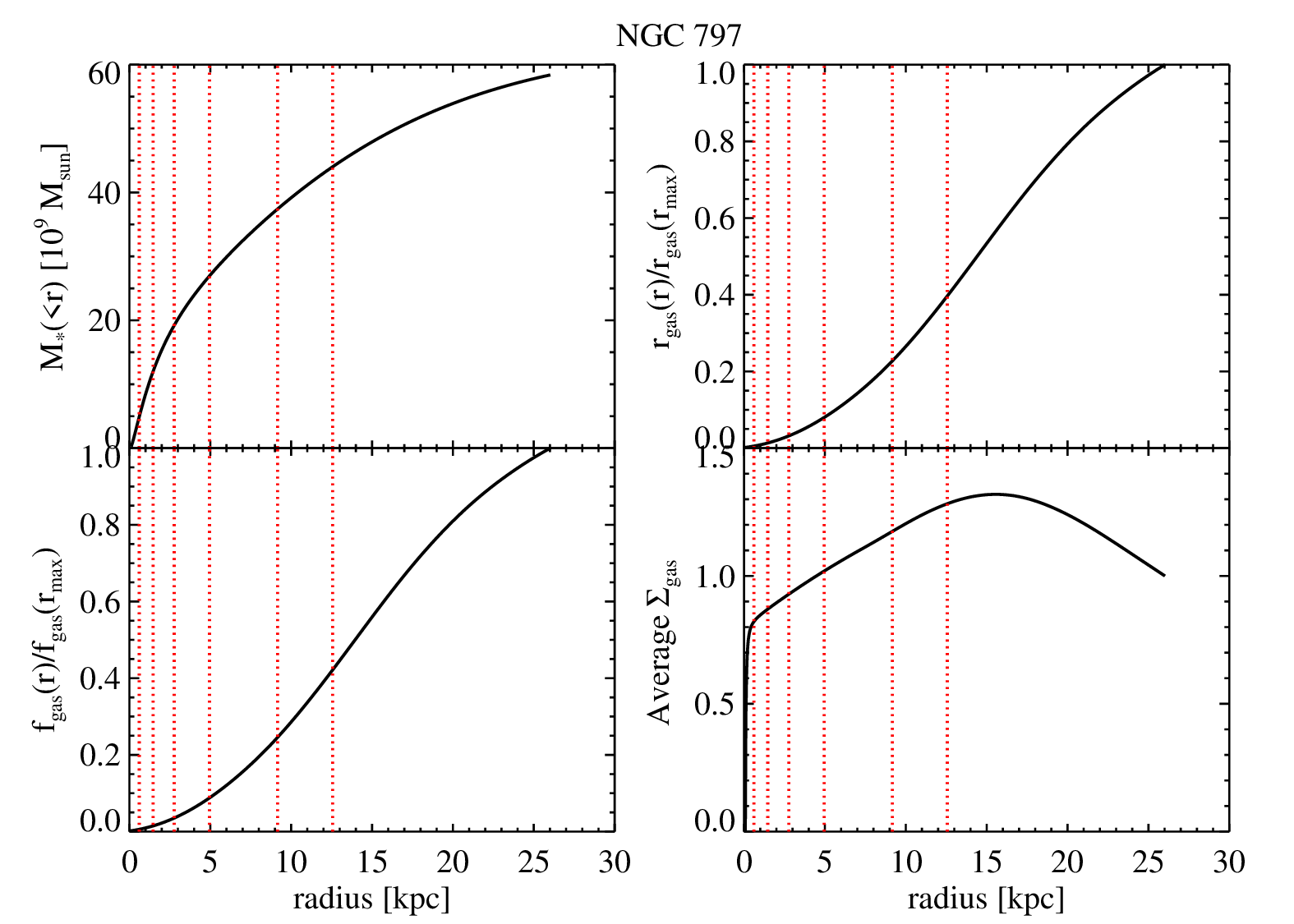}
  \caption{An illustration of the relevant profiles for aperture
    correction. The galaxy is NGC 797 from the WHISP sample and note
    that 'gas' here refers to \hi\ only. The top left panel shows the
    enclosed stellar mass as a function of radius, using a fixed
    $\Mstar/L_r=1.78$. The dotted vertical lines which are the same in
    all panels, indicate the radius corresponding to the radius of the
    SDSS 3'' fibres at $z=0.02, 0.05, 0.1, 0.2, 0.5$ and 2.0. The top
    right panel shows $A_r$ from equation~(\ref{eq:ap_r}), the bottom
    left the corresponding $A_f$ for the gas fraction, and the bottom
    right the average gas column density inside radius $r$,
    \mugas(r). From the figure it is clear that \mugas does not show a
    monotonous trend so cannot be (easily) aperture corrected, while
    both $A_r$ and $A_f$ are relatively well-behaved.}
  \label{fig:apcorr_example_ngc797}
\end{figure}

These quantities have quite different radial dependence, as
illustrated in Fig.~\ref{fig:apcorr_example_ngc797}.  As this figure
shows (lower right panel), the behaviour of \mugas\ might be far from
monotonic which means that this quantity is not suited for aperture
corrections. Fig.~\ref{fig:apcorr_example_ngc797} also shows a series of
dotted vertical lines. These show the radius sampled by the SDSS fibre
at redshifts 0.02, 0.05, 0.1, 0.2, 0.5 and 2.0, respectively. Since
the median redshift of the SDSS is close to 0.1, we can clearly see
that aperture corrections will be substantial regardless of indicator
chosen. 

It turns out that both \rgas\ and \fgas\ are suitable for aperture
correction, but as \rgas\ does not depend on the absolute value of the
$M_*/L$ ratio we choose to focus on this here and we write 
\begin{equation}
  \label{eq:ap_r}
  A_r = \rgas(R)/\rgas(R_{\mathrm{tot}}),
\end{equation}
where $R$ is the radius you wish to aperture correct \emph{to}, and
\rgas($R_{\mathrm{tot}}$) is the radius corresponding to the total
$\rgas$. 

In the following we will derive a method to estimate $A_r$ for
galaxies of different Hubble types, but it is worth commenting on
$R_{\mathrm{tot}}$ in equation~(\ref{eq:ap_r}). This cannot be related
to any quantity in the observed plane (such as the Arecibo) beam as
that would give different aperture corrections for the same galaxy at
different redshifts, so it has to be related to the physical
properties of the galaxy. We will take it to be related to the radii
containing 50 per cent and 90 per cent of the light below, but this is
a significant source of systematic uncertainty which we will see
below.

We will also ignore $M_*/L$ variations with radius, as well as the
molecular contribution to the total gas mass here. This is a
reasonable approximation in view of the systematic uncertainty just
mentioned, but also because their effects are opposite, including
molecular gas typically decreases $A_r$, while including $M_*/L$
variations typically increases $A_r$.

\begin{figure}
  \centering
  \includegraphics[width=84mm]{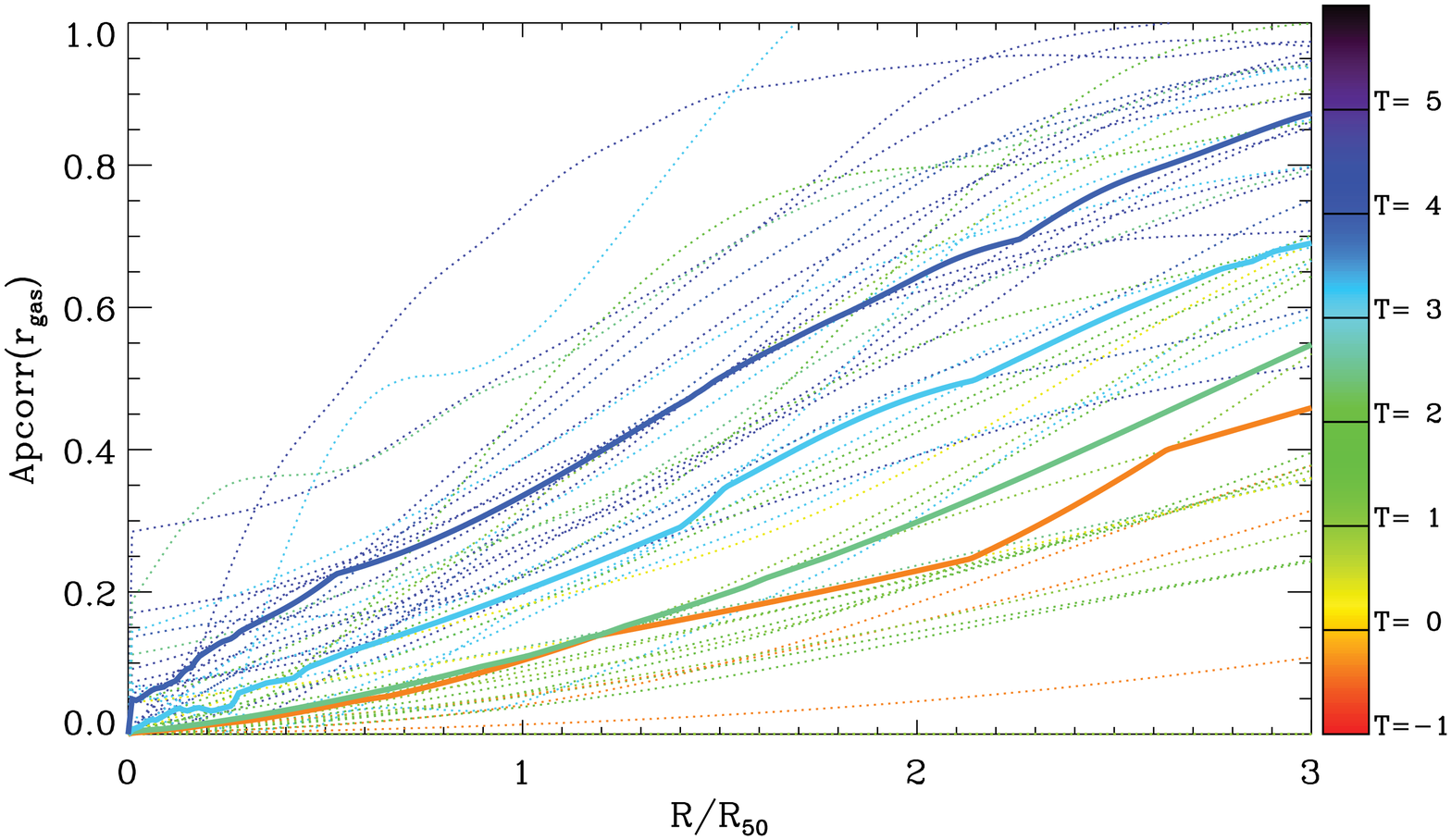}
  \caption{The aperture correction for $\rgas$ as a function of the
    radius and morphological type (indicated by colouring) for the
    combined WHISP and UMa samples. The thick solid lines show the
    average aperture correction profiles for four bins in T-type.}
  \label{fig:ap_corr_vs_T_type}
\end{figure}

However $A_r$ shows a significant dependence on galaxy
morphology. This is not unexpected as the gas content of galaxies has
long been known to show a significant variation with Hubble type
\citep[e.g.][]{1994ARA&A..32..115R}. This also extends to the
distribution of gas in galaxies, and in
Fig.~\ref{fig:ap_corr_vs_T_type} which shows all aperture correction
profiles, $A_r$ for the combined WHISP and UMa sample.

The thick solid lines show for clarity the average profiles in four
bins in T-type, which are from the top down, $[-1, 1)$, $[0, 2)$, $[2,
5)$ and $[6, 10)$. As is clear from this figure there is a very
noticeable type-dependence. However it is also a very clear spread
around the mean.  To explore the dependence on outer radius we also
extrapolate the trends shown by fitting polynomials to the thick solid
curves in Fig.~\ref{fig:ap_corr_vs_T_type}.

To take the scatter properly into account we calculate the likelihood
of a given aperture correction at a given radius for a given T-type,
$P(A_r | R/R_{50}, T)$. The likelihood distribution is derived from
the observed distribution of UMa and WHISP galaxies assuming an error
of 0.1 in $A_r(R)$ and a type uncertainty of 1 class. When we apply
this to the SDSS galaxies we convert the concentration, R90/R50, to
T-type using a linear fit to the data in \citet{2001AJ....122.1238S}
propagating the uncertainties on R90 and R50 to provide a likelihood
distribution for $T$. We use this to marginalise $P(A_r | R/R_{50},
T)$ over $T$.

\begin{figure*}
  \centering
  \includegraphics[width=184mm]{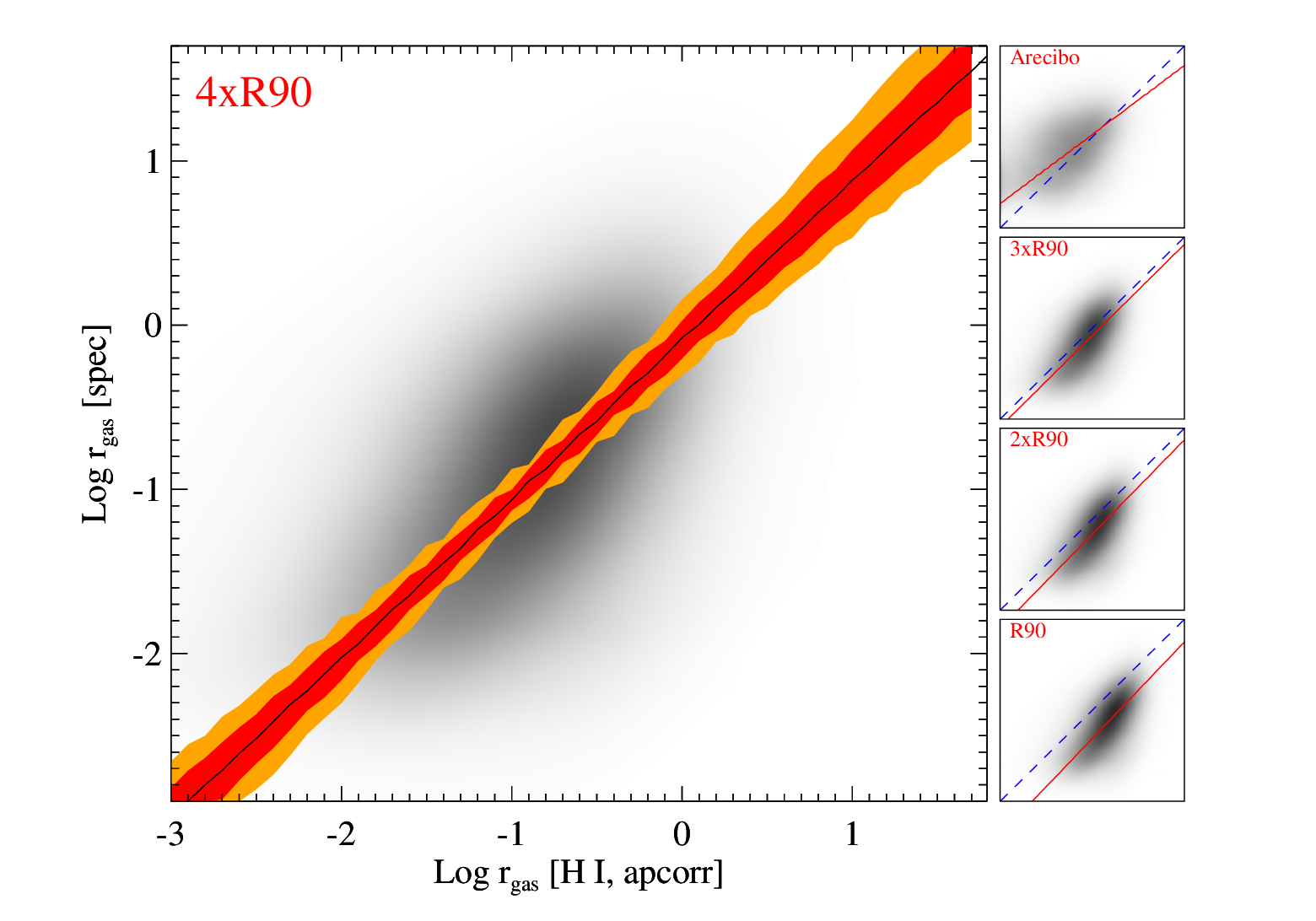}
  \caption{The main panel shows a comparison of $M_{\subhi}/M_*$,
    aperture corrected down to a 3" aperture, with our spectroscopic
    gas estimates. The grey scale shows the sum of the 2D PDF for this
    comparison, assuming independence. The shaded region shows the
    best fit linear fit to the data with the shading indicating 68 per cent
    and 95 per cent confidence intervals on the posterior predicted values
    (see Appendix~\ref{sec:aperture_corr} for details). For the main
    panel it was assumed that the total \hi\ resides within 4 times
    the radius containing 90 per cent of the light, $R_{90}$, while the small
    panels on the side show the same for an outer radius of $R_{90}$,
    $2\times R_{90}$, $3\times R_{90}$ and that the outer radius is
    identical to the Arecibo beam. The solid red line in each panel is
    the median fit using the Bayesian method described in
    Appendix~\ref{sec:bayes-line-regr}, while the dashed line shows
    the 1--1 line for comparison.}
  \label{fig:apcorr_hi_vs_specgass}
\end{figure*}

The result of applying this methodology to the galaxies in the SDSS
with $z>0.01$ and \hi\ detections from either S05, ALFALFA, or GASS,
can be seen in Fig.~\ref{fig:apcorr_hi_vs_specgass}. This figure
compares the aperture corrected values on the x-axis with our
spectroscopic measures on the y-axis. To construct the main panel we
have created a 2D PDF assuming the PDF of the quantity on the x and y
axis are independent, a very reasonable assumption, and then added
those 2D PDFs together. The shaded region shows the best linear fit to
the data using the Bayesian fitting technique described in
Appendix~\ref{sec:bayes-line-regr} with 68 per cent and 95 per cent posterior
confidence intervals indicated. In the main panel we have assumed an
outer radius for the \hi\ equal to $4\times R_{90}$, where $R_{90}$ is
the radius containing 90 per cent of the $r$-band light. 

The smaller panels show the effect of using other outer radii, with
the top panel using the 1.75' radius of the Arecibo beam, and the
three panels below using the indicated multiple of $R_{90}$. All
aperture correction schemes using an outer radius equal to a multiple
of $R_{90}$ results in a final relation that has a slope close to 1
(from 0.97 in the main panel to 1.06 for an outer radius equal to
$R_{90}$), but with a systematic variation in the zero point. In
contrast, assuming an outer radius equal to the Arecibo beam leads to
a poor aperture correction and a best-fit line with slope 0.76. 

We conclude from this that while the unit slope is indicative of a
good agreement between the aperture corrected \hi\ content and our
spectroscopic gas mass estimates, the systematic uncertainties in this
comparison are sufficient to render this a rather weak test of our gas
estimator.

\section{Bayesian linear regression with non-Gaussian error bars}
\label{sec:bayes-line-regr}

In Appendix~\ref{sec:aperture_corr} we need to fit a linear regression
line for data with heteroscedastic and non-Gaussian error bars. In the
case of heteroscedastic and Gaussian error bars, this is well known in
the literature as fitting a linear functional relationship and has
been used in astronomy for quite some time
\citep[e.g.][]{1996ApJ...470..706A,2007ApJ...665.1489K,2010MNRAS.tmp.1154A}. 

Thus we write our equations as 
\begin{eqnarray}
  \label{eq:9}
  x  & = & U + \epsilon_x \\
  y & = & V + \epsilon_y = a + b U + \mathrm{noise} \\
\end{eqnarray}
and where we assume that we know $\mathrm{PDF}(U+\epsilon_x) =
\mathrm{PDF}_x(x)$ and $\mathrm{PDF}(V + \epsilon_t) =
\mathrm{PDF}_y(a+bx)$. We can then write the probability distribution
for the slope and intercept as:
\begin{equation}
  \label{eq:10}
  P(a, b| x, y) \propto P(x, y| a, b) = \mathrm{PDF}_x(x)
  \mathrm{PDF}_y(a + b x).
\end{equation}
Where we have assumed uniform priors on $a$ and $b$, alternatively it
might be better to use a uniform prior on the angle
\citep[e.g.][]{2010MNRAS.tmp.1154A} but the constraints are tight on
$b$ so it is not crucial here.

Given the small samples required we calculate the parameters on a grid
rather than using Markov Chain Monte Carlo. We marginalise over $U$
and $V$ to get the PDFs for $a$ and $b$.  To get the confidence
interval in Fig.~\ref{fig:apcorr_hi_vs_specgass} we insert the PDFs for
$a$ and $b$ in equation~(\ref{eq:9}).

\section{The practicalities of creating 2D maps of gas content}
\label{sec:pract-creat-2d}

In section~\ref{sec:trends} we show a number of 2D maps of gas content
and depletion time in nearby galaxies. For completeness and
reproducibility this section describes in detail how these were
constructed. 

\begin{table}
  \centering
  \begin{tabular}{lcc}
\multicolumn{1}{c}{Map}                             &  Bin size & Kernel bandwidth \\
$\log M_*$ vs $\log \Sigma_*$                         &  [0.05, 0.04]  &  [0.20, 0.20] \\
$\log M_*$ vs $g-r$                                         &  [0.05, 0.02]  &  [0.20, 0.10] \\
$\log M_*$ vs $12 + \log \mathrm{O/H}$        &  [0.05, 0.02]  &  [0.20, 0.20] \\
$\log M_*$ vs $\log \mathrm{SFR}/M_*$           &  [0.05, 0.03]  &  [0.20, 0.20] \\
log \nii/\ha\ vs log \oiii/\hb  &  [0.02, 0.02]  &  [0.10, 0.10] \\
D4000$_\mathrm{N}$ vs H$\delta_A$                &  [0.01, 0.17]  &  [0.10, 0.40] \\
  \end{tabular}
  \caption{Bin sizes and kernel bandwidths used for the 2D gas maps in
  Figures~\ref{fig:rgas_vs_various}--~\ref{fig:d4_hda_check_vs_a40_z09}. Note
  that for ALFALFA the bin sizes is twice the width indicated here.}
  \label{tab:bandwidths}
\end{table}

For each panel all objects that were classified as star-forming using
the BPT diagram and for which we have \mugas\ estimates from the
spectra and that satisfies the specific redshift or $R/R_{50}$ cuts
are identified. For each diagram of $x$ versus $y$ we created a 2D
histogram for these objects with 100 bins in each direction where we
calculate the mean log \rgas\ or $\log t_R$ in each bin. The resulting
bin sizes are given in Table~\ref{tab:bandwidths}. We identify the
bins with less than $N+\delta$ objects and for the objects that fall
in these bins we use a 2D kernel density estimator using a Gaussian
kernel with the bandwidths given in Table~\ref{tab:bandwidths}, again
calculating the mean in each bin. We then replace all bins with less
than $N$ objects with this kernel estimates. We use $N=100$ and
$\delta=5$ for the plots in this paper. This combination of techniques
gives good resolution where there are many objects while improving the
S/N in the regions with few galaxies.  For display purposes we show
the maps only where the sum of the kernel contributions to that bin is
larger than 1.  We use the same method also for the maps using the Z09
photometric estimator and using the data from ALFALFA, but in the
latter case we use 50 bins in each direction given the much smaller
sample.

\end{document}